\documentclass[preprint,aps,amsmath,nofootinbib,tightenlines]{revtex4}

\usepackage{graphicx}
\usepackage{amsmath}
\usepackage{slashed}
\usepackage{bm}
\usepackage{xspace}
\usepackage{epsfig}

  \newcount\hour \newcount\minute
  \hour=\time \divide \hour by 60
  \minute=\time
  \newcommand{\mydate}{\ \today \ - \number\hour :\ifnum \minute<10 0\fi 
\number\minute}

\newcommand{\pb}{\overline{\mathcal{P}}}

\newcommand{\nb}{\bar{n}}
\newcommand{\nbs}{\slashed{\nb}}

\newcommand{\cB}{\mathcal{B}}
\newcommand{\cBp}{\mathcal{B}_{\! \bot}}
\newcommand{\cBn}{\mathcal{B}_n}
\newcommand{\cBpn}{\mathcal{B}_{\!n \bot}}
\newcommand{\cBps}{\slashed{\mathcal{B}}_{\! \bot}}

\newcommand{\cD}{{\mathcal D}}
\newcommand{\cP}{{\mathcal P}}

\newcommand{\mcdot}{\!\cdot\!}
\newcommand{\bn}{{\bar n}}
\newcommand{\bnP}{\overline {\mathcal P}}
\newcommand{\cPslash}{ {\mathcal P}\!\!\!\!\slash}

\newcommand{\Dslash}{D\!\!\!\!\slash}
\newcommand{\nslash}{n\!\!\!\slash}
\newcommand{\bnslash}{\bar n\!\!\!\slash}

\newcommand{\plus}{\ensuremath{\!+\!}}
\newcommand{\minus}{\ensuremath{\!-\!}}

\newcommand{\beq}{\begin{equation}}
\newcommand{\eeq}{\end{equation}}
\newcommand{\bal}{\begin{align}}
\newcommand{\eal}{\end{align}}
\newcommand{\bsp}{\begin{split}}
\newcommand{\esp}{\end{split}}
\newcommand{\ben}{\begin{enumerate}}
\newcommand{\een}{\end{enumerate}}
\newcommand{\bca}{\begin{cases}}
\newcommand{\eca}{\end{cases}}
\newcommand{\bpm}{\begin{pmatrix}}
\newcommand{\epm}{\end{pmatrix}}
\newcommand{\bbm}{\begin{bmatrix}}
\newcommand{\ebm}{\end{bmatrix}}
\newcommand{\bsm}{\begin{smallmatrix}}
\newcommand{\esm}{\end{smallmatrix}}
\newcommand{\bvm}{\begin{vmatrix}}
\newcommand{\evm}{\end{vmatrix}}
\newcommand{\bmx}[1]{\left(\begin{array}{*{#1}{c}}}
\newcommand{\emx}{\end{array}\right)}
\newcommand{\bbmx}[2]{\renewcommand{\arraystretch}{#2}\left[\begin{array}{*{#1}{c}}}
\newcommand{\ebmx}{\end{array}\right]}
\newcommand{\bpmx}[2]{\renewcommand{\arraystretch}{#2}\left(\begin{array}{*{#1}{c}}}
\newcommand{\epmx}{\end{array}\right)}
\newcommand{\bmxw}[1]{\renewcommand{\arraystretch}{2}\left(\begin{array}{*{#1}{c}}}
\newcommand{\bmxww}[1]{\renewcommand{\arraystretch}{2.5}\left(\begin{array}{*{#1}{c}}}
\newcommand{\bdet}[1]{\renewcommand{\arraystretch}{2.5}
    \left|\begin{array}{*{#1}{c}}}
\newcommand{\bndet}[1]{\renewcommand{\arraystretch}{1.5}
      \left|\begin{array}{*{#1}{c}}}
\newcommand{\edet}{\end{array}\right|\renewcommand{\arraystretch}{1}}
\newcommand{\bea}{\begin{eqnarray}}
\newcommand{\eea}{\end{eqnarray}}
\newcommand{\bit}{\begin{itemize}}
\newcommand{\eit}{\end{itemize}}

\newcommand{\nn}{\nonumber}

\def\qslash{q\!\!\!\slash}

\def\epsslash{\varepsilon\!\!\!\slash}
\def\Delslash{\Delta\!\!\!\slash}
\newcommand{\mklg}[1]{\mbox{\large $#1$}}
\newcommand{\mklga}[1]{\mbox{\Large $#1$}}

\newcommand{\SCETa}{\ensuremath{{\rm SCET}_{\rm I}}\xspace}
\newcommand{\SCETb}{\ensuremath{{\rm SCET}_{\rm II}}\xspace}
\newcommand{\OMIT}[1]{}
\newcommand{\OMITapp}[1]{}

\begin{document}
\setlength\baselineskip{15pt}

\preprint{ \vbox{ 
\hbox{arXiv:0809.1093} 
\hbox{MIT-CTP-3909} 
} }

\title{\phantom{x}\vspace{0.5cm}
Reparametrization Invariant Collinear Operators
\vspace{0.3cm}
}

\author{Claudio Marcantonini}
\email{cmarcant@mit.edu}
\affiliation{Center for Theoretical Physics, Massachusetts
Institute of Technology, Cambridge, MA 02139, USA
\vspace{0.3cm}}

\author{Iain W. Stewart\vspace{0.2cm}}
\email{iains@mit.edu}
\affiliation{Center for Theoretical Physics, Massachusetts
Institute of Technology, Cambridge, MA 02139, USA
\vspace{0.3cm}}

\begin{abstract}
\vspace{0.3cm}

In constructing collinear operators, which describe the production of energetic
jets or energetic hadrons, important constraints are provided by
reparametrization invariance (RPI). RPI encodes Lorentz invariance in a power
expansion about a collinear direction, and connects the Wilson coefficients of
operators at different orders in this expansion to all orders in $\alpha_s$.  We
construct reparametrization invariant collinear objects.  The expansion of
operators built from these objects provides an efficient way of deriving RPI
relations and finding a minimal basis of operators, particularly when one has an
observable with multiple collinear directions and/or soft particles.  Complete
basis of operators are constructed for pure glue currents at twist-4, and for
operators with multiple collinear directions, including those appearing in
$e^+e^-\to 3\,{\rm jets}$, and for $pp\to 2\,{\rm jets}$ initiated via
gluon-fusion.
 

\end{abstract}

\maketitle

\section{Introduction}

Factorization theorems play a crucial role in our understanding of
QCD~\cite{Collins:1989gx,Sterman:1995fz}.  For processes with large momentum
transfer or energy release they provide a separation of the high-energy
perturbative contributions from the low energy process independent functions
describing non-perturbative dynamics.  The soft-collinear effective theory
(SCET) provides a systematic approach to the separation of hard, soft, and
collinear dynamics in processes with energetic hadrons or
jets~\cite{Bauer:2000ew,Bauer:2000yr,Bauer:2001ct,Bauer:2001yt}. It has an
operator based approach to hard-collinear factorization which provides a simple
framework for deriving the convolution formulae connecting Wilson coefficients
and collinear operators. The hard Wilson coefficients describe the short
distance process dependent contributions, and the operators built out of
collinear and soft fields encode the longer distance hadronization into
individual energetic hadrons, energetic jets, or hadrons with soft momenta.
With more than one collinear direction the factorization for SCET operators was
first considered in Ref.~\cite{Bauer:2002nz}, and it was demonstrated that the leading
order operators efficiently encode traditional factorization theorems for
processes like Deep-Inelastic Scattering (DIS), Drell-Yan, Deeply-Virtual
Compton Scattering (DVCS), and exclusive form factors with hard momentum
transfer.  Compared to more traditional methods, an advantage of the effective
theory approach to high-energy factorization is the systematic description of
power corrections by higher order operators and effective
Lagrangians~\cite{Chay:2002vy,Beneke:2002ph,Hill:2002vw,Bauer:2003mg}.

An important constraint on the construction of both leading and power suppressed
operators in SCET is provided by reparametrization invariance (RPI).  The
utility of reparametrization invariance was first discussed in
Ref.~\cite{Luke:1992cs} in the context of heavy quark effective theory (HQET).
In HQET there are 3 generators for RPI, and the transformations involve a
time-like vector $v^\mu$ where $v^2=1$. For collinear operators in SCET, RPI
transformations act on null vectors $n^\mu$ and $\bn^\mu$ where $n\mcdot \bn=2$
and there are 5 generators for each type of collinear field.  Reparametrization
invariance in SCET was first discussed in Ref.~\cite{Chay:2002vy} and
generalized to the complete set of RPI transformations in
Ref.~\cite{Manohar:2002fd}.

\begin{figure} 
\begin{center}
\includegraphics[height=5cm]{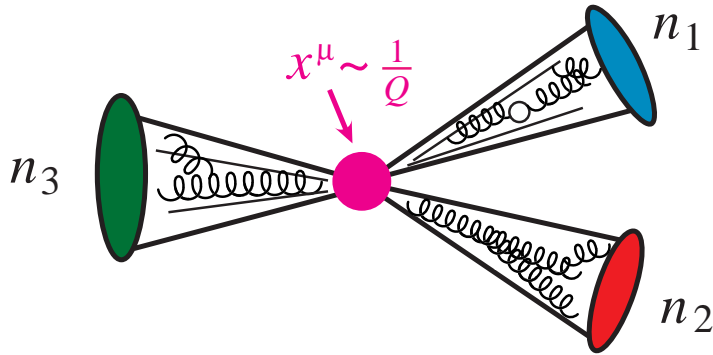} 
\end{center}
\vskip-18pt
\caption[1]{
Three collinear jets labeled by vectors $n_i^\mu$.
\label{fig:jets}}
\end{figure}

To see how reparametrization constraints come about, let's consider a process
with multiple energetic jets defined by an infrared safe jet algorithm, as pictured in
Fig.~\ref{fig:jets}. We assign labels $n_1^\mu$, $n_2^\mu$, $n_3^\mu$, $\ldots$
to the jets, which are null $n_i^2=0$, and whose vector components identify the
directions $\vec n_i$ of the total momentum vector of all hadrons in the jet.
The hadronization in each jet takes place in a collinear cone about each $\vec
n_i$, and we refer to the energetic particles in this jet as $n_i$-collinear.
Interactions between particles in different jets can take place only by hard
exchange at short distance or by soft exchange at long distance. The description
of the physics of a jet is simplified by a suitable set of coordinates, which
are provided by $n_i^\mu$, and a complementary null vector $\bn_i^\mu$ where
$\bn_i^2=0$ and $n_i \mcdot \bn_i =2$.  The momentum of a particle in the $i$'th
jet can be decomposed in these coordinates as
\begin{align} \label{p}
 p^\mu= n_i\mcdot p\: \frac{\bn_i^\mu}{2}+ \bn_i\mcdot p\: \frac{n_i^\mu}{2} 
   + p_{n_i\perp}^\mu\,.
\end{align}
The collinear modes for the jet have momentum scaling as $(n_i\cdot p,
\bn_i\cdot p, p_{n_i\perp}) \sim Q(\lambda^2, 1, \lambda)$ where $\lambda \ll 1$
and $Q^2$ is a large perturbative momentum scale (the jet-energy). In cases
where our discussion is generic to any one jet we will leave off the subscript
$i$, so $n_i\to n$ and $\bn_i\to \bn$. The definition of $\perp$ in
Eq.~(\ref{p}) is relative to $n$ and $\bn$, and for this reason we use the
notation $p_{n\perp}^\mu$, with $n\mcdot p_{n\perp}=\bn\mcdot p_{n\perp}=0$.
When it is clear which $n$ and $\bn$ we are referring to we will sometimes write
$p_{\perp}^\mu$ for $p_{n\perp}^\mu$.  For each $n$, collinear operators are
built up from quark $\xi_n$ and gluon $A_n^\mu$ fields, which are labeled by
their collinear direction, and describe quantum fluctuations close to the
direction $n$ with offshellness $p^2\ll Q^2$.  Two collinear directions are
described by distinct collinear fields when $n_i\cdot n_j \gg \lambda^2$ for
$i\ne j$~\cite{Bauer:2002nz}.  In Eq.~(\ref{p}) $\bn_i$ is introduced solely to
provide a basis vector for the decomposition, unlike $n_i$ which has a physical
association. For multiple collinear directions we have the freedom to introduce
multiple $\bn_i^\mu$ vectors.

Reparametrization constraints arise because the decomposition in Eq.~(\ref{p})
is not unique.  We can shift $n_i$ by a small amount and still have a suitable
basis vector for the $i$'th jet.  We also have a large amount of freedom in the choice
of $\bn_i$. For each $\{n,\bn\}$ pair the most general set of RPI
transformations which preserves the relations $n^2 = 0$, $\nb^2 = 0$, and $n
\mcdot \nb = 2$ are
\begin{eqnarray}\label{repinv}
\text{(I)} \left\{
\begin{tabular}{l}
$n_\mu \to n_\mu + \Delta_\mu^\perp$ \\
$\nb_\mu \to \nb_\mu$
\end{tabular}
\right.\qquad \text{(II)} \left\{
\begin{tabular}{l}
$n_\mu \to n_\mu$ \\
$\nb_\mu \to \nb_\mu + \varepsilon_\mu^\perp$
\end{tabular}
\right.\qquad \text{(III)} \left\{
\begin{tabular}{l}
$n_\mu \to (1+\alpha)\, n_\mu$ \\
$\nb_\mu \to (1-\alpha)\, \nb_\mu$
\end{tabular}
\right. \,,
\end{eqnarray}
where the five infinitesimal parameters are $\{\Delta^\bot_\mu, \epsilon^\bot_\mu,
\alpha\}$, and satisfy $\nb \mcdot \epsilon^\bot = n \mcdot \epsilon^\bot = \nb
\mcdot \Delta^\bot = n \mcdot \Delta^\bot = 0$. To ensure that $n$ provides an
equivalent physical description of the collinear direction for these particles
requires the power counting $\{\Delta^\bot_\mu, \epsilon^\bot_\mu, \alpha\}\sim
\{\lambda^1,\lambda^0,\lambda^0 \}$~\cite{Manohar:2002fd}. Thus $n$ can only be
shifted by a small amount, while parametrically large values of $\alpha$ and
$\varepsilon^\bot_\mu$ are allowed.  In B-meson decays, constraints from
reparametrization invariance in SCET have been derived for heavy-to-light
currents with parameters $v$ and $n$, at the first subleading order in
Refs.~\cite{Chay:2002vy,Beneke:2002ph,Pirjol:2002km}, and to second order in
Ref.~\cite{Arnesen:2005nk}. Results for light-light SCET currents with one
collinear direction $n$, were derived at first subleading order in
Ref.~\cite{Hardmeier:2003ig}. The extension of RPI relations to collinear
operators involving light quark masses was developed in Ref.~\cite{Chay:2005ck}.

The goal of our paper is to provide a simple procedure for constructing the
RPI-completion of operators $O(n_i,\bn_i,v_i)$ that depend on multiple
light-like vectors $\{n_i,\bn_i\}$ and time-like vectors $v_i$. The procedure
should be sufficiently general to be used for any hard-scattering process, and
also easy to extend to any desired order in the twist or $\lambda$ expansion. To
achieve this we must deal with a technical obstacle: so far all applications of
RPI to hard-scattering in the SCET and in other factorization literature have
constructed a complete basis of operators first and then dealt with deriving
connections between the operators order by order in the $\lambda$ expansion.
This approach quickly becomes cumbersome at higher orders or when dealing with
operators with multiple directions. For example, in this approach the RPI
completion of a basis of three jet operators
$O(n_1,n_2,n_3,\bn_1,\bn_2,\bn_3)$, would require studying three copies of
Eq.~(\ref{repinv}) or nine transformations.\footnote{ In Ref.~\cite{Hill:2004if}
  it was shown that the construction of heavy-to-light operators can be
  simplified if only operators in a particular frame are required, by taking
  linear combinations of the RPI transformations that only act in this frame. In
  Ref.~\cite{Arnesen:2005nk} this was described as the derivation of RPI
  conditions on a projected surface, and the complete set of such
  transformations was used for the ${\cal O}(\lambda^2)$ analysis done there.
  The formalism derived here makes a full analysis sufficiently simple that the
  consideration of projected surfaces becomes unnecessary.}

For cases with multiple time-like vectors, an alternative approach is known from
HQET~\cite{Manohar:2002fd}. Here an RPI heavy quark field is constructed at the
beginning, $H_v$, which has an expansion that starts with the standard HQET
field, $H_v = h_v + \ldots$.  A basis of reparametrization invariant operators
built from $H_v$ automatically encodes the RPI relations at any order in the
power expansion, and when expanded generates a series of operators with
connected Wilson coefficients. In this paper we develop a suitable set of RPI
and gauge invariant objects for SCET.  These objects include a quark field
operator $\Psi_n$, a gluon field strength operator ${\cal G}_n^{\mu\nu}$, and
$\delta$-function operators which pick out the large momenta of collinear
fields.  The gauge invariance of these objects is ensured using a
``reparametrization invariant Wilson line'' operator ${\cal W}_n$. These
objects allow us to extend the invariant operator procedure to processes that
depend on null-vectors.

In hard-scattering processes, DIS provides a familiar context where the
construction of a minimal operator basis requires judicial use of the quark and
gluon equations of motion, and an invariance under reparametrizations of a
light-like
direction~\cite{Politzer:1980me,Jaffe:1981td,Jaffe:1982pm,Ellis:1982wd,Ellis:1982cd},
for a review see~\cite{Jaffe:1996zw}. The invariance under reparametrizations
becomes more valuable at higher orders in the expansion, being particularly
constraining on the basis of twist-4 operators derived in

Refs.~\cite{Jaffe:1981td,Jaffe:1982pm,Ellis:1982wd,Ellis:1982cd}. We derive RPI
constraints for collinear operators in DIS and compare to these classic results
as a test of our setup. For DIS the minimization of the basis of RPI operators
is quite similar to the reduction of operators in Ref.~\cite{Jaffe:1982pm}.  On
the other hand the basis of SCET operators are comprised entirely of analogs of
``good'' quark and gluon fields, namely a two-component quark field $\chi_n$ and
just two components of the gluon field strength, ${\cal B}_{n\perp}^\mu$. These
objects both incorporate Wilson lines, and for these operators it is easier
to find a minimal basis. The RPI relations provide Lorentz invariance
connections between the Wilson coefficients in this basis.  These constraints
carry a process independence, they depend on the type of operators being
considered, but not on the precise process in which they will be used.  It
should be emphasized that when matrix elements are considered for a particular
process, a further reduction in the number of independent hadronic functions
becomes possible. For twist-4 quark operators in DIS this type of further
reduction was discussed in detail in Ref.~\cite{Ellis:1982cd} and for inclusive B-decays in.~\cite{Tackmann:2005ub}, but this type of
reduction is not our focus here.

Our construction is general enough that it applies not just to DIS like
processes, but to operators with multiple collinear directions, which are useful
for processes with multiple hadrons and jets.  These operator bases provide a
starting point for deriving appropriate factorization theorems for different
processes. The invariant operator procedure becomes more and more efficient as
the number of directions grows. 
 
The outline of our paper is as follows. In section~\ref{sec:SCET} we review
ingredients from SCET needed for our analysis.  We divide hard interactions into
two categories, those with an external hard leptonic reference vector $q^\mu$,
and those where the hard interaction is between strongly interacting particles.
Since most SCET applications focus on the former case, we address some of the
additional notational complications that occur for the latter. Section III
introduces a formalism for using reparametrization invariant objects in SCET.
A set of RPI invariant collinear objects is constructed in
section~\ref{sect:RPIGI}, followed by a summary of identities that can be used
to reduce the operator basis in section~\ref{sect:const}. The inclusion of mass
effects is considered in section~\ref{sect:rpim}, and the expansion of the RPI
objects is carried out in section~\ref{sect:R}.  Applications for constructing
operators are considered in section~\ref{sec:app}.  In section~\ref{sec:appS} we
verify that our approach provides a simple way to reproduce the known RPI result
for the chiral-even scalar current given in Ref.~\cite{Hardmeier:2003ig}.  In
section~\ref{sec:genQG} we construct a general basis of field structures
involving up to four active quark or gluon operators, and with up to four
distinct collinear directions. In section~\ref{sec:DISq} we consider the special
case of quark operators for DIS at twist-4 with one collinear direction, and
compare with the literature. In section~\ref{sec:DISg} we derive a basis of
operators for pure gluon scattering in DIS up to twist-4. Finally we apply the
formalism to jet production.  In section~\ref{sec:2jets} we demonstrate that
very little information is gained about the operator basis describing $e^+e^-
\to 2\,{\rm jets}$.  In section~\ref{sec:3jets} we show that RPI turns out to be
quite powerful for constraining the $e^+e^-\to 3\,{\rm jet}$ operators. Finally
we show that RPI is also useful for two jet production from gluon-fusion,
$gg\to q\bar q$, and we construct a basis of operators for this process in
section~\ref{sec:gfusion}.  Conclusions are given in section~\ref{sect:conclusion}.

\section{Review of SCET} 
\label{sec:SCET}

In sections~\ref{sect:SCETcol} and \ref{sect:SCETgi} below, we introduce some
basic definitions and properties of SCET that we will need for our computations.
In section~\ref{sect:RPI} a brief review of the null and time-like RPI
transformations is given, and in section~\ref{sect:convolutions} a review of
hard-collinear convolutions is given since they play an important role in
subsequent sections.

\subsection{Fields, Wilson lines, and Power Counting}
\label{sect:SCETcol}

SCET fields include collinear gluons $A_n^\mu$ and collinear quarks $\xi_n$ for
each distinct direction $n$.  An important attribute of the collinear fields is
that they carry both a large label momentum $p$ and a coordinate $x$, such as
$\xi_{n,p}(x)$.  The label momenta are picked out by momentum operators, $\pb_n
\xi_{n,p} = \bn\mcdot p\, \xi_{n,p}$ and ${\mathcal P}_{n\perp}^\mu \xi_{n,p}=
p_\perp^\mu \xi_{n,p}$, while derivatives $i\partial^\mu$ act on the coordinate
$x$ and scale as $i\partial^\mu\sim \lambda^2$ (see Ref.~\cite{Bauer:2001ct}).
Having two types of derivatives makes it simple to couple collinear and
ultrasoft particles for \SCETa, including ultrasoft gluons $A_{us}^\mu$ and quarks
$q_{us}$, and when appropriate, heavy quarks $h_v^{us}$ as well. The soft fields
for \SCETb are $A_s^\mu$, $q_s$, and a heavy quark $h_v^{s}$.

We define collinear covariant derivatives as
\begin{eqnarray}
  i\nb\mcdot D_n=\pb_n+g\nb\mcdot A_{n,p}\,,\qquad
  iD_n^{\perp\mu} = \mathcal{P}_{n\bot}^\mu +
  gA_{n,p}^{\perp\mu}\,,\qquad
 in\mcdot D_n = i n\mcdot \partial + g n\mcdot A_{n,p} \,.
\end{eqnarray}
When integrating out hard offshell fluctuations and constructing gauge invariant
structures in SCET, it is necessary to include collinear Wilson lines, $W_n$,
defined by
\begin{eqnarray} \label{W}
  W_n(x) & = & \Big[ \sum_{\textrm{perms}} 
   \exp\Big( \frac{-g}{\pb}\: \nb\mcdot A_{n,p}(x) \: \Big) \Big] \,.
\end{eqnarray}
The collinear fields $A_{n,p}^\mu$ are defined with the zero-bin
procedure~\cite{Manohar:2006nz}.
To couple ultrasoft degrees of freedom we define an ultrasoft covariant derivative
\begin{eqnarray}
  iD_{us}^\mu = i\partial^\mu + gA_{us}^\mu \,,
\end{eqnarray}
that can act on collinear fields. At lowest order the coupling to $n$-collinear
fields involves $n\mcdot D_{us}$ and can be removed from the Lagrangian by the
BPS field redefinition~\cite{Bauer:2001yt}
\begin{eqnarray} \label{fd}
 \xi_{n,p}(x) \to Y_n(x) \xi_{n,p}(x) \,,\qquad 
 A_{n,q}(x) \to Y_n(x) A_{n,q}(x) Y_n^\dagger(x) \,,
\end{eqnarray} 
with the ultrasoft Wilson line 
\begin{eqnarray} \label{Y}
 Y_n(x^\mu) = {\rm P} 
  \exp \Big( i\, g \int^0_{-\infty}
 \!\!
  ds \ n \mcdot
 A_{us}(x^\mu + s n^\mu ) \Big) \,.
\end{eqnarray}
This field redefinition allows us to organize power corrections as gauge
invariant products of collinear and ultrasoft fields as we discuss in the next
section. In describing $e^+e^-\to {\rm jets}$ in SCET it is convenient to make a
field redefinition with Wilson lines over $(0,\infty)$ rather than the $Y_n$ shown
in Eq.~(\ref{Y})~\cite{Bauer:2003di,Chay:2004zn}. For $q\bar q\to {\rm jets}$
one can use lines over $(0,\infty)$ or $(-\infty,0)$.  The final results are
always independent of the choice of the reference point for $Y$ in the field
redefinition (the $-\infty$ in Eq.~(\ref{Y})) since it does not dictate the
direction of the lines in the final result~\cite{Arnesen:2005nk} (though the
same choice should be used in all parts of the computation).

Operators are formed from products of the above fields, and the power counting
for an operator is determined by adding up contributions from its constituents.
The power counting for the fields and derivatives in \SCETa is\footnote{We will
  often suppress the labels on collinear fields when writing them out is not
  essential.}
\begin{align} \label{pc1}
 & \ \xi_n \sim \lambda, 
 &(n\mcdot A_n,\bn\mcdot A_n,A_n^{\perp}) & \sim (\lambda^2, 1,\lambda) \,,
 & q_{us} & \sim h_v^{us} \sim \lambda^3, 
 & A_{us} & \sim \lambda^2 \,, \nn\\
 & i\partial^\mu \sim \lambda^2 \,,
 & (in\mcdot \partial, \bn\mcdot \cP, \cP_{n\perp}) & \sim (\lambda^2,
 1,\lambda)\,,
 & W_n & \sim Y_n \sim \lambda^0 \,.
\end{align}
Here the ultrasoft fields describe fluctuations with offshellness much less
than the collinear particles.  These objects can be used to construct operators
for processes with multiple jets.  For a collinear jet we have $\lambda\sim
\Delta/Q$ with $\Lambda_{\rm QCD} \ll \Delta \ll Q$.  For a collinear hadron we
have a smaller $\lambda$, namely $\lambda\sim \Lambda_{\rm QCD}/Q$. For
processes with two or more hadrons the interactions in the theory \SCETb must be
considered. With a small parameter $\eta \sim \Lambda_{\rm QCD}/Q\ll 1$ the
power counting of fields in this theory are
\begin{align} \label{pc2}
 & \ \xi_n \sim \eta, 
 &(n\mcdot A_n,\bn\mcdot A_n,A_n^{\perp}) & \sim (\eta^2, 1,\eta) \,,
 & q_{s} & \sim h_v^{s} \sim \eta^{3/2}, 
 & A_{s} & \sim \eta \,, \nn\\
 & i\partial_s^\mu \sim \eta \,,
 & (in\mcdot \partial, \bn\mcdot \cP, \cP_{n\perp}) & \sim (\eta^2, 1,\eta)\,,
 & W_n & \sim S_n \sim \eta^0 \,.
\end{align}
Here the soft fields describe fluctuations with similar offshellness to the
collinear fields.  In cases with jets and energetic hadrons a succession of
\SCETa and \SCETb theories needs to be considered.

Our article focuses on building reparametrization invariant operators from
products of collinear fields that describe an underlying hard interaction, since
this is the most involved part of the construction.  The simple strategy we
follow to incorporate ``ultrasoft'' and ``soft'' fields into the analysis is
summarized in sections~\ref{sect:SCETgi} and \ref{sect:RPI} below.

\subsection{Gauge Invariant Field Products and Convolutions} \label{sect:SCETgi}

To build operators in SCET we want to use structures which are gauge invariant
and homogeneous in the power counting. Although the precise manner in which the
Wilson lines $W_n$ appears is determined by matching, and the precise manner in
which Wilson lines $Y_n$ appear is determined by ultrasoft-collinear
factorization, some general structures can be identified. For \SCETa a
convenient set of structures are:
\begin{align} \label{objs}
 & \chi_n \equiv W_n^\dagger \xi_n , \qquad 
 {\mathcal D}_n^\mu \equiv W_n^\dagger D_n^\mu W_n \,,  \\[3pt]
 &  q_{us}^n \equiv Y_n^\dagger q_{us} , \qquad
  {\mathcal D}_{us}^{n\mu} \equiv Y_n^\dagger D_{us}^\mu Y_n \,,\qquad 
  {\mathcal H}_v^n \equiv Y_n^\dagger h_v,  \nn
\end{align}
together with the $\cP_{n}^\mu$ label momentum operator and derivative operator
$i\partial^\mu$ acting on these gauge invariant structures.  The collinear
fields in Eq.~(\ref{objs}) are the ones obtained after the field redefinition in
Eq.~(\ref{fd}).  It is convenient to be able to switch the collinear derivatives
multiplied by Wilson lines for gauge invariant field strengths, for which we use
\begin{eqnarray} \label{toBD}
   i{\mathcal D}_n^{\!\perp\mu} 
  &=& \cP_{n\perp}^\mu  + g \cBpn^\mu \,, \qquad\qquad\!\!\!
  i\overleftarrow {\mathcal D}_n^{\!\perp\mu} 
  = - \cP_{n\perp}^{\,\dagger\mu} - g \cBpn^\mu   \,, \nn\\[6pt]
  i n\mcdot {\mathcal D}_n &=& in\mcdot\partial + g n\mcdot \cBn \,,\qquad
  i n\mcdot \overleftarrow{\mathcal D}_n 
   =in\mcdot\overleftarrow\partial - g n\mcdot \cBn \,,
\end{eqnarray}
and note that $\bn\mcdot {\cal D}_n = \bar \cP_n$.  Here the field strength
tensors are
\begin{equation}
 g \cBpn^\mu \equiv \Big[\frac{1}{\pb_n} [i\nb\mcdot {\mathcal D}_n,i
{\mathcal D}_n^{\perp\mu}] \Big] \,, \qquad\qquad 
 g n \mcdot \cBn \equiv \Big[\frac{1}{\pb_n}  [i\nb\mcdot {\mathcal D}_n,i n
\mcdot {\mathcal D}_n] \Big],
\end{equation}
where the label operators and derivatives act only on fields inside the outer
square brackets, and $g {\cal B}_{n\perp}^\mu$ and $gn\mcdot {\cal B}_n$ are
Hermitian.

For \SCETb with hadrons we have the same collinear invariant objects as in
Eq.~(\ref{sh}), and similar soft invariant objects, that are obtained by
replacing the ultrasoft fields by their soft counterparts, ${\cal
  H}_v^n=(Y_n^\dagger h_v^{us})\to {\cal H}_v^n=(S_n^\dagger h_v^s)$, ${\cal
  D}_{us}^{n\mu}\to {\cal D}^{n\mu}_s$, and $q_{us}^n=(Y_n^\dagger q_{us})\to
q_s^n=(S_n^\dagger q_s)$. The soft Wilson line $S_n^\dagger$ is generated by
integrating out offshell fluctuations which determine its direction $n$, and
outgoing/incoming boundary conditions. Most often these operators can be
constructed by a matching calculation from \SCETa, in which case the properties
of the soft Wilson lines are directly inherited from the ultrasoft ones in
\SCETa~\cite{Bauer:2002aj}, and the product of $C(Q^2,\omega_i)$ from
Eq.~(\ref{A}) and $J(\omega_i,k_j)$ from Eq.~(\ref{melt}) becomes the Wilson
coefficient of the factorized operator in \SCETb. In this paper we focus on
\SCETa examples.

\subsection{Reparametrization invariance} \label{sect:RPI}

When a set of fields have their largest momentum component in a light-like or
time-like direction then the structure of operators built from these fields is
constrained by reparametrization invariance. This invariance appears due to the
ambiguity in the decomposition of momenta in terms of basis vectors and in terms
of large and small components. For a collinear momentum, the set of five
transformations on the light-like basis vectors $n_i^\mu$ and $\bn_i^\mu$ were
given in Eq.~(\ref{repinv}). These infinitesimal changes preserve the relations
$n_i^2 = 0$, $\nb_i^2 = 0$, $n_i \mcdot \nb_i = 2$, and with the power
counting $\{\Delta^\perp,\varepsilon^\perp,\alpha\}\sim
\{\lambda,\lambda^0,\lambda^0\}$ can have no physical consequences on the
description of an observable. The type-III boost simply ensures that
$(\#Nn_i)-(\#N\bn_i)-(\#Dn_i)+(\#D\bn_i)=0$ for each $i$, where $(\#Nn_i)$
counts the number of $n_i$ factors in the numerator of an operator,
$(\#D\bn_i)$ counts the numbers of $\bn_i$ factors in the denominator, etc. With
three collinear directions an example of a type-III RPI invariant parameter is
\begin{eqnarray}
 \frac{ n_1\mcdot \bn_2 \ \bn_1\mcdot \bn_3 }{\bn_2\mcdot \bn_3} \,.
\end{eqnarray} 
The type-I and type-II transformations of collinear objects are more interesting
and are summarized in Table~\ref{table123}, which we take from
Ref.~\cite{Manohar:2002fd}. Since the factors induced by these transformations
occur at different orders in $\lambda$, demanding overall invariance of a
physical process provides connections between the Wilson coefficients of
operators at different orders in the expansion.

\begin{table}[t!]
\begin{center}
\begin{tabular}{|rcl|rcl|}
\hline
 \multicolumn{3}{|c|}{Type (I)} & \multicolumn{3}{c|}{Type (II)} 
   \\ \hline
 $n$ & $\to$ & $n+\Delta^\perp$ & $n$ & $\to$ & $n$ 
 \\
 $\bn$ & $\to$ & $\bn$ & $\bn$ & $\to$ & $\bn+\varepsilon^\perp$ 
 \\
 \ $n\mcdot D_n$ & $\to$ & $n\mcdot D_n+\Delta^\perp\mcdot 
 D_n^\perp$ 
  & $n\mcdot D_n$ & $\to$ & $n\mcdot D_n$ 
 \\
 $D_{n\perp}^\mu$ & $\to$ & $D_{n\perp}^\mu - 
 \mklg{\frac{\Delta_\perp^\mu}2} \: \bn\mcdot D_n - \mklga{\frac{\bn^\mu}2}
 \Delta^\perp \mcdot D_{n\perp}$ 
 \ &\  
  $D_{n\perp}^\mu$ & $\to$ & $D_{n\perp}^\mu - 
 \mklga{\frac{\varepsilon_{\perp}^{\mu}}2}\,  \:
  n\mcdot D_n-\mklga{\frac{n^\mu}2} \, \varepsilon^\perp\! \mcdot 
 D_n^\perp$
 \\
 $\bn\mcdot D_n$ & $\to$ & $ \bn\mcdot D_n$ 
 &\ $\bn\mcdot D_n$ & $\to$ & $ \bn\mcdot D_n + 
  \varepsilon^\perp\mcdot D_n^\perp$ 
 \\
  $\xi_n$ & $\to$ & $\left(1 + \frac14\,\Delslash^\perp\bnslash \right)\xi_n$&
  $\xi_n$ & $\to$ & $\left(1 + \frac12\, \epsslash^\perp 
  \frac{1}{\textstyle  \bn\mcdot D_n}\: {\Dslash}\,{}_n^\perp \right)\,\xi_n$
 \\[3pt]
 $W$ & $\to$ & $W$ 
 & $W$ & $\to$ & $\Big[ \big(1-\frac{1}{\textstyle \bn\mcdot D_n} 
  \,\, \epsilon^\perp\mcdot D_n^\perp \big) W \Big]$  
 \\[4pt]
 \hline
\end{tabular}
\end{center}
{\caption{Summary of infinitesimal type I and II transformations from 
Ref.~\cite{Manohar:2002fd}. With multiple collinear directions these transformations 
exist for each $\{n_i,\bn_i\}$ pair.}
\label{table123} }
\end{table}

When we couple collinear and ultrasoft particles there is another ambiguity,
associated with the decomposition of a collinear momentum into large and small
pieces.  If the total momentum $P^\mu$ of a collinear particle is decomposed
into the sum of a large collinear $p^\mu$ and a small ultrasoft momentum $k^\mu$:
\begin{eqnarray}
P^\mu &=& p^\mu + k^\mu 
  = \frac{n^\mu}{2} \nb \mcdot (p + k) + \frac{\nb^\mu}{2} n \mcdot
k + (p_\bot + k_\bot)^\mu,
\end{eqnarray}
then operators must be invariant under a transformation that takes $\bn\mcdot
p\to \bn\mcdot p + \bn\mcdot \ell$, $p_\perp^\mu\to p_\perp^\mu+\ell_\perp^\mu$,
$\bn\mcdot k\to \bn\mcdot k-\bn\mcdot \ell$, and $k_\perp^\mu \to k_\perp^\mu -
\ell_\perp^\mu$.  To construct invariant objects that have nice gauge
transformation properties we use the combined covariant
derivatives~\cite{Bauer:2003mg,Beneke:2002ni}, 
\begin{align}
 & i D_{n\perp}^\mu + W_n i D_{us\perp}^\mu W_n^\dagger \,,
 & & i \bn\mcdot D_{n} + W_n i\bn\mcdot D_{us} W_n^\dagger \,.
\end{align}
This can be implemented by taking
\begin{eqnarray} \label{rmD}
  i {\cal D}_{n\perp}^\mu \to 
  i {\rm D}_{\rm full}^{\perp\mu} =  i {\cal D}_{n\perp}^\mu +  i D_{\rm us}^{\perp\mu} \,,
  \qquad
  \pb_n \to i \bn\mcdot {\rm D}_{\rm full} =  \pb_n + i \bn\mcdot D_{\rm us} \,,
\end{eqnarray}
and then expanding in $\lambda$. The results in Eq.~(\ref{rmD}) give powerful
relations as they relate the coefficients of operators involving collinear
fields to those involving ultrasoft fields. These relations are quite easy to
derive order by order in $\lambda$. Note that reparametrization constraints
associated with transformation of the ultrasoft Wilson line $Y_n$ are
automatically enforced by the other constraints.\footnote{For example, prior to
  the field redefinition only the combination $in\mcdot {\rm D}=in\mcdot\partial
  +g n\mcdot A_{us} + gn\mcdot A_n$ appears acting on collinear fields. A type-I
  transformation connects this to a ${\cal D}^\perp_n$, and Eq.~(\ref{rmD}) then
  connects this to the same $i D_{\rm us}^{\perp}$ that one would find by direct
  transformation of $Y_n$.}  

Finally we review RPI for a time-like vector from HQET~\cite{Luke:1992cs}.  The
momentum $P^\mu$ of a heavy quark is decomposed as $P^\mu = m v^\mu + k^\mu,$
where $m$ is the heavy quark's mass, $v^\mu$ is its velocity, and $k^\mu$ is a
residual momentum of order $m\lambda^2$.  For an infinitesimal $\beta^\mu\sim
\lambda^2$ with $v\mcdot \beta=0$, the shifts
\begin{equation}
v^\mu \rightarrow v^\mu + \beta^\mu \quad\text{and}\quad k^\mu
\rightarrow k^\mu - m \beta^\mu,
\end{equation}
can have no physical consequences. This implies invariance under the
infinitesimal change $h_v\to h_v+\delta h_v$ with $\delta h_v = (im \beta\mcdot x
+ \beta\!\!\!\slash/2) h_v$. A superfield can be constructed which is invariant
under the full transformation~\cite{Luke:1992cs}
\begin{eqnarray}
  H_v(x) =  e^{-imv\cdot x} \bigg[ \frac{1}{\sqrt{2}(1+v\mcdot{\cal V}/|{\cal V}|)} 
     \Big( 1+ \frac{\slash\!\!\!{\cal V}\, \slash\!\!\! v}{|{\cal V}|}\Big) \bigg]
    h_v(x)  \label{HRPI}
\end{eqnarray}
where
\begin{align}
 {\cal V}^\mu&=v^\mu + i D_{\rm us}^\mu/m\, .\label{VH}
\end{align}
Using this superfield one can build operators $O=O[H_v(x),D^\mu]$ that are
invariant under reparametrizations of the time-like vector. Here $H_v =
e^{-imv\cdot x}[1 + i\Dslash/(2m)+\ldots ] h_v$ at the first non-trivial order.
Note that for heavy quarks, no dynamic component of the momentum is the same size
as the hard fluctuations, so there is no analog of the $\delta$-functions in
Eq.~(\ref{sh}).  This is the main complication we face in constructing invariant
operators in SCET. The closest one gets in HQET is when we have two
auxiliary time-like vectors, $v$ and $v'$, such as in $B\to D^{(*)}$ decays.
Here the invariant Wilson coefficients must be functions $C({\cal V}\mcdot {\cal
  V}')$~\cite{Neubert:1993mb}.

\subsection{Convolutions} \label{sect:convolutions}

In the presence of collinear fields a hard interaction can introduce
convolutions in variables $\omega_i$ between the perturbatively calculable
Wilson coefficient $C(Q^2,\omega_i)$ and the matrix element of the collinear
operators. In this case the amplitude, cross-section, or decay rate has the form
\begin{eqnarray} \label{A}
  A = \int\! [d\omega_1\cdots d\omega_k]\: 
   C(Q^2,\omega_i)\:  \langle O(\omega_i) \rangle \,.
\end{eqnarray}
The convolutions occur because a component of the hard momentum and of one or
more collinear momenta are ${\cal O}(\lambda^0)$. The exchange of momentum
between the hard and collinear components yields a convolution in variables
$\omega_i$, where the number of such variables is constrained by gauge
invariance and by momentum conservation in the matrix element. A gauge
invariant momentum from the collinear fields can be picked out by a delta
function acting on one of the collinear objects in Eq.~(\ref{objs}), such as
$[\delta(\omega-\bn\mcdot {\cal P}_n) \chi_n]$, and traditionally in SCET a
subscript notation is used for these products, 
\begin{align} \label{sh0}
 \chi_{n,\omega}
  &\equiv \Big[ \: \delta\big(\omega- \pb_n\big) \chi_n \Big]\,, 
 & (i{\cal D}^\mu_{n\perp})_\omega &\equiv 
   \Big[  i{\cal D}_{n\perp}^\mu \, 
   \delta\big(\omega- \pb_n^\dagger\big)\Big]
  \,, \nn\\[3pt]
 ( g \cB_{n\perp}^\mu)_{\omega}
 &\equiv \Big[ g \cB_{n\perp}^\mu\:
  \delta\big(\omega- \pb_n^\dagger \big) \Big]\,, 
 & ( g n \mcdot \cB_n)_{\omega}
 &\equiv \Big[ g n \mcdot \cB_n\:
  \delta\big(\omega - \pb_n^\dagger\big) \Big]\,.
\end{align}
We will refer to these as homogeneous objects since they have a definite order
in $\lambda$, and call the operators build from these objects homogeneous
operators.  As an example we have the bilinear scalar operator,
\begin{eqnarray} \label{Oeg}
  O(\omega_1,\omega_2) =
    \bar\chi_{n,\omega_1} \,\frac{\bnslash}{2}\, \chi_{n,\omega_2} \,.
\end{eqnarray}

When we consider RPI it will be convenient to use different $\delta$
functions and convolution variables $\hat\omega$, that are type-III invariant.
Essentially each $\pb_n=\bn\mcdot {\cal P}_n$ must be multiplied by a scalar
transforming as $n$ under RPI type-III.  There are two cases to consider:
\begin{itemize}
\item[i)] situations where there is a reference vector $q^\mu$
for the hard interaction, $|q^2| = Q^2 \gg\Lambda_{\rm QCD}$, which is external
to the QCD dynamics, 
\item[ii)] situations where the hard interactions are purely from
strongly interacting particles.
\end{itemize}

Case i) applies to examples such as DIS where $q^\mu$ is the momentum transfer
from the virtual photon, or $e^+e^-\to \ {\rm jets}$ where $q^\mu$ is the four
momentum of the $e^+e^-$ pair. Here we can use $n\mcdot q\sim \lambda^0$ to make
the $\delta$-function type-III invariant for $n$-collinear fields. Since $Q^2
\gg \Delta \Lambda_{\rm QCD}\gg \Lambda_{\rm QCD}^2$ we know that $n\mcdot q \gg
n\mcdot p$, where $p$ is the momentum of a collinear particle in the jet. Thus
we use a variable $\hat\omega$ with mass dimension two, and will find
$\delta$-functions of the form\footnote{ For $B$-decays these type-III invariant
  $\delta$-functions were used in Ref.~\cite{Pirjol:2002km}, with $q^\mu\simeq
  m_b v^\mu$, $\delta(\hat\omega-n\mcdot q\, \pb_n) = \delta(\hat\omega-m_b
  n\mcdot v\, \pb_n) = 1/m_b\,\delta(\hat\omega'-n\mcdot v\,\pb_n)$, where
  $\hat\omega=m_b\hat\omega'$. This form of invariant $\delta$-function was also
quite useful for analyzing the factorization theorem for $e^+e^-\to J/\psi X$ in
Ref.~\cite{Fleming:2003gt}.}
\begin{align} \label{deltaw}
  \delta\big(\hat\omega - n\mcdot q\, \pb_n\big) \,.
\end{align}
We also introduce a subscript notation with hatted variables,
\begin{align} \label{sh}
 \chi_{n,\hat\omega}
  &\equiv \Big[ \: \delta\big(\hat\omega\!-\!n\mcdot q\, \pb_n\big) \chi_n \Big]\,, 
 & (i{\cal D}^\mu_{n\perp})_{\hat\omega} &\equiv 
   \Big[  i{\cal D}_{n\perp}^\mu \, 
   \delta\big(\hat\omega\!-\!n\mcdot q\, \pb_n^\dagger\big)\Big]
  \,, \nn\\[3pt]
 ( g \cB_{n\perp}^\mu)_{\hat\omega}
 &\equiv \Big[ g \cB_{n\perp}^\mu\:
  \delta\big(\hat\omega\!-\!n\mcdot q\, \pb_n^\dagger \big) \Big]\,, 
 & ( g n \mcdot \cB_n)_{\hat\omega}
 &\equiv \Big[ g n \mcdot \cB_n\:
  \delta\big(\hat\omega\!-\!n\mcdot q\, \pb_n^\dagger\big) \Big]\,.
\end{align}
Since $\delta(\hat\omega-n\cdot q \pb_n)\sim \lambda^0$, it is leading order in
the power counting. Furthermore, we have $\delta(\hat \omega - n\cdot q \, \pb)=
\delta(\hat\omega/n\cdot q - \pb)/n\cdot q$, so identifying $\hat\omega =
n\mcdot q\: \omega$ there is no real change to the structure of Eq.~(\ref{A}).
An operator built out of the components given in Eq.~(\ref{sh}) has multiple
labels, $O(\hat\omega_1,\hat\omega_2,\ldots)$, and the Wilson coefficient for
the operator will be a function of the same parameters,
$C(\hat\omega_1,\hat\omega_2,\ldots)$, yielding Eq.~(\ref{A}) with
$\hat\omega$'s replacing $\omega$'s.

For processes in case ii) there is no analog of the external $q^\mu$. Examples
here include $pp\to {\rm jets}$, or any other hard process that does not involve
external leptons or photons. The key difference with case i) is that here the hard
interaction must involve two or more collinear directions, so we are guaranteed
that there are scalar products $n_i\mcdot n_j\sim\lambda^0$. For this type of
reaction the type-III invariant $\delta$-functions which are convoluted with
Wilson coefficients always involve large momenta for two different collinear
directions,
\begin{align} \label{deltaij}
  \Delta_{ij} \equiv
  \delta\Big(\hat\omega_{ij} - \frac12 n_i\mcdot n_j \, \pb_{n_i} \pb_{n_j} \Big) \,.
\end{align}
Here $\pb_{n_i}$ acts on a gauge invariant block of $n_i$-collinear fields, and
$\pb_{n_j}$ acts on a block of $n_j$-collinear fields. Since this
$\delta$-operator does not act on a single block of collinear fields we will not
use a subscript notation like Eq.~(\ref{sh}) for $\hat\omega_{ij}$. In this case
 the structure of the factorization theorem between operators and Wilson
 coefficients is a bit different than in Eq.~(\ref{A}). For example, consider an
 operator with collinear objects for four directions, where the convolution is
\begin{align} \label{AA}
  \int \big[\prod_{ij}\,  d\hat\omega_{ij} \big]\:  C(\hat\omega_{ij})
  [ \prod_{km}\, \Delta_{km}]  \ 
  \bar\chi_{n_1} (g \cB_{n_3}^\perp) (g \cB_{n_4}^\perp) \chi_{n_2}
  \,.
\end{align}
Here the products are over the six unique pairs $ij$ with $i\ne j$, and
$\pb_{n_i}$ in the $\Delta_{km}$ acts on the $n_i$-collinear field(s).  The
convolutions in Eq.~(\ref{AA}) can be manipulated into the form of Eq.~(\ref{A})
by inserting four factors of $1=\int d\omega_i\: \delta(\omega_i-\pb_{n_i})$,
writing $\hat\delta_{ij}=\delta(\hat\omega_{ij} - n_i\cdot n_j\, \omega_i\omega_j/2)$
and carrying out the integrals over the six $\hat\omega_{ij}$'s to give
\begin{align} \label{AAA}
  \int \big[d\omega_1\cdots d\omega_4 \big]\:  C\big(n_i\mcdot n_j\, \omega_i
  \omega_j \big)
  \bar\chi_{n_1,\omega_1} (g \cB_{n_3,\omega_3}^\perp) (g
  \cB_{n_4,\omega_4}^\perp) \chi_{n_2,\omega_2}
  \,.
\end{align}
Here the RPI-III transformation of the measure cancels against that of the
$\delta$-functions in the operator, and RPI has constrained the Wilson
coefficients to only depend on invariant products $n_1\mcdot
n_2\omega_1\omega_2$, $n_1\mcdot n_3\omega_1\omega_3$, etc.

Due to the simplicity of the ultrasoft-collinear coupling at leading order in
SCET a further factorization of the EFT matrix element can be made into
collinear pieces $J$, and ultrasoft pieces $S$ at each order in the power
counting:
\begin{eqnarray} \label{melt}
  \langle O(\omega_i) \rangle = \int dk_j \: J(\omega_i,k_j)\: S(k_j) \,.
\end{eqnarray}
However it is the factorization in Eq.~(\ref{A}) that will be central to our
discussion of reparametrization invariant operators.

\section{Reparametrization Invariant Objects for SCET} \label{sec:INV}

To construct an expansion in operators in SCET the standard procedure is to
build a gauge invariant basis of operators with definite power counting, order
$\lambda^k$, and to assign a Wilson coefficient to each one.  Afterwards one can
impose RPI order by order and find relations among Wilson coefficients. On the
contrary what we will do is to start with RPI and gauge invariant objects, to be
constructed in section~\ref{sect:RPIGI}.  These objects do not have a definite
power counting order, in particular we will know the order in the
$\lambda$-expansion where they start, but they will contain terms at all higher
orders as well. We build a basis with these RPI and gauge invariant objects,
which is made minimal using equations of motion and kinematic constraints as
discussed below in section~\ref{sect:const}. (Equation of motion constraints for
homogeneous operators are also summarized in this section.) Each element of this basis is
assigned a Wilson coefficient, and then the elements are expanded to find the
final basis with elements of a definite power counting.  In this way we
immediately obtain relations between Wilson coefficients of operators at
different orders. Once we expand and check for redundancy, the number of
independent Wilson coefficients is equal to the number of independent RPI
operators in the reduced basis.

\subsection{Construction of RPI and Gauge Invariant objects} \label{sect:RPIGI}

We now construct reparametrization invariant objects in SCET whose leading
terms give the fields in Eq.~(\ref{pc1}). These are then generalized to objects
that are simultaneously RPI and gauge invariant whose leading terms give the
objects in Eqs.~(\ref{objs},\ref{sh}). For simplicity only collinear objects are
considered in this section. Pulling out the large phases from the collinear
quark field and gluon field strength, and decomposing the full theory field into
independent collinear sectors we have at tree level,
\begin{eqnarray} \label{psiG}
  \psi(x) = \sum_n e^{-ix\cdot \cP_n} \psi_n(x) \,,\qquad
  G^{\mu\nu}(x)  = \sum_n  e^{-ix\cdot \cP_n} G_n^{\mu\nu}(x) \,. \label{RPdef}
\end{eqnarray}
Full Lorentz invariance act on the fields $\psi(x)$ and $G^{\mu\nu}(x)$, but
the RPI transformations that we are interested acts independently on each
collinear sector labeled by $n$.  Two sectors $i$, $j$ are independent if
$n_i\cdot n_j\gg \lambda^2$, and the sums in Eq.~(\ref{psiG}) are really over
equivalence classes, $\{n\}$, where a class consists of vectors related by RPI.
From the discussion in section~\ref{sect:RPI} the $n$-reparametrization
invariant collinear quark and field strength are easy to identify
\begin{eqnarray} \label{psin}
  \psi_{n} &\equiv &\left(1+\frac{1}{\bn\cdot
    D_n} \slash\!\!\!\! D_n^{\perp}\frac{\bn\!\!\!\slash}{2}\right)\xi_{n}\,,
  \qquad\qquad
 ig \, G_{\mu\nu}^n \equiv \left[iD^n_{\mu},iD^n_{\nu}\right]\,.
\end{eqnarray}
Under the transformations in Table~\ref{table123} for $\{n,\bn\}$, the quark
field $\psi_n$ remains invariant~\cite{Manohar:2002fd}, while the gluon tensor is
invariant because the vector $D_n^\mu$ is invariant. To make the fields in
Eq.~(\ref{psin}) invariant under the additional reparametrization
transformations that link collinear and ultrasoft derivatives we replace
$in\cdot D_{n} \to in\cdot D_n + g n\cdot A_{us}$, $iD_{n\perp}^\mu\to
iD_{n\perp}^\mu + W_n i D_{us\perp}^\mu W_n^\dagger$, and $i \bn\cdot D_{n}\to i
\bn\cdot D_{n} + W_n i\bn\cdot D_{us} W_n^\dagger$. After this replacement the
decoupling field redefinitions in Eq.~(\ref{fd}) can be made.  In
Eq.~(\ref{psin}) $\nslash \xi_n =0$, and the term in $\psi_n$ with a
$\perp$-covariant derivative corresponds to the two components of the full
fermion field that are small when $p_\perp/\bn\cdot p \ll 1$. Since $\nslash
\psi_n\ne 0$, the $\psi_n$ field does not provide a definite power counting for
operators. For example, $\bar\psi_n \bnslash\psi_n\sim \lambda^0$ whereas
$\bar\psi_n \nslash \psi_n\sim \lambda^4$.

We also need reparametrization invariant $\delta$-functions whose expansions
reproduce Eqs.~(\ref{deltaw}) and (\ref{deltaij}) at lowest order. For example,
these are needed to construct an RPI operator which when expanded gives
$\bar\chi_{n,\omega_1}\bnslash \chi_{n,\omega_2}$ at lowest order.  For
situations where there is an external hard vector $q^\mu$ the invariant
$\delta$-function is
\begin{align} \label{DEL}
 \hat\Delta_i \equiv \delta(\hat\omega - 2 q\cdot i\partial_{n_i}) 
    &= \delta(\hat\omega -  n_i\mcdot q\, \pb_{n_i}) + \ldots \,,
\end{align} 
where as described in section~\ref{sect:SCETgi}, $q^\mu$ is a parameter specific
to the kinematics of the process being studied.  Notice that $\delta(\hat\omega- 2
q\cdot i\partial_n)$ starts at ${\cal O}(\lambda^0)$, is RPI, and is gauge invariant
when acting on singlet operators.  Here
\begin{eqnarray} \label{delta}
 i\partial_n^\mu
    \equiv  \frac{n^\mu}{2} \pb_n +  \cP_{n\perp}^\mu 
    +  \frac{\bn^\mu}{2}  in\mcdot\partial_n \,,
\end{eqnarray}
and functions of $i\partial_n^\mu\sim (\lambda^2,1,\lambda)$ can be expanded in
powers of $\lambda$. Note that $\pb_n$ and $\cP_{n\perp}^\mu$ are only non-zero
when they act on $n$-collinear fields. It is useful to extend this property to
the full $i\partial_n^\mu$, which we can do by distributing an $i\partial^\mu$
derivative across all fields that it acts on, writing for example $i\partial^\mu
\bar\psi_{n_1} \psi_{n_2}= (i\partial_{n_1}^\mu\bar\psi_{n_1})\psi_{n_2} +
\bar\psi_{n_1} (i\partial_{n_2}^\mu \psi_{n_2})$. In some hard processes there
is more than one external hard vector, and a natural question arises as to
whether $q^\mu$ provides a unique choice for this construction.  For example, in
DVCS, $\gamma^* p\to \gamma^{(*)} p'$ we have the momentum $q^\mu$ of the
incoming $\gamma^*$ and the momentum $q^{\prime\mu}$ of the outgoing
$\gamma^{(*)}$. In Appendix~\ref{app_qqp} we show that as long as
$q_\perp-q'_\perp\sim \lambda$ or smaller, the choice $q$ suffices, since for
the purpose of constructing a basis of operators it is equivalent to the choice
of any linear combination of $q$ and $q'$. On the other hand, for situations
where there is no external hard vector $q^\mu$, the appropriate RPI
$\delta$-function is
\begin{align} \label{Delijexpn}
 \hat\Delta_{ij} \equiv \delta(\hat\omega_{ij} - 2 i\partial_{n_i} \cdot
 i\partial_{n_j}) 
  =  \delta\Big(\hat\omega_{ij} - \frac12 n_i\cdot n_j\: \pb_{n_i}\pb_{n_j}\Big) + \ldots\,.
\end{align}
This $\delta$-function operator acts on two independent collinear
directions. In general we must include in an operator a set of $\hat\Delta_i$
and $\hat \Delta_{ij}$ which are linearly independent.  Once we expand, the first
term in the series for $\hat\Delta_{ij}$ is not independent of the first term
from $\hat\Delta_i$, so the $\delta$-function shown on the RHS of
Eq.~(\ref{Delijexpn}) can always be eliminated, as we did in Eq.~(\ref{AAA}).

We will also make use of a reparametrization invariant
Wilson line, ${\cal W}_n$, which has the same gauge transformation properties as
$W_n$,
\begin{eqnarray}
  {\cal W}_n = W_n \: e^{-i R_n} \label{defW}\,.
\end{eqnarray}
Here the operator $R_n$ starts with a term at ${\cal O}(\lambda)$ and is built
of $n$-collinear gluon fields,
\begin{align}
  R_n = R_n\big[\, \pb_n\,, \cP_{n\perp}^\mu\,, 
     g {\cal B}_{n\perp}^\mu \,,
    t^\mu \big] \,,
\end{align}
where the vector $t^\mu$ is either $q^\mu$ or $i\partial_{n'}^\mu$ with $n\cdot
n' \sim \lambda^0$.  Furthermore, $R_n$ is Hermitian, dimensionless, and
collinear gauge invariant.  We leave the explicit construction of $R_n$ to
section~\ref{sect:R} below, and for the remainder of this section take these
properties as given.

Under collinear gauge transformations, $\psi_n$ and ${\cal W}_n$ transform the
same way as $\xi_n$ and $W_n$, and $G_n^{\mu\nu}$ transforms as a nonabelian
field strength. Thus using ${\cal W}_n$ we can form analogs of the results in
Eq.~(\ref{sh}) that are simultaneously RPI and gauge invariant, namely the
superfields
\begin{align}
  \Psi_{n} \equiv 
   {\cal W}_n^\dagger \psi_n  \,,\qquad
  {\cal G}^{\mu\nu}_{n} \equiv  
      {\cal W}_n^\dagger G^{\mu\nu}_n {\cal W}_n \,. \label{SF}
\end{align}
For cases with an external $q^\mu$ we also introduce a subscript notation,
\begin{align}\label{SF2}
 &  \Psi_{n,\hat\omega} \equiv \big[ \delta(\hat\omega-2q\mcdot i\partial_n) 
  \Psi_{n}
 \big] \,,\qquad
  {\cal G}^{\mu\nu}_{n,\hat\omega} \equiv \Big[ 
{\cal G}^{\mu\nu}_n
   \,  \delta(\hat\omega + 2q\mcdot i\, \overleftarrow{\partial_n})  \Big] \,. 
\end{align}
Operators built out of the superfields $\Psi_n$ and ${\cal G}_n^{\mu\nu}$ are
simultaneously RPI and gauge invariant. They are not homogeneous in the power
counting, but the superfields reduce to the objects in Eq.~(\ref{sh}) at lowest
order in the $\lambda$ expansion.  For example, the superfield for the fermion
\begin{align}
  \Psi_{n} &= e^{i R_n} \, W_n^\dagger \left(1+\frac{1}{i\bn\cdot {D_n}}
     {iD_n}\!\!\!\!\!\!\slash^{\perp}\frac{\bn\!\!\!\slash}{2}\right)\xi_{n}
   = e^{i R_n[i\partial_n,{\cal B}_n]}
   \left(1+\frac{1}{\pb_n}
   i\mathcal{D}_n\!\!\!\!\!\!\slash^{\perp}\frac{\bn\!\!\!\slash}{2}\right)
  \chi_{n}\, \nn \\
  &= \chi_{n} + \ldots \,.
\end{align}
Similarly, $(g^\perp_{\nu\nu'} \bn_\mu) ig( {\cal G}^{\mu\nu'}_{n})
= \pb_n\, g {\cal B}_{n\nu}^{\perp} +\ldots$.  Thus to form a RPI
version of the bilinear fermion operator $O(\omega_1,\omega_2)$ in
Eq.~(\ref{Oeg}) we simply take
\begin{align}
 {\bf Q}(\hat\omega_1,\hat\omega_2) &= \bar \Psi_{n,\hat\omega_1} \: \qslash\: \Psi_{n,\hat\omega_2}\,,
\end{align}
and note that expanding in $\lambda$ gives ${\bf Q}(\hat\omega_1,\hat\omega_2) =
(n\mcdot q)^{-1} \: O(\omega_1,\omega_2) + \ldots$.

We will also need the equations of motion for the RPI quark and gauge
superfields in Eq.~(\ref{SF}). The $n$-collinear Lagrangian for the quark field
is \cite{Bauer:2000yr}
\begin{equation} \label{Lagr1}
 \mathcal{L}_{qn} =\bar{\xi}_n \left( i n \cdot D_n + i D\!\!\!\!\slash ^{\perp}_n \frac{1}{i \bar{n} \cdot D_n} i D\!\!\!\!\slash^\perp_n  \right) \frac{\bar{n}\!\!\!\slash}{2} \xi_n \, , 
\end{equation}
We can write Eq.~(\ref{Lagr1}) in terms of $\psi_n$ as a simple Dirac Lagrangian
\begin{equation} \label{lagr2}
 \mathcal{L}_{qn} =\bar{\psi}_n\,  i D\!\!\!\!\slash_n \,  \psi_n \,, 
\end{equation}
The equation of motion for $\psi_n $ is a simple Dirac equation $
D\!\!\!\!\slash_n \, \psi_n =0$.  Using $ \mathcal{W}_n
\mathcal{W}_n^\dagger=1$, we can write $\mathcal{W}_n^\dagger i
D\!\!\!\!\slash_n \, \mathcal{W}_n \mathcal{W}_n^\dagger \psi_n =0$, and thus
obtain the equation of motion for $\Psi_n$
\begin{align} \label{Eom2}
 \hat{\mathcal{D}}\!\!\!\!\slash_n  \, \Psi_n & =0  \, .
\end{align} 
Here $\hat{\mathcal{D}}_n^\mu$ is the RPI and gauge invariant derivative
\begin{equation} \label{hatD}
  \hat{\mathcal{D}}^\mu_n \equiv {\cal W}_n^\dagger D_n^\mu\, {\cal W}_n =
  e^{iR_n}\, \mathcal{D}^\mu_n\, e^{-iR_n} \, .
\end{equation} 
For the gluon field we have the equation of motion $ [ i{D}^n_\nu, 
G_n^{\mu\nu}] = ig T^A \sum_f \bar\psi_n^f T^A \gamma^\mu \psi_n^f $, and for
the superfield
\begin{align} \label{Eom3}
 i \hat {\cal D}^n_\nu {\cal G}_n^{\nu\mu}= [i \hat {\cal D}^n_\nu, {\cal G}_n^{\nu\mu}] = -ig\, T^A
\bar{\Psi}_n^f T^A \gamma^\mu \Psi_n^f \,.
\end{align}
Note that $ig {\cal G}_n^{\mu\nu}= [i\hat {\cal D}_n^\mu, i\hat {\cal D}_n^\nu]$.

\subsection{Reducing the Operator Basis} \label{sect:const}

In general there are three steps that one can consider to reduce the
perturbative and nonperturbative information in the EFT to its minimal form:
\begin{itemize}
\item[a)] Find a minimal basis of homogeneous operators and of RPI operators
  that suffice at the desired order in $\lambda$. The homogeneous operators can
  be written entirely in terms of $\chi_n$, ${\cal B}_{n\perp}^\mu$, and
  $\cP_\perp^\mu$.
  \item[b)] Compare the homogeneous and RPI basis to determine which perturbative
    Wilson coefficients are fixed by RPI.
 \item[c)] Consider the decomposition of matrix elements of operators in the
   homogeneous basis, and derive further relations between the resulting
   non-perturbative functions.
\end{itemize}
Generically the relation between the operator basis looks like
\begin{align} \label{Oexpn}
  \sum_{n_i} \sum_\ell \int\! [\prod_j d\hat{\omega}_j]
   \: \hat{C}_\ell(\hat{\omega}_j)\:  [\textbf{Q}_\ell(\hat{\omega}_j)]=
   \sum_{n_i}\sum_\ell \int\!  [ \prod_j d\omega_j]
   \: C_\ell(\omega_j)\:  [ O_\ell(\omega_j)] + \ldots\,,
\end{align}
where $\textbf{Q}_\ell(\hat\omega_j)$ are RPI operators and $O_\ell(\omega_j)$
are homogeneous operators, and the ellipse denotes higher order terms in the
power expansion.  In general our focus in this article is to carry out b) which
is still largely process independent. For the most part we give no discussion of
item c), which obviously must be considered process by process. In order to
consider b) we must first determine a) which is the focus of this section. We
will discuss the equations of motion and other relations that allow a reduction
in the basis of operators at each order in $\lambda$.

First we consider the gauge invariant objects with homogeneous power counting.
We would like to demonstrate that all operators can be reduced to a form that
only involves the basic building blocks $\chi_n$, $g {\cal B}_{n\perp}^\mu$, and
$\cP_\perp^\mu$. All other homogeneous objects can be reduced to these. For
example, one might think that the objects $g {\cal B}_{\perp\perp}^{\mu\nu}
\equiv [ 1/\pb\, W^\dagger [iD_{n\perp}^\mu,iD_{n\perp}^{\nu}] W]$ and $g {\cal
  B}_{\perp 2}^{\mu} \equiv [ 1/\pb\, W^\dagger [iD_{n\perp}^\mu,in\mcdot D_{n}]
W]$ are independent.  However they are related to the building blocks by
\begin{align} \label{BI0}
  g {\cal B}_{\perp\perp}^{\mu\nu} 
  &= \frac{1}{\pb} {\cal P}_\perp^\mu (g {\cal B}_\perp^\nu) 
    - \frac{1}{\pb} {\cal P}_\perp^\nu (g {\cal B}_\perp^\mu) 
    +\frac{1}{\pb} 
      \big[ g {\cal B}_\perp^\mu ,   g {\cal B}_\perp^\nu \big]
  \,, \\[4pt]
  g {\cal B}_{\perp 2}^{\mu} 
  &= \frac{1}{\pb} {\cal P}_\perp^\mu (gn\mcdot {\cal B}) 
    - \frac{1}{\pb} in\mcdot \partial_n (g {\cal B}_\perp^\mu) 
    +\frac{1}{\pb} 
      \big[ g {\cal B}_\perp^\mu ,   g n\mcdot {\cal B} \big] \,, \nn
\end{align}
where we will see below that $n\cdot {\cal B}$ and $in\cdot\partial_n {\cal
  B}_\perp^\mu$ can also be reduced using the gluon equation of motion.  For
$\chi_n$ the equation of motion is
\begin{align} \label{eom}
  in\mcdot \partial_n \chi_n &= -(g n\mcdot {\mathcal B}_n) \chi_n 
  - i\,{\slash\!\!\!\!\mathcal D}_n^\perp  \frac{1}{\pb_n}
    i\,{\slash\!\!\!\!\mathcal D}_n^\perp \: \chi_n \,,
\end{align}
which allows us to eliminate $in\mcdot \partial_n$ derivatives on $\chi_n$.  To
obtain the equations of motion for the gluon objects we consider $-g^2 T^A
\sum_f \bar\psi_n^f W_n T^A W^\dagger_n \gamma^\mu \psi_n^f = [ i{\cal D}^n_\nu,
[i{\cal D}_n^\mu,i{\cal D}_n^\nu]]$. Expanding in $\lambda$ and multiplying on
the right with $\delta(\omega-\pb_n^\dagger)$ gives three equations
\begin{align} \label{eomg}
  &\quad\ \:  \omega\, (g n\mcdot {\mathcal B})_{\omega}  =
    2 {\mathcal P}^\perp_\nu  (g {\mathcal B}^\nu_{\perp})_{\omega}
   +  \frac{2\omega'}{\omega} \big[ (g{\mathcal B}_\perp^\nu)_{\omega-\omega'}, 
         (g {\mathcal B}^\perp_{\nu})_{\omega'} \big]
  -\frac{2}{\omega}\, g^2  T^A \sum_{f} \big[ {\bar\chi}_{n}^f\: T^A \nbs 
     \chi_{n}^f \big]_{\omega} 
     \,, 
  \nn 
 \\[0pt]
 &   \omega \big[ i n\mcdot \partial_n g \cBp^\mu \big]_\omega
  = -\big[ \cP^\perp_\nu [ g \cBp^\mu,g \cBp^\nu]\big]_\omega
   - \big[ g\cB^\perp_\nu, [ \cP_\perp^{[\mu} g \cBp^{\nu]} ]\big]_\omega 
   - \big[ g\cB^\perp_\nu, [ g \cBp^\mu, g \cBp^\nu ]\big]_\omega 
 \nn\\[3pt]
&\quad  + \frac{\omega}{2} \big[ \cP_\perp^\mu gn\mcdot \cB \big]_\omega
 - \big[ \cP^\perp_\nu \cP_\perp^{[\mu}\, g \cBp^{\nu]}\big]_\omega
  + \frac{\omega}{2} \big[ g \cBp^\mu, g n\mcdot \cB\big]_\omega
  - \frac{\omega'}{2} \big[( g n\mcdot \cB)_{\omega-\omega'}, (g
  \cBp^\mu)_{\omega'} \big] \nn \\[3pt]
 &\quad  
-g^2  T^A \sum_{f} \Big[ \bar\chi_n^f T^A \gamma_\perp^\mu \frac{1}{\pb} 
  ( \cPslash_\perp \plus g  \cBps) \frac{\bnslash}{2} \chi_n^f \Big]_\omega
   g^2  T^A \sum_{f} \Big[ \bar\chi_n^f \frac{\bnslash}{2}
  ( \cPslash_\perp^\dagger \plus g  \cBps)   \frac{1}{\pb^\dagger}  
   T^A \gamma_\perp^\mu \chi_n^f \Big]_\omega
,
 \nn\\[6pt]
& g^2 T^A \sum_{f} \Big[ \bar\chi_n^f (\cPslash_\perp^\dagger\plus g \cBps)
\frac{1}{\pb^\dagger} T^A \frac{1}{\pb} (\cPslash_\perp \plus g \cBps)
\bnslash \chi_n^f \Big]_\omega 
  \nn\\[3pt]
 &\quad = \frac{\omega}2 \big[ in\mcdot \partial_n g n\mcdot \cB \big]_\omega
    - \big[ (\cP_\perp)^2 g n\mcdot \cB \big]_\omega 
    -\big[ \cP^\perp_\nu [g \cBp^\nu, gn\mcdot \cB]\big]_\omega
    -\big[ g {\cBp}_\nu, [\cP_\perp^\nu g n\mcdot\cB ]\big]_\omega
  \nn\\[5pt]
 &\qquad  
   +\big[ g {\cBp}_\nu, [in\mcdot \partial_n g \cBp^\nu]\big]_\omega
   -\big[ g {\cBp}_\nu, [ g \cBp^\nu, g n\mcdot \cB] \big]_\omega
   +\big[ in\mcdot \partial_n \cP_\nu^\perp g \cBp^\nu \big]_\omega
  \nn\\[5pt]
&\qquad
  +  \frac{\omega'}{2}  \big[ (g n\mcdot \cB)_{\omega-\omega'},(g n\mcdot
 \cB)_{\omega'} \big]
  \,.
\end{align}
Here we sum over the color $A$, over the flavors $f$, and integrate over the repeated
index $\omega'$. In our analysis the first two equations will be used to eliminate
$g n\mcdot {\cal B}_n$ and $in\mcdot \partial_n\, g {\cal B}_\perp^\mu$
respectively.  The last relation only becomes relevant at higher orders than
those we consider here. The above relations imply that when building a
homogeneous basis of operators we do not need to consider the objects
\begin{align}
 in\mcdot\partial_n \chi_n\, , 
 & &  n\mcdot \cB_n \,, && in\mcdot \partial_n \cB_{n\perp}^\mu \,, 
 & &  {\cal B}_{\perp\perp}^{\mu\nu} \,, && {\cal B}_{\perp 2}^\mu \,. 
\end{align}

Next we derive relations that can be used to reduce RPI operators to a minimal
form.  Given the definition in Eq.~(\ref{hatD}), we can write
$i\hat{\mathcal{D}}^\mu_n = i\partial_n^\mu + [i\hat{ \mathcal{D}}_n^{ \mu}]$,
and it is straightforward using Eq.~(\ref{DWpartial}) below to prove that
\begin{align} \label{qdotD}
 [ q\mcdot i\partial_n i \hat{\mathcal{D}}^\nu_n ]
 =  q_\mu ig \mathcal{G}_n^{\mu\nu} \,,
\end{align} 
and hence that $q_\mu [ \hat{\mathcal{D}}_n^\mu]=0$. (The results here and below
apply equally well for $t=q$ and $t= i\partial_{n'}$ with $n\cdot n'\sim
\lambda^0$. For simplicity we use the notation with $t=q$.)
Eq.~(\ref{qdotD}) can be used to rewrite the quark superfields equation of motion in
Eq.~(\ref{Eom2}) as
 \begin{align}\label{EOM1}
i\partial_n \!\!\!\!\!\!\slash\ \, \Psi_n 
  &= - \Big[ \frac{1}{q\mcdot i\partial_n} \,q_\mu \gamma_\nu
   ig\mathcal{G}_n^{\mu \nu} \Big]  \Psi_n \,,
\end{align}
Since $q\cdot i\partial_n\: \delta(\hat \omega-2 q\cdot i\partial_n) = \frac12
\hat\omega\: \delta(\hat\omega-2 q\cdot i\partial_n)$ we also have the result
\begin{align} \label{EOM1a}
  q\mcdot i\partial_n \Psi_{n,\hat\omega} &= \frac{\hat\omega}{2} \: \Psi_{n,\hat\omega} \,.
\end{align} 
In a similar way, $q\cdot i\partial_n {\cal G}_{n,\hat\omega}^{\mu\nu} =
(-\hat\omega/2) {\cal G}_{n,\hat\omega}^{\mu\nu}$. The collinear gluon equation
of motion for $\mathcal{G}_n^{\mu\nu}$ in Eq.~(\ref{Eom3}) can be rewritten as
\begin{align} \label{EOM2}
\left[ i \partial^n_{\nu}  \mathcal{G}_n^{\nu\mu} \right] 
   = -ig\, T^A \bar{\Psi}_n^f T^A \gamma^\mu \Psi_n^f
   + \Big[  \Big[\frac{1}{q\mcdot i\partial_n} q_\alpha \mathcal{G}_{n\,
     \nu}^{\alpha}\Big]
  \,,\,
 ig\mathcal{G}_n^{\mu\nu} \Big]  \, . 
\end{align}
The quark and gluon operators will have $\hat\omega$ subscripts,
$\Psi_{n,\hat\omega}$ and ${\cal G}_{n,\hat\omega}$, so only the equations of
motion in Eqs.~(\ref{EOM1},\ref{EOM2}) should be used to remove derivatives
since the $i\partial_n$ derivatives commute with the presence of the
$\delta$-function denoted by the $\hat\omega$ subscript.  The QCD Bianchi identity, $D_\mu
G_{\nu\sigma}+D_\nu G_{\sigma\mu} + D_{\sigma} G_{\mu\nu}=0$, also gives a
relation for $\mathcal{G}_n^{\mu\nu}$, namely $\hat{\mathcal{D}}^\mu_n
\mathcal{G}_n^{\nu\sigma} + \hat{\mathcal{D}}^\nu_n \mathcal{G}_n^{\sigma
  \mu}+\hat{\mathcal{D}}^\sigma_n \mathcal{G}_n^{\mu\nu} = 0$.  Rearranging it
gives the following relation
 \begin{align} \label{BI}
i \partial_n^\alpha  \mathcal{G}^{\mu\nu}_n & = 
  q_\beta \bigg\{ \!\bigg[\! \Big[\frac{ig}{q\mcdot i\partial_n} \mathcal{G}^{\alpha
      \beta}_n\Big] , \mathcal{G}^{\mu \nu}_n \bigg]
    \!-\! \bigg[\! \Big[ \frac{ig}{q\mcdot i\partial_n}  \mathcal{G}^{\beta\mu}_n \Big],
    \mathcal{G}^{ \nu \alpha}_n \bigg]
    \!+\! \bigg[\! \Big[ \frac{ig}{q\mcdot i\partial_n}  \mathcal{G}^{\beta\nu}_n \Big],
    \mathcal{G}^{ \mu \alpha}_n \bigg] \bigg\} 
  - i \partial_n^{[\mu} \mathcal{G}^{\nu] \alpha}_n    \,, 
\end{align} 
which implies that $i \partial_n^\alpha  \mathcal{G}^{\mu\nu}_n$, $i
\partial_n^\mu \mathcal{G}^{\nu\alpha}_n$, and $i \partial_n^\nu
\mathcal{G}^{\alpha\mu}_n$ are not all independent.
Closing Eq.~(\ref{BI}) with $\gamma^\mu$ allows us to remove $i
\partial\!\!\!\slash_n \mathcal{G}^{\mu\nu}_n$, which is how we will choose to
use this identity in quark operators. An analog of the Bianchi identity does not
occur for the building block $g {\cal B}_n^\mu$ in homogeneous operators; it
easy to verify that when expanded in $\lambda$, Eq.~(\ref{BI}) is trivially
satisfied.  Eqs.~(\ref{EOM1}--\ref{BI}) are the RPI equivalent of the results in
Eqs.~(\ref{eom},\ref{eomg}), and can be used to reduce the RPI operator basis.
 
The above results imply that when building an RPI operator basis we do not need
to consider the objects
\begin{align}
  & i\partial\!\!\!\slash_n \Psi_n \,,
  \qquad q\mcdot i \partial_n \Psi_{n,\hat\omega} \,,
  \qquad [ i\partial^n_\nu {\cal G}_n^{\nu\mu}] \,,
  \qquad  i\partial_n \!\!\!\!\!\!\slash\, \ {\cal G}_n^{\mu\nu}  \,,
  \qquad [ q\mcdot i\partial_n\, {\cal G}_{n,\hat\omega}^{\nu\mu}]
   \,.
\end{align}
This list is not exhaustive. By manipulating operators in specific situations
further structures can be eliminated using a combination of the above
identities.  For example, for in sections~\ref{sec:DISq} and \ref{sec:DISg} below we will see
that $q_\mu \mathcal{G}^{\mu\nu}_{n} i\partial^n_{ \nu}$, with the
$i\partial^n_{\nu}$ acts on a $n$-collinear quark or gluon field, can be
eliminated.

In principle one can just count the number of RPI operators and compare to the
number of operators in a homogeneous operator basis with definite power counting
to determine whether there are any RPI constraints on the Wilson coefficients.
The key issue here is that of linear independence, even if one has the the same
number of operators in the RPI and homogeneous basis, it could be that two RPI
operators constrain the same linear combination of operators in the homogeneous
basis.

\subsection{Extension to Massive Collinear Fields} \label{sect:rpim}

Massive collinear quarks in SCET were first studied in
Refs.~\cite{Rothstein:2003wh,Leibovich:2003jd}. After the field redefinition in
Eq.~(\ref{fd}) they have the LO Lagrangian
\begin{equation} \label{Lagrm}
 \mathcal{L}_{qn} =\bar{\xi}_n \left[ i n \cdot D_n + 
  (i D\!\!\!\!\slash ^{\perp}_n-m) \frac{1}{i \bar{n} \cdot D_n} 
  (i D\!\!\!\!\slash^\perp_n+m)  \right] \frac{\bar{n}\!\!\!\slash}{2} \xi_n \, .
\end{equation}
The appropriate RPI transformations with massive quarks were determined in
Ref.~\cite{Chay:2005ck}. The only change is in the type-II transformation of the
fermion field, where one has to add a mass dependent term:
\begin{equation} \label{ximass}
\xi_{n}\overset{\mathrm{II}}{\longrightarrow}\left[1+
  \frac{\varepsilon\!\!\!\slash^{\perp}}{2}\frac{1}{\bar{n}\cdot i D_n}
  (i D\!\!\!\!\slash_{n\perp}-m)\right]\xi_{n}.
\end{equation}
Under this transformation the Lagrangian in Eq.~(\ref{fd}) falls into two
invariant parts, one fixed by the leading order kinetic term and one whose
coefficient encodes the choice of mass scheme.  Note that the RPI transformation
itself is not modified by the presence of a mass term, the transformation of
$\bn$ is still exactly as in Eq.~(\ref{repinv}). 

We can now build an analog of the RPI superfield for a massive collinear quark.
The reparametrization invariant quark field is
\begin{equation}
\psi_{n}=\left(1+\frac{1}{\bar{n}\cdot i D_n}(i D\!\!\!\!\slash\:_n^{\perp}+m)
  \frac{\bar{n}\!\!\!\slash}{2}\right)\xi_{n}.
\end{equation}
This leads to the modified RPI superfield for a massive collinear quark
\begin{align}
\Psi_{n} 
  &=  e^{i R_n}  \left[1+\frac{1}{\pb_n} ( {i\cD_n}\!\!\!\!\!\!\slash^{\perp} + m
  )\frac{\bn\!\!\!\slash}{2}\right] \chi_{n}
\,.
\end{align}
This result is included for completeness. Our focus in the remainder of the
paper will be on massless collinear quark fields.

\subsection{Determination of $R_n$ and Expansion of $\Psi_n$ and ${\cal
    G}_n^{\mu\nu}$} \label{sect:R}
  
In this section we derive an expression for $R_n$ appearing in the RPI Wilson
line, and then expand the invariant objects $\Psi_n$, ${\cal G}_n^{\mu\nu}$,
$\delta(\hat\omega - 2q\cdot i\partial_n)$, and $\delta(\hat\omega_{12} -
2 i\partial_{n_1}\cdot i\partial_{n_2})$.  We can define the collinear Wilson
line $W_n$ by the equation:
\begin{equation}
\left[ (\bar{n} \cdot D_n) W_n \right] = 0 \, . \label{defWcol}
\end{equation}
We define the RPI $\mathcal{W}_n$ generalizing (\ref{defWcol}) to a covariant
derivative $D_n$ along a (non light-like) direction $t$ as: 
\begin{equation}
\left[ (t \cdot D_n) \mathcal{W}_n \right] = 0 \, , \label{defIRPW}
\end{equation}
where $t$ is such that $n\cdot t \sim \lambda^0$. This implies the
momentum space representation:
\begin{equation}
\mathcal{W}_n = \left[  \sum_{\mathrm{perms}} 
  \mathrm{exp} \left(  \frac{-g}{(t \cdot i\partial_n)}  t \cdot  A_n \right) \right] \,.
\end{equation}
We would like to find $R_n$ such that ${\cal W}_n=W_n e^{-i R_n}$. Thus
$e^{-iR_n}$ is the operator that rotates $W_n$ from the light-like direction $n$
to the direction $t$. ${\cal W}_n$ is reparametrization invariant to the choice
of the basis vector $n$, which labels the $n$-collinear fields $A_n^\mu$, since
such reparametrizations cannot change the fact that $n\cdot t\sim \lambda^0$.
Recall that the subscript $n$ on $\mathcal{W}_n$ labels the equivalence class
$\{ n\}$ of vectors that are related by type-I and type-III RPI transformations.
For any $t$ such that $n\cdot t\sim \lambda^0$ we have
\begin{align} \label{tt}
 \frac{1}{t\cdot i\partial_n} \, t\cdot A_n 
   = \frac{1}{\pb_n}\:  \bn\mcdot A_n + \ldots \,,
\end{align}
and thus 
\begin{align} \label{WW}
  {\cal W}_n = W_n + \ldots \,,
\end{align} 
where the ellipses represent power suppressed terms.  In Eq.~(\ref{tt}) the
$n\cdot t$'s in the numerator and denominator cancel out in the leading term,
leaving a $t$ independent result.

For situations where we have an external hard vector $q^\mu$, we can simply take
$t^\mu=q^\mu$ and use the corresponding ${\cal W}_n$ as the RPI invariant Wilson
line. 

For situations where there is no external $q^\mu$, the choice for $t^\mu$ in
${\cal W}_n$ is less obvious since the only available RPI vectors are operators
themselves, $i\partial_{n'}^\mu$, where $n'$ is a distinct collinear direction
from $n$. In this situation, any choice $t^\mu=i\partial^\mu_{n'}$ satisfying
$n\cdot t= n\cdot n'\: \pb_{n'} + \ldots \sim \lambda^0 +\ldots$ is equally
good, and the existence of the hard interaction guarantees that such an $n'$
exists. In this case ${\cal W}_n$ still yields $W_n$ at lowest order, and hence
only behaves like an operator in the $n'$ direction through terms in the power
corrections, namely the ellipsis in Eq.~(\ref{WW}). In these ellipse terms the
$i\partial_{n'}$'s appear linearly order by order. Since the derivative
$i\partial_{n'}$ does not act on $n$-collinear fields it behaves just like an
external vector $q$ as far as manipulations related to the $n$-collinear fields
are concerned.

In the remainder of this section we adopt the notation $t=q$, even though the
algebra applies equally well to both cases mentioned above, with the
substitution $q\to t=i\partial_{n'}$ in appropriate
places. The only complication for the case $t=i\partial_{n'}$ is that the dot
product $n\cdot i\partial_{n'}$ must be expanded using
\begin{align}
  2 i\partial_{n}\mcdot i\partial_{n'} 
 &  = \frac{n\mcdot n'}{2} \,\pb_{n} \pb_{n'} 
  + n'\mcdot i\partial_{n\perp}\pb_{n'} + n\mcdot i\partial_{n'\perp} \pb_n
  + 2 i \partial_{n\perp} \mcdot i\partial_{n'\perp} 
  + \frac{\bn\cdot n'}{2}\, n\mcdot i\partial_n\, \pb_{n'} 
\nn\\[3pt]
 &\  
  + \frac{\bn'\cdot n}{2}\, n'\mcdot i\partial_{n'} \pb_{n}
  + \bn\mcdot i\partial_{n'}^\perp \, n\mcdot i\partial_n
  + \bn'\mcdot i\partial_{n}^\perp \, n'\mcdot i\partial_{n'}
  + \frac{\bn\cdot \bn'}{2}\,  n\mcdot i \partial_{n}\: n'\mcdot i\partial_{n'}
 \,,
\end{align}
where the first term is $\sim \lambda^0$, the next two $\sim \lambda$, the
following three are $\sim \lambda^2$, then the next two are $\sim \lambda^3$, and the
last one is $\sim \lambda^4$.

Adopting $t=q$, Eq.~(\ref{defIRPW}) can be used to prove that
\begin{align} \label{DWpartial}
 (q \cdot i D_n)
   &= \mathcal{W}_n \, (q \cdot i \partial_n) \, \mathcal{W}_n^\dagger \, . 
\end{align}
To calculate $iR_n$ we exploit Eq.~(\ref{DWpartial}) and calculate $iR_n$ order
by order in $\lambda$. Substituting Eq.~(\ref{defW}) into Eq.~(\ref{DWpartial}) we find
\begin{align}\label{DRpartial}
 (q \cdot i \mathcal{D}_n)
  &= e^{-iR_n} \, (q \cdot i \partial_n ) \, e^{iR_n} \, . 
\end{align}
Because of the Hermicity of $i {\cal D}_n^\mu$ and $ i \partial_n^\mu$, $R_n$ is
Hermitian.  Applying the Hadamard formula to Eq.~(\ref{DRpartial}) we obtain
\begin{align} \label{qdR}
 (q \cdot i \mathcal{D}_n)&=  (q \cdot  i\partial_n)
  + \sum_{j=1}^\infty \frac{1}{j!} \{\!\{ (q \cdot i\partial_n), (iR_n)^j\, \}\!\} \, ,
\end{align}
where $\{\!\{ A,B\}\!\} = [A,B]$ and
\begin{align}
 \{\!\{ A, B^{j} \}\!\}  =   \{\!\{ [A,B] , B^{j-1}\}\!\} 
    =[[\cdots[A,\overbrace{B,]B,]\ldots,]B]}^{j}\,.
\end{align}
Expanding $R_n$ in terms with $R_n^{(k)}\sim \lambda^k$ we can expand all the
objects in Eq.~(\ref{qdR}) in $\lambda$ and solve the resulting equations order
by order for $R_n^{(k)}$. Thus we write
\begin{align} \label{Lex}
 i R_n  &= \sum_{k=1}^\infty  i R_n^{(k)} \,,  \nonumber \\[4pt]
( q \cdot i \mathcal{D}_n )&= \frac{n \cdot q}{2} \bnP_n
+ (q_\perp \cdot \mathcal{P}_{n\perp}) + (q_\perp \cdot g\mathcal{B}_{n\perp})
 +  \frac{\bn \cdot q}{2} (n \cdot i\partial_n)
 + \frac{\bn \cdot q}{2} (g n \mcdot \mathcal{B}_n) \, , \nonumber \\
 (q \cdot i \partial_n )&= \frac{{n} \cdot q}{2} \bnP_n
  +(q_\perp\cdot \mathcal{P}_{n\perp} )+ \frac{\bn \cdot q}{2} (n \cdot i\partial_n) 
  \, . 
\end{align}
$(q \cdot i\partial_n)$ is a derivative operator, so when it acts in a commutator
with $( g \mathcal{B}_n^\mu)$ we have
\begin{align}\label{dercom}
  [(q \cdot i\partial_n) , ( g \mathcal{B}_n^\mu) ] & 
 = \big[q \cdot i\partial_n \, ( g
  \mathcal{B}_n^\mu) \big] \, , 
\end{align}
where the last set of square brackets means that the derivative acts only
inside. Substituting Eq.~(\ref{Lex}) into (\ref{qdR}) we can solve for
$iR_n^{(k)}$. The first two terms are
\begin{align}  \label{R12}
  i R_n^{(1)} &= \Big[\frac{2}{n\mcdot q\,\overline{\cP}_n}\, q_\perp\mcdot (g {\cal B}_n^\perp)
  \Big] \, , \\ 
  i R_n^{(2)} &= \Big[ \frac{1}{n\mcdot q\,\overline{\cP}_n}\,
   (\bn\mcdot q)\,(  g n\mcdot {\cal B}_n)
  \Big] 
   \!-\! \Big[ \frac{4 q_\perp\mcdot {\cal P}_{n\perp} }{(n\mcdot
     q\,\overline{\cP}_n)^2} \, q_\perp\mcdot ( g {\cal B}_n^\perp ) \Big]
\nn\\
 &\quad 
 +
  \Big[ \frac{2}{n\mcdot q\,\overline{\cP}_n} \Big[\Big[ \frac{1}{n\mcdot
    q\,\overline{\cP}_n} \, q_\perp\mcdot ( g {\cal B}_n^\perp ) \Big], 
   \, q_\perp\mcdot (g  {\cal B}_n^\perp ) \Big]\Big]\,. \nn
\end{align}
The $n\cdot \cB_n$ term should be further reduced with the equation of motion in
Eq.~(\ref{eomg}) to terms involving $\chi_n$ and $\cB_{n\perp}^\mu$.  In terms of
the $i R_n^{(k)}$ we can determine the $\lambda$ expansion of the invariant
Wilson line
\begin{equation}
 \mathcal{W}_n=\sum_{k=0}^{\infty} \mathcal{W} ^{(k)}_n \label{lambdaW}\,. 
\end{equation}
Using the definition in Eq.~(\ref{defW}) the first few terms are
\begin{align}
 \mathcal{W} ^{(0)}_n & =  W_n \, ,
 & \mathcal{W} ^{(1)}_n & =  - W_n (iR_n^{(1)})\, , 
 & \mathcal{W} ^{(2)}_n & =   \Big[ \frac12 (iR_n^{(1)})^2 - (iR_n^{(2)})\Big] \, . 
\end{align}
The expansion of the invariant Wilson line is therefore
\begin{align}
  {\cal W}_n = W_n - W_n (iR_n^{(1)}) + W_n 
   \Big[ \frac12 (iR_n^{(1)})^2 - (iR_n^{(2)})\Big]
  + \ldots \,.
\end{align}
Using these $R_n^{(k)}$'s and Table~\ref{table123} it is simple to check
explicitly that ${\cal W}_n$ is RPI up to order $O(\lambda^3)$.  Note that we
did not assign a suppression for $q_\perp$ anywhere above (ie, we took
$q_\perp\sim \lambda^0$). Taking $q_\perp\sim\lambda$ causes further suppression
of some of the terms in Eq.~(\ref{R12}). For cases where $q_\perp=0$ the
expansion of $\mathcal{W}_n$ starts at $O(\lambda^2)$.

We will also need the $\lambda$ expansion of the invariant $\delta$-functions,
$\delta(\hat\omega-2 q\cdot i\partial_n)$ and $\delta(\hat\omega_{12}- 2
i\partial_{n_1}\cdot i\partial_{n_2})$. For the former we have
\begin{align}
  \delta(\hat\omega - 2 q\cdot i\partial_n) 
   &= \delta(\hat\omega -  n\mcdot q\, \pb_n -2 q_\perp\mcdot \cP_{n\perp} 
    - \bn\mcdot q \, in\mcdot\partial_n )
\\[5pt]
  &=\frac{1}{n\mcdot q}
  \Big[ \Big(1 + \sum_{k=1}^\infty p^{(k)}_n \Big) \delta(\omega-  \pb_n) \Big]
 \,, \nn
\end{align}
where the first two terms are
\begin{align} \label{pnsin}
  p_n^{(1)} &= - \frac{2q_{\perp}\mcdot \mathcal{P}_{n\perp}}{n\mcdot q} \frac{d}{d\omega}\,,
  & p_n^{(2)} &= 2 \Big(\frac{q_{\perp}\mcdot \mathcal{P}_{n\perp}}{n\mcdot q}\Big)^2 
   \frac{d^2}{d\omega^2}
   - \frac{\bn\mcdot q}{n\mcdot q} (in\mcdot\partial_n) \frac{d}{d\omega} \,.
\end{align}
When combining the operator with the Wilson coefficient $C(\omega_i)$ we can
integrate by parts to move these derivatives onto the $C(\omega_i)$ and
leave a simple $\delta$-function in the operator. For the $\delta$-function with
two collinear directions we have
\begin{align}
 & \delta(\hat\omega_{12} - 2 i\partial_{n_1}\mcdot i\partial_{n_2} ) 
\\
 &  = \delta\Big(\hat\omega_{12} - \frac{n_1\mcdot n_2}{2} \pb_{n_1} \pb_{n_2} 
  - \pb_{n_1} n_1\mcdot i\partial_{n_2\perp} - \pb_{n_2} n_2\mcdot i\partial_{n_1\perp}  
  -2 i\partial_{n_1\perp}\mcdot i\partial_{n_2\perp} 
 -\frac{\bn_1 \mcdot n_2}{2}\, n_1\mcdot i\partial_{n_1} \pb_{n_2} 
\nn\\[4pt]
 &\
  -\frac{\bn_2 \mcdot n_1}{2}\, n_2\mcdot i\partial_{n_2} \pb_{n_1}
  - n_1\mcdot i\partial_{n_1} \bn_1\mcdot i\partial_{n_2\perp} 
  - n_2\mcdot i\partial_{n_2} \bn_2\mcdot i\partial_{n_1\perp} 
  - \frac{\bn_1\mcdot \bn_2}{2} n_1\mcdot i\partial_{n_1} n_2\mcdot i\partial_{n_2} 
\Big) \nn \\[5pt]
  &= \Big[ \Big(1 + \sum_{k=1}^\infty p^{(k)}_{n_1n_2} \Big) 
 \delta\Big(\hat\omega_{12}- \frac{n_1\mcdot n_2}{2} \, \pb_{n_1} \pb_{n_2} \Big)\Big]
 \,, \nn
\end{align}
where the first two terms are
\begin{align} \label{ppsin}
  p_{n_1 n_2}^{(1)} &= - \Big[  \pb_{n_1} n_1\mcdot i\partial_{n_2\perp} 
  + \pb_{n_2} n_2\mcdot i\partial_{n_1\perp} \Big]\frac{d}{d\omega}\,, 
 \nn\\[5pt]
  p_{n_1n_2}^{(2)} &= \Big[  \pb_{n_1} n_1\mcdot i\partial_{n_2\perp} 
  + \pb_{n_2} n_2\mcdot i\partial_{n_1\perp} \Big]^2 \frac{d^2}{d\omega^2} \nn\\
 &\   - \Big[ 2 i\partial_{n_1\perp}\mcdot i\partial_{n_2\perp} 
 +\frac{\bn_1 \mcdot n_2}{2}\, n_1\mcdot i\partial_{n_1} \pb_{n_2} 
 +\frac{\bn_2 \mcdot n_1}{2}\, n_2\mcdot i\partial_{n_2} \pb_{n_1} \Big]  \frac{d}{d\omega} \,.
\end{align}
All terms with $n\cdot i\partial_n$ in Eqs.~(\ref{pnsin}) and (\ref{ppsin}) will
be further reduced by the equations of motion in Eqs.~(\ref{eom}) and
(\ref{eomg}) when they appear in operators.

Finally we expand the superfields in Eq.~(\ref{SF}) in $\lambda$, writing
\begin{align}
 &  \Psi_{n,\hat\omega}= \sum_{k=1}^{\infty} \Psi_{n,\hat\omega}^{(k)} ,\qquad \qquad
  {\cal G}^{\mu\nu}_{n,\hat\omega}= \sum_{k=1}^{\infty} {\cal G}^{(k)\mu\nu}_{n,\hat\omega} \,, \label{ESF}
\end{align}
where $\Psi^{(k)}_{n,\hat\omega}\sim \lambda^k$ and ${\cal
  G}_{n,\hat\omega}^{(k)\mu\nu}\sim \lambda^k$.
The expansion of the quark superfield is straightforward, the first few orders are
\begin{align}  \label{Psin}
\Psi_{n,\hat\omega}^{(1)} & = \frac{1}{n\mcdot q}\,  \chi_{n,\omega} \, , \\
\Psi_{n,\hat\omega}^{(2)} & =   
  \frac{1}{n\mcdot q} \Big(
 \frac{1}{\omega}i\mathcal{D}\!\!\!\!\slash_{n,\omega_a-\omega}^{\perp}
   \frac{\bn\!\!\!/}{2}\chi_{n,\omega_a} + 
  iR_{n,\omega_a-\omega}^{(1)}  \chi_{n,\omega_a} + \big[ p_n^{(1)}
  \chi_{n,\omega} \big]
  \Big) \, ,\nn \\
\Psi_{n,\hat\omega}^{(3)} & =  
 \frac{1}{n\mcdot q} \Big(
  iR_{n,\omega_a-\omega}^{(2)} \chi_{n,\omega_a} 
  +\big[ p_n^{(2)} \chi_{n,\omega} \big] +
 \frac{1}{\omega_a\plus \omega_b}\, iR_{n,\omega_a-\omega_b-\omega}^{(1)} \,
 i\mathcal{D}\!\!\!\!\slash_{n,\omega_b}^{\perp}
   \frac{\bn\!\!\!/}{2}  \chi_{n,\omega_a}  
 \nn \\[4pt]
   & +\Big[
  p_{n}^{(1)} \,\frac{1}{\omega}\,
 i\mathcal{D}\!\!\!\!\slash_{n,\omega_a-\omega}^{\perp}
   \frac{\bn\!\!\!/}{2}  \chi_{n,\omega_a}  \Big]
   + 
   \big[ p_n^{(1)}  iR_{n,\omega_a-\omega}^{(1)}  \chi_{n,\omega_a} \big] + \frac{1}{2}
   iR^{(1)}_{n,\omega_a-\omega_b-\omega} iR^{(1)}_{n,\omega_b} \chi_{n,\omega_a}
  \Big) \, . \nn
\end{align}
Here there is an implicit integration over the repeated indices $\omega_a$ and
$\omega_b$.
For the gluon superfield first it is useful to expand $W^{\dagger}G_{\mu\nu}W$:
\begin{align}  \label{EG}
W^{\dagger}ig G_{\mu\nu}W
 &=\frac{n_{\mu}}{2}\left[i\bar{n}\cdot\mathcal{D},i\mathcal{D}_{\perp\nu}\right]
  -\frac{n_{\nu}}{2}\left[i\bar{n}\cdot\mathcal{D},i\mathcal{D}_{\perp\mu}\right]
  +\left[i\mathcal{D}_{\perp\mu},i\mathcal{D}_{\perp\nu}\right]
  +\frac{\bar{n}_{\mu}}{2}\frac{n_{\nu}}{2}\left[in\cdot\mathcal{D},i\bar{n}\cdot\mathcal{D}\right]
  \nn\\
&\qquad
  +\frac{n_{\mu}}{2}\frac{\bar{n}_{\nu}}{2}\left[i\bar{n}\cdot\mathcal{D},in\cdot\mathcal{D}\right]
  +\frac{\bar{n}_\mu}{2}[in\cdot \mathcal{D},i\mathcal{D}_{\perp \nu}]
  - \frac{\bar{n}_\nu}{2}[in\cdot \mathcal{D},i\mathcal{D}_{\perp \mu}]
\nn\\[5pt]
& =\frac{n_{\mu}}{2}\left[\mathcal{\bar{P}}g \mathcal{B}_{\perp\nu}\right]
  \!- \frac{n_{\nu}}{2}\left[\mathcal{\bar{P}}g \mathcal{B}^\perp_{\mu}\right]
  \plus\left[\mathcal{\bar{P}}g \mathcal{B}^{\perp\perp}_{\mu\nu}\right]
  \!-\frac{\bar{n}_{\mu}n_{\nu}}{4}\left[\mathcal{\bar{P}}g n \mcdot\mathcal{B}\right]
  \!+\frac{n_{\mu}\bar{n}_{\nu}}{4} \left[\mathcal{\bar{P}}g n  \mcdot
    \mathcal{B}\right]
 \nn\\
 &\qquad 
  - \frac{\bn_\mu}{2} \left[ \pb g \cB^{\perp 2}_\nu \right] 
  + \frac{\bn_\nu}{2} \left[ \pb g \cB^{\perp 2}_\mu \right] 
 \,,
\end{align}
where $g \cB^{\perp\perp}_{\mu\nu}$ and $g \cB^{\perp 2}_\mu$ are given by the
combinations of fields in Eq.~(\ref{BI0}).  Using this result to determine the
first few terms ${\cal G}_{n,\hat\omega}^{(k)\mu\nu}$ from expanding Eq.~(\ref{SF}),
we find
\begin{align} \label{Gsin}
  ig {\cal G}^{(1)\mu\nu}_{n,\hat\omega} & = 
    \frac{\omega}{2(n \cdot q)} \big[ 
    n^{\nu} (g \mathcal{B}^{\mu}_{ n \perp })_{\omega}
  -  n^{\mu} (g \mathcal{B}^{\nu}_{ n \perp })_{\omega} \big]
  \, ,\\[5pt]
 ig {\cal G}^{(2)\mu\nu}_{n,\hat\omega} & =
  \frac{1}{n\mcdot q} \Big\{
   \big[ \mathcal{P}_\perp^\mu (g\mathcal{B}_{n
        \perp}^\nu)_{\omega} \big] 
  -  \big[ \mathcal{P}_\perp^\nu 
  (g\mathcal{B}_{n\perp}^\mu)_{\omega} \big]  
  + \big[ ( g \mathcal{B}_{n
      \perp}^\mu) ,(g\mathcal{B}_{n\perp}^\nu) \big]_{\omega} 
\nonumber \\[4pt]
 &  +\frac{\omega }{4} (\bar{n}^\mu n^\nu \minus n^{\mu}\bar{n}^\nu)
    (g n\mcdot \mathcal{B}_n)_{\omega} 
 +   \frac{\omega}{2} \Big[
  iR_{n,\omega-\omega_a}^{(1)}, n^{\nu} (g \mathcal{B}^{\mu}_{ n \perp })_{\omega_a}
    \minus n^{\mu} (g \mathcal{B}^{\nu}_{ n \perp })_{\omega_a} \Big] 
  \nonumber \\[4pt]
  & - \frac{1}{2}  \Big[ n^{\nu} p_n^{(1)}\omega (g \mathcal{B}^{\mu}_{ n
      \perp })_{\omega} - n^{\mu} p_n^{(1)} \omega (g \mathcal{B}^{\nu}_{
      n \perp })_{\omega} \Big] \Big\} 
  \,,\nn\\[4pt]
 ig {\cal G}^{(3)\mu\nu}_{n,\hat\omega} & =
    \frac{-1}{2(n \cdot q)} \Big\{ \big[
    (\bn^{\mu} \cP_{\perp}^\nu  -  \bn^{\nu} \cP_{\perp}^\mu ) 
    (g n\mcdot \mathcal{B}_{ n})_{\omega}\big]
  - \big[ in\mcdot\partial_n \big( \bn^\mu g \mathcal{B}^{\nu}_{\perp,\omega } 
   - \bn^\nu g \mathcal{B}^{\mu}_{\perp,\omega } \big) \big] \nn\\[4pt]
  & +\! \big[ \bn^\mu g \mathcal{B}^{\nu}_{\perp,\omega } , gn\mcdot \cB_n \big]
    \!-\! \big[ \bn^\nu g \mathcal{B}^{\mu}_{\perp,\omega } , gn\mcdot \cB_n \big]
  \!-\!  \Big[ n^{\nu} \omega(g \mathcal{B}^{\mu}_{ n \perp })_{\omega}p_n^{(2)\dagger}
  -  n^{\mu}\omega (g \mathcal{B}^{\nu}_{ n \perp })_{\omega} p_n^{(2)\dagger} \Big]
  \nn\\
  &+ \ldots \Big\} \,, \nn 
\end{align} 
where again there is an implicit integration over $\omega_a$ in terms where it
appears.  Here the ellipsis denotes terms in ${\cal G}^{(3)\mu\nu}_n$ with an
$iR_n^{(1,2)}$ or $p_n^{(1)}$ which were not needed for our analysis. The $(g
n\mcdot \cB_n)$ and $[in\mcdot\partial_n \, g {\cal B}_\perp^\mu]$ terms are
further reduced to $\cP_\perp$'s, $(g {\cal B}_\perp^\mu)$'s, and $\chi_n$'s by
using the equation of motion in Eq.~(\ref{eomg}). Finally, recall that the
expansion coefficients in Eqs.~(\ref{ppsin},\ref{Psin},\ref{Gsin}) do not encode
the RPI relations between collinear and ultrasoft fields which can be determined
using Eq.~(\ref{rmD}).

The above results can be used to expand the RPI basis of operators in terms of
operators in the homogeneous basis as in Eq.~(\ref{Oexpn}).


\section{Applications} \label{sec:app}




\subsection{Scalar Current}  \label{sec:appS}

As a first example to show how the expansion of a RPI current works, we expand
the scalar chiral-even bilinear currents (LL+RR), for processes with a hard
external vector $q^\mu$ up to order $\lambda^3$. In the basis built from
superfields there is only one current that satisfies these conditions
\begin{align}   \label{SCE}
\bar{\Psi}_{n,\hat{\omega}_1}q\!\!\!\slash\: \Psi_{n,\hat{\omega}_2} \, .
\end{align}
All the other possible currents (for example $\bar{\Psi}_{n,\hat{\omega}_1}
\gamma_\mu q_\nu
\mathcal{G}_{n,\hat{\omega}_3}^{\mu\nu}\Psi_{n,\hat{\omega}_2}$) have expansions
that start at ${\cal O}(\lambda^4)$ or beyond.  To recover the basis with a
homogeneous power counting, all we have to do is to expand~(\ref{SCE}) using
Eq.~(\ref{Psin}),
\begin{align} \label{scalarcurrent}
& \bar{\Psi}_{n,\hat{\omega}_1}q\!\!\!\slash\:\Psi_{n,\hat{\omega}_2}
 = \frac{1}{(n \mcdot q)}\bar{\chi}_{n,\omega_{1}}\frac{\bar{n}\!\!\!\slash}{2}
\chi_{n,\omega_{2}}
 \\[3pt]
  &
 + \frac{1}{2 \, \omega_1 (n \mcdot q)^2} \bar{\chi}_{n, \omega_a}
   i\overleftarrow{\mathcal{D}}\!\!\!\!\slash_{\perp \omega_1-\omega_a}
\bar{n}\!\!\!\slash\: q\!\!\!\slash_{\perp} \chi_{n,\omega_{2}}
 - \frac{1}{2\, \omega_2 (n \mcdot q)^2 } \bar{\chi}_{n,\omega_{1}}
    q\!\!\!\slash_{\perp}\bn\!\!\!\slash
\, i\mathcal{D}\!\!\!\!\slash_{\perp \omega_a-\omega_2}
  \chi_{n, \omega_{a}} 
  \nonumber\\[3pt]
& + \frac{1}{ (\omega_1\minus \omega_a)(n \mcdot q)^2}\bar{\chi}_{n, \omega_a}
  (q_{\perp} \mcdot  g \mathcal{B}_\perp  )_{ \omega_1- \omega_a }
  \bar{n}\!\!\!\slash \chi_{n,\omega_{2}} 
  +\frac{1}{(\omega_2\minus\omega_a) (n \mcdot q)^2 }\bar{\chi}_{n,\omega_{1}} \bar{n}\!\!\!\slash
  (q_{\perp} \mcdot g\mathcal{B}_{\perp})_{\omega_a-\omega_2}
  \chi_{n, \omega_{a}}
  \nonumber\\[3pt]
  & -\frac{1}{(n \mcdot q)^2}\frac{\partial }{\partial\omega_{1}} \:
  \bar{\chi}_{n,\omega_{1}}\mathcal{P}_{\perp}^{\dagger}\mcdot q_{\perp}
  \bar{n}\!\!\!\slash \chi_{n,\omega_{2}}
 - \frac{1}{(n \mcdot q)^2}\frac{\partial }{\partial\omega_{2}} \:
 \bar{\chi}_{n,\omega_{1}} \bar{n}\!\!\!\slash \mathcal{P}_{\perp}\mcdot
   q_{\perp}\chi_{n,\omega_2} \, . \nn
\end{align} 
Thus all the ${\cal O}(\lambda^3)$ terms (the twist-3 terms on the last three
lines) are connected.  Eq.~(\ref{scalarcurrent}) agrees with the original
derivation of these constraints given in Eqs.~(122-126) of
Ref.~\cite{Hardmeier:2003ig}. The ease at which Eq.~(\ref{scalarcurrent}) was
derived demonstrates the power of the invariant operator formalism.  In this
example there is only one supercurrent to ${\cal O}(\lambda^3)$, so all Wilson
coefficients are connected to the coefficient of the leading operator
$\bar\chi_{n,\omega_1} \bnslash \chi_{n,\omega_2}$. Note that here all of the
connected operators involve a $q_\perp$, which we have counted as ${\cal
  O}(\lambda^0)$. We will see below that for situations with two collinear
directions, where in the end its natural to specialize to a frame where
$q_\perp=0$, the connections tend to appear at higher twist. For situations with
three or more collinear directions RPI will provide useful constraints on the
basis already at lowest order.

\subsection{General Quark and Gluon Operators} \label{sec:genQG}

In this section we enumerate an operator basis for the general set of collinear
quark and gluon operators up to ${\cal O}(\lambda^4)$.  This basis is useful
for many applications, and we keep our notation as general as possible.  In
particular we consider up to 4 distinct collinear directions (which for example
could be used for $e^+e^-\to 4{\rm jets}$, or $gg, qg, q\bar q\to 2\,{\rm
  jets}$).  We also discuss a basis both for the homogeneous operators with a
definite power counting, and for the RPI operators.

For processes with a hard $q^\mu$, the most general basis of homogeneous quark
operators in SCET up to ${\cal O}(\lambda^4)$ is
\begin{align} \label{GB}
 O^{(0a)} &= \bar{\chi}_{n_1,\omega_1}\, \Gamma
  \, \chi_{n_2,\omega_2} \,,
 & O^{(1a)} &= \bar{\chi}_{n_1,\omega_1}  \Gamma_{\alpha}\,
 \mathcal{P}^{\dagger\alpha}_{\!n_1\perp}\,  \chi_{n_2,\omega_2} \, ,
  \\[3pt]
 O^{(1b)}  &= \bar{\chi}_{n_1,\omega_1} \, \Gamma_{\alpha} \,
   \mathcal{P}^{\alpha}_{\!n_2\perp}\,  \chi_{n_2,\omega_2} \,,
 & O^{(1c)}&= \bar{\chi}_{n_1,\omega_1}\, \Gamma_\alpha
  \big( i  g \mathcal{B}^{\alpha}_{n_3\perp } \big)_{\omega_3}\, \chi_{n_2,\omega_2} \, , 
 \nonumber \\[3pt]
 O^{(2a)}  &= \bar{\chi}_{n_1,\omega_1}  \Gamma_{\alpha\beta} \,
   \mathcal{P}^{\dagger\alpha}_{\!n_1\perp }  \mathcal{P}^{\beta}_{\!n_2\perp}\,
  \chi_{n_2,\omega_2}  \,, 
& O^{(2b)} &= \bar{\chi}_{n_1,\omega_1}  \Gamma_{\alpha\beta} \,
 \mathcal{P}^{\dagger\alpha}_{\!n_1\perp }  \mathcal{P}^{\dagger
   \beta}_{\!n_1\perp} \, \chi_{n_2,\omega_2} \, ,
  \nonumber \\[3pt]
O^{(2c)} &= \bar{\chi}_{n_1,\omega_1}  \Gamma_{\alpha\beta}
  \, \mathcal{P}^{\alpha}_{\!n_2\perp}  \mathcal{P}^{\beta}_{\!n_2\perp}  \chi_{n_2,\omega_2} \,,
& O^{(2d)}&= \bar{\chi}_{n_1,\omega_1} \Gamma_{\alpha\beta}\,
  \mathcal{P}^{\dagger \alpha}_{\!n_1\perp} 
  \big( g  \mathcal{B}^{ \beta}_{n_3\perp}\big)_{\omega_3} \,\chi_{n_2,\omega_2} \, , 
 \nonumber \\[3pt]
 O^{(2e)}&= \bar{\chi}_{n_1,\omega_1} \Gamma_{\beta\alpha}
   \big( g  \mathcal{B}^{\beta}_{n_3\perp }\big)_{\omega_3}  \mathcal{P}^{\perp  \alpha}_{n_2
    } \chi_{n_2,\omega_2} \, , 
&  O^{(2f)}&= \bar{\chi}_{n_1,\omega_1} \Gamma_{\alpha\beta}
   \big[\mathcal{P}^{\perp\alpha }_{n_3}  \big(g  \mathcal{B}^{\perp\beta}_{n_3 
   }\big)_{\omega_3} \big]  \chi_{n_2,\omega_2} \, ,
 \nonumber \\[3pt]
  O^{(2g)}&= \bar{\chi}_{n_1,\omega_1}  \Gamma_{\alpha\beta}\, 
   \big(g \mathcal{B}^{\alpha } _{n_3\perp }\big)_{\omega_3} \big( g
   \mathcal{B}^{\beta}_{n_4\perp}\big)_{\omega_4}\,  \chi_{n_2,\omega_2} \, , 
& O^{(2h)} &= \big(\bar{\chi}_{n_1,\omega_1}\Gamma_1
 \chi_{n_2,\omega_2}\big)\big(\bar{\chi}_{n_3,\omega_3}\Gamma_2
 \chi_{n_4,\omega_4}\big) . 
 \nn
\end{align} 
If we need to specify the subscripts we write for example
$O^{(2g)}(\omega_1,\omega_3,\omega_4,\omega_2)$, with the $\omega_i$ listed from
left to right. Due to the equations of motion in Eqs.~(\ref{eom},\ref{eomg}) we
did not need to consider $in\cdot\partial_n \chi_n$ or $g n\cdot \cB_n$.  For
each operator there may be a set of different Dirac, flavor, and color
structures $\Gamma^j_{\alpha_1\cdots\alpha_n}$ which depend on the particular
phenomena being studied (including also two choices for color for the $\Gamma_i$
in the four-quark operators $O^{(2i)}$).  In general for each independent
$\Gamma^j_{\alpha_1\cdots\alpha_n}$ structure the operator has a Wilson
coefficient that must be determined order by order in perturbation theory.  We
included in Eq.~(\ref{GB}) the mixed quark and gluon operators. For pure gluon
operators up ${\cal O}(\lambda^4)$ we have the homogeneous basis
\begin{align} \label{GB2}
 O^{(0b)} &= {\cal B}^{\perp\mu}_{n_1,\omega_1}
  \, {\cal B}^{\perp\nu}_{n_2,\omega_2} \,,
& O^{(1d)} &= {\cal B}^{\perp\mu}_{n_1,\omega_1}
  \, {\cal P}_{n_1\perp}^{\dagger\alpha}
  \, {\cal B}^{\perp\nu}_{n_2,\omega_2} \,,
  \\[4pt]
 O^{(1e)} &= {\cal B}^{\perp\mu}_{n_1,\omega_1}
  \, {\cal P}_{n_2\perp}^{\alpha}
  \, {\cal B}^{\perp\nu}_{n_2,\omega_2} \,,
 & O^{(1f)} &=   {\cal B}^{\perp\mu}_{n_1,\omega_1}
  \, {\cal B}^{\perp\nu}_{n_2,\omega_2} \, {\cal
    B}^{\perp\tau}_{n_3,\omega_3} \,,
  \nn\\[4pt]
 O^{(2i)} &= {\cal B}^{\perp\mu}_{n_1,\omega_1}
  \, {\cal P}_{n_1\perp}^{\dagger\alpha}  {\cal P}_{n_1\perp}^{\dagger\beta}
  \, {\cal B}^{\perp\nu}_{n_2,\omega_2} \,,
 & O^{(2j)} &= {\cal B}^{\perp\mu}_{n_1,\omega_1}
  \,  {\cal P}_{n_1\perp}^{\dagger\alpha} {\cal P}_{n_2\perp}^{\beta}
  \, {\cal B}^{\perp\nu}_{n_2,\omega_2} \,,
 \nn \\[4pt]
 O^{(2k)} &= {\cal B}^{\perp\mu}_{n_1,\omega_1}
  \, {\cal P}_{n_2\perp}^{\alpha} {\cal P}_{n_2\perp}^{\beta} 
  \, {\cal B}^{\perp\nu}_{n_2,\omega_2}\,,
 & O^{(2l)} &=   \big[ {\cal P}_{n_1\perp}^{\alpha}  
  {\cal B}^{\perp\mu}_{n_1,\omega_1} \big]
  \, {\cal B}^{\perp\nu}_{n_2,\omega_2} \, {\cal
    B}^{\perp\tau}_{n_3,\omega_3} \,,
  \nn\\[6pt]
 O^{(2m)} &= {\cal B}^{\perp\mu}_{n_1,\omega_1}
  \, \big[  {\cal P}_{n_2\perp}^{\alpha}  
   {\cal B}^{\perp\nu}_{n_2,\omega_2} \big] 
  \, {\cal B}^{\perp\tau}_{n_3,\omega_3}\,,
 & O^{(2n)} &=   {\cal B}^{\perp\mu}_{n_1,\omega_1}
  \, {\cal B}^{\perp\nu}_{n_2,\omega_2} \, \big[
   {\cal P}_{n_3\perp}^{\alpha}  {\cal B}^{\perp\tau}_{n_3,\omega_3} \big]\,,
  \nn\\[4pt]
 O^{(2o)} &=   {\cal B}_{n_1,\omega_1}^{\perp\mu}
  \,  {\cal B}^{\perp\nu}_{n_2,\omega_2} \,
  {\cal B}^{\perp\sigma}_{n_3,\omega_3} \,
    {\cal B}^{\perp\tau}_{n_4,\omega_4} \,.
  \nn
\end{align}
Here we do not need to consider operators with $gn \mcdot \mathcal{B}_{n}$ and
$gn \mcdot \partial_n \mathcal{B}^\mu_{n\perp}$ because using the equations of
motion in Eq.~(\ref{eomg}) they can be written in terms of the operators in
Eq.~(\ref{GB2}), and are hence redundant.  

To setup the computation of constraints on Wilson coefficients we also need to
build an RPI basis of operators using the objects in Eq.~(\ref{RPdef}) and
$i\partial_n^\mu$. Because each operator will be RPI, its Wilson coefficient is
truly independent of those for other operators in the basis.  The RPI operators
can then be expanded in terms of homogeneous operators made out of of gauge
invariant objects, and doing so we obtain operators in the homogeneous basis
with all the constraints coming from reparametrization invariance.  The number
of constraints on Wilson coefficients is equal to the number of homogeneous
operators minus the number of RPI operators, once we have accounted for linear
dependencies~\cite{Finkemeier:1997re,Lee:1997bi}.

Let's construct the RPI basis of operators which is the analog of those in
Eqs.~(\ref{GB}) and (\ref{GB2}).  The operators with no $i\partial_n^\mu$
derivatives are
%
%
\begin{align} \label{0RPIC}
\textbf{Q}^{(0q)}
   = &\bar{\Psi}_{n_1,\hat{\omega}_1}\Gamma \Psi_{n_2,\hat{\omega}_2} \, , 
 & \textbf{Q}^{(0g)} & 
   = {\cal G}_{n_1,\hat{\omega}_1}^{\mu\nu} \, {\cal G}_{n_2,\hat{\omega}_2}^{\sigma\tau}\,,
\end{align}
where the basis of Dirac structures $\Gamma$, and contraction of indices
$\mu\nu\sigma\tau$ in ${\bf Q}^{(0g)}$ depends on the kind of current we are
studying.  For cases without a $q^\mu$ the subscripts $\hat \omega_i$ are erased
and RPI operators are multiplied by the $\hat\Delta_{ij}$ factors shown in
Eq.~(\ref{Delijexpn}).  Recall that we do not have a good power counting in the
RPI basis, this basis makes the RPI properties transparent but the power
counting more tricky. When ${\bf Q}^{(aq)}$ and ${\bf Q}^{(ag)}$ are expanded in
terms of operators that are homogeneous in the power counting, they contain a
leading order term, so they are relevant operators to consider at LO.  Of the
RPI objects only $i
\partial_n^\mu$ starts at leading order, so theoretically we can construct an
infinite set of LO operators using $(i\partial_n)^k$ for any $k$.  However, the
structure of this operator provides additional constraints. In particular the
${\cal O}(\lambda^0)$ term is $i\partial_n^\mu = (n^\mu/2)
\bnP_n+\cdots$, and the collinear momentum $\bnP_n$
acting on a $n$-collinear field such as $\chi_{n,\omega_1}$ just gives a number,
$\omega_1$, which can be absorbed into the Wilson coefficient
$C({\omega}_1,{\omega}_2)$. For cases with a $q^\mu$ this implies that adding $i
\partial_n^\mu$'s in a scalar operator (where all vector indices are contracted)
most often gives an operator that differs from one we already have only at
${\cal O}(\lambda)$.  For these scalar operators we can count
$i\partial_n^\mu\sim {\cal O}(\lambda)$ when determining which RPI operators are
required, and for simplicity we follow this counting in the remainder of this
section. If we have an operator with a free vector index $\mu$, then this index
can be carried by $i\partial_n^\mu = (n^\mu/2) \bnP_n +\cdots$, and
the partial derivative does count as ${\cal O}(\lambda^0)$.

The expansion of the RPI operators in Eq.~(\ref{0RPIC}) in terms of homogeneous
operators up to ${\cal O}(\lambda^4)$ is
\begin{align}
 \textbf{Q}^{(0q)} &=
  \bar{\Psi}^{(1)}_{n_1,\hat{\omega}_1}\Gamma\Psi^{(1)}_{n_2,\hat{\omega}_2}
  +\bar{\Psi}^{(2)}_{n_1,\hat{\omega}_1}\Gamma\Psi^{(1)}_{n_2,\hat{\omega}_2}
  +\bar{\Psi}^{(1)}_{n_1,\hat{\omega}_1}\Gamma\Psi^{(2)}_{n_2,\hat{\omega}_2} 
 \\[4pt]
  &\ \  +\bar{\Psi}^{(2)}_{n_1,\hat{\omega}_1}\Gamma\Psi^{(2)}_{n_2,\hat{\omega}_2}
  +\bar{\Psi}^{(1)}_{n_1,\hat{\omega}_1}\Gamma\Psi^{(3)}_{n_2,\hat{\omega}_2}
  +\bar{\Psi}^{(3)}_{n_1,\hat{\omega}_1}\Gamma\Psi^{(1)}_{n_2,\hat{\omega}_2}
  + O(\lambda^5) \,,
  \nonumber \\[6pt]
  \textbf{Q}^{(0g)} &=
   {\cal G}_{n_1,\hat{\omega}_1}^{(1)\mu\nu} \, {\cal G}_{n_2,\hat{\omega}_2}^{(1)\sigma\tau}
 + {\cal G}_{n_1,\hat{\omega}_1}^{(2)\mu\nu} \, {\cal G}_{n_2,\hat{\omega}_2}^{(1)\sigma\tau}
 + {\cal G}_{n_1,\hat{\omega}_1}^{(1)\mu\nu} \, {\cal G}_{n_2,\hat{\omega}_2}^{(2)\sigma\tau}
  \nn\\[4pt]
&\ \  + {\cal G}_{n_1,\hat{\omega}_1}^{(2)\mu\nu} \, {\cal G}_{n_2,\hat{\omega}_2}^{(2)\sigma\tau}
 + {\cal G}_{n_1,\hat{\omega}_1}^{(3)\mu\nu} \, {\cal G}_{n_2,\hat{\omega}_2}^{(1)\sigma\tau}
 + {\cal G}_{n_1,\hat{\omega}_1}^{(1)\mu\nu} \, {\cal G}_{n_2,\hat{\omega}_2}^{(3)\sigma\tau}
 + O(\lambda^5) \,, \nn
\end{align}
where the $\Psi^{(k)}_{n,\hat{\omega}}$ and ${\cal G}_{n,\hat{\omega}}^{(k)}$
are given in Eqs.~(\ref{Psin}) and (\ref{Gsin}).  To look for RPI relations the
results of this expansion must be compared to power suppressed operators which
also can generate ${\cal O}(\lambda^3)$ and ${\cal O}(\lambda^4)$ terms. Up to
this order the power suppressed operators involving two or more quark fields are
\begin{align} \label{RPIop}
\textbf{Q}^{(1a)}
 &=  \bar{\Psi}_{n_1,\hat{\omega}_1}\Gamma_{\alpha}\,  i \partial_{n_2}^{\alpha} \,
 \Psi_{n_2,\hat{\omega}_2} \, ,  
%
%
& \textbf{Q}^{(1b)}
  &=  \bar{\Psi}_{n_1,\hat{\omega}_1} \Gamma_{\alpha} 
  \: i\!\overleftarrow{\partial}_{\!\!n_1}^{\!\alpha} \: \Psi_{n_2,\hat{\omega}_2}
  \,,
 \\[3pt]
%
%
%
\textbf{Q}^{(1c)}
  &=  \bar{\Psi}_{n_1,\hat{\omega}_1}\Gamma_{\beta {\beta^\prime}}\,
  \mathcal{G}^{\beta{\beta^\prime}}_{n_3,\hat{\omega}_3}\, \Psi_{n_2,\hat{\omega}_2}  \, ,
%
%
%
& \textbf{Q}^{(2a)}
 &= 
\bar{\Psi}_{n_1,\hat{\omega}_1} \Gamma_{\alpha\alpha^\prime} \: i\!\overleftarrow{
  \partial}_{\!\!n_1}^{\!\alpha}\, i \partial_{n_2}^{\alpha^\prime}\: \Psi_{n_2,\hat{\omega}_2} \,
, 
 \nonumber \\[3pt]
%
%
 \textbf{Q}^{(2b)}
&= 
  \bar{\Psi}_{n_1,\hat{\omega}_1} \Gamma_{\alpha\alpha^\prime}
\: i\!\overleftarrow{\partial}_{\!\!n_1}^{\!\alpha} i\! \overleftarrow{
  \partial}_{\!\!n_1}^{\!\alpha^\prime} \: \Psi_{n_2,\hat{\omega}_2} \, ,
%
%
%
& \textbf{Q}^{(2c)} 
 &= 
\bar{\Psi}_{n_1,\hat{\omega}_1} \Gamma_{\alpha\alpha^\prime} \: i
\partial_{n_2}^{\alpha} \, i \partial_{n_2}^{\alpha^\prime}\: \Psi_{n_2,\hat{\omega}_2} \, , 
 \nonumber \\[3pt]
%
%
%
\textbf{Q}^{(2d)}
 &=
  \bar{\Psi}_{n_1,\hat{\omega}_1} \Gamma_{\alpha \beta\beta^\prime}  \: i\!
\overleftarrow{\partial}_{\!\! n_1}^{\!\alpha}\:
\mathcal{G}_{n_3,\hat{\omega}_3}^{\beta\beta^\prime}  \Psi_{n_2,\hat{\omega}_2}  \, , 
%
%
& \textbf{Q}^{(2e)}
 &=
  \bar{\Psi}_{n_1,\hat{\omega}_1} \Gamma_{\alpha \beta\beta^\prime}  \big[ i
  \partial_{n_3}^{\alpha} \mathcal{G}_{n_3,\hat{\omega}_3}^{\beta\beta^\prime} \big]
\Psi_{n_2,\hat{\omega}_2}  \, , 
  \nonumber \\[3pt]
%
%
\textbf{Q}^{(2f)}  
&=
  \bar{\Psi}_{n_1,\hat{\omega}_1} \Gamma_{\alpha \beta\beta^\prime}\,
\mathcal{G}_{n_3,\hat{\omega}_3}^{\beta\beta^\prime} \: i \partial_{n_2}^{\alpha}\:
\Psi_{n_2,\hat{\omega}_2}  \, , 
%
%
& \textbf{Q}^{(2g)}
 &=  \bar{\Psi}_{n_1,\hat{\omega}_1}\Gamma_{\alpha\beta\gamma\delta}
 \,\mathcal{G}^{\alpha\beta}_{n_3,\hat{\omega}_3}
 \, \mathcal{G}^{\gamma\delta}_{n_4,\hat{\omega}_4} \Psi_{n_2, \hat{\omega}_2}  \, , 
  \nonumber \\[3pt]
%
 \textbf{Q}^{(2h)}
 &= \left[\bar{\Psi}_{n_1,\hat{\omega}_1}\Gamma_{1}\Psi_{n_2,\hat{\omega}_2}\right] 
 \left[\bar{\Psi}_{n_3,\hat{\omega}_3}\Gamma_{2}\Psi_{n_4,\hat{\omega}_4}\right] 
   \,. \hspace{-0.5cm} \nn
\end{align}   
Again a minimal basis for Dirac structures $\Gamma$ will depend on the process
being studied and may differ between the various ${\mathbf Q}^{(ix)}$ operators. Such a
basis will also in general differ from the one for the homogeneous operators in Eq.~(\ref{GB}).
We will adopt notation such as
${\mathbf Q}^{(2g)}(\hat\omega_1,\hat\omega_3,\hat\omega_4,\hat\omega_2)$ when we wish to
specify these subscripts.  For a field basis for the higher order operators with gluon
fields (whose expansion starts at ${\cal O}(\lambda^3)$ or ${\cal O}(\lambda^4)$) we have
\begin{align} \label{RPIop2}
  \textbf{Q}^{(1d)} & 
   = {\cal G}_{n_1,\hat{\omega}_1}^{\mu\nu}  i\partial_{n_2}^\alpha\,
       {\cal G}_{n_2,\hat{\omega}_2}^{\sigma\tau}\,,
  & \textbf{Q}^{(1e)} & 
   = {\cal G}_{n_1,\hat{\omega}_1}^{\mu\nu}  i\!\overleftarrow \partial \!{}_{n_1}^\alpha
      \,  {\cal G}_{n_2,\hat{\omega}_2}^{\sigma\tau}\,,
 \\
  \textbf{Q}^{(1f)} & 
   = {\cal G}_{n_1,\hat{\omega}_1}^{\mu\nu} 
       {\cal G}_{n_2,\hat{\omega}_2}^{\sigma\tau} {\cal G}_{n_3,\hat{\omega}_3}^{\alpha\beta}\,,
& \textbf{Q}^{(2i)} & 
   = {\cal G}_{n_1,\hat{\omega}_1}^{\mu\nu}  i\partial_{n_2}^\alpha\,i\partial_{n_2}^\beta\,
       {\cal G}_{n_2,\hat{\omega}_2}^{\sigma\tau}\,,
\nn\\
  \textbf{Q}^{(2j)} & 
   = {\cal G}_{n_1,\hat{\omega}_1}^{\mu\nu}  
    i\!\overleftarrow \partial \!{}_{n_1}^\alpha 
    i\!\overleftarrow \partial \!{}_{n_1}^\beta
      \,  {\cal G}_{n_2,\hat{\omega}_2}^{\sigma\tau}\,,
 & \textbf{Q}^{(2k)} & 
   = {\cal G}_{n_1,\hat{\omega}_1}^{\mu\nu}  
    i\!\overleftarrow \partial \!{}_{n_1}^\alpha  i\partial_{n_2}^\beta
      \,  {\cal G}_{n_2,\hat{\omega}_2}^{\sigma\tau}\,,
 \nn\\
\textbf{Q}^{(2l)} & 
   = [i\partial_{n_1}^\gamma {\cal G}_{n_1,\hat{\omega}_1}^{\mu\nu} ]
       {\cal G}_{n_2,\hat{\omega}_2}^{\sigma\tau} {\cal G}_{n_3,\hat{\omega}_3}^{\alpha\beta}\,,
 & \textbf{Q}^{(2m)} & 
   =  {\cal G}_{n_1,\hat{\omega}_1}^{\mu\nu} 
      [i\partial_{n_2}^\gamma {\cal G}_{n_2,\hat{\omega}_2}^{\sigma\tau} ]
     {\cal G}_{n_3,\hat{\omega}_3}^{\alpha\beta}\,,
\nn\\ 
 \textbf{Q}^{(2n)} & 
   =  {\cal G}_{n_1,\hat{\omega}_1}^{\mu\nu} 
       {\cal G}_{n_2,\hat{\omega}_2}^{\sigma\tau} 
     [i\partial_{n_3}^\gamma {\cal G}_{n_3,\hat{\omega}_3}^{\alpha\beta}] \,,
 & \textbf{Q}^{(2o)} & 
   =  {\cal G}_{n_1,\hat{\omega}_1}^{\mu\nu} 
       {\cal G}_{n_2,\hat{\omega}_2}^{\sigma\tau} 
      {\cal G}_{n_3,\hat{\omega}_3}^{\alpha\beta} 
     {\cal G}_{n_4,\hat{\omega}_4}^{\gamma\delta} \,.
 \nn
\end{align}
We will include a basis of Dirac structures and expand the RPI operators in
Eqs.~(\ref{RPIop}) and (\ref{RPIop2}) in terms of the homogeneous ones in
several of the examples below, and consider whether there are non-trivial RPI
relations on a case-by-case basis.

\subsection{Deep Inelastic Scattering for Quarks at Twist-4}
\label{sec:DISq}

In this section we consider spin-averaged DIS at twist-4. This provides a test
of our technique of constructing a minimal basis, for an example where the basis
is already well known~\cite{Jaffe:1981td,Jaffe:1982pm,Ellis:1982cd}. We will see
that RPI constrains the Wilson coefficients of the homogeneous collinear
operators. Our analysis is really of scalar operators with one collinear
direction, $q_\perp=0$, with overall derivatives set to zero. DIS is the most
popular application for these operators, so we frame our discussion in that
language. For simplicity we consider the QCD electromagnetic current $J^\mu=
\bar q \gamma^\mu q$ for one-flavor of quark.  (We briefly discuss the
generalization to non-singlet operators in a footnote.)  The study of higher
twist in DIS and related processes is an active field of research, for
example~\cite{Geyer:2000ma,Retey:2000nq,Brodsky:2002cx,Collins:2002kn,Ji:2004wu,Belitsky:2005qn}.
In the language of SCET, DIS was first studied in~\cite{Bauer:2002nz}, whose
notation we follow.  The virtual photon has momentum transfer $q^2=-Q^2$, and $x
= Q^2/(2p\cdot q ) $ is the Bjorken variable.

In the Breit frame the momentum of the virtual photon is $q^\mu = Q(\bar{n}^\mu
- n^\mu)/2$, and the incoming proton momentum is $p^\mu =n^\mu \bar{n} \cdot p
/2 + \bar{n}^\mu m_p^2/(2\bar{n} \cdot p)$ where $m_p$ is the mass of the
proton.  Expanding in $m_p/Q$ we have $\bn\cdot p = Q /x -x m_p^2/Q + \ldots$.
The energetic proton has a small invariant mass $p^2 = m^2_p \sim \Lambda_{\rm
  QCD}^2$, and in the Breit frame it is described by collinear fields in the
effective theory with a power counting in $\lambda=\Lambda_{\rm QCD}/Q$. It is
convenient to pick this frame in order to be able to assign definite power
counting to momentum components. What reparametrization invariance enforces is
that all results are invariant to small perturbations about this frame, encoded by
changes to the collinear reference vector $n^\mu$. Since these changes are small
we are free to use the same power counting when studying the RPI relations.
There is a larger class of frame independence, which says for example that the
same results would be found if we compare an analysis in the Breit-frame with an
analysis made about the initial proton rest frame, but this set of ``big'' frame
transformations does not encode non-trivial dynamic information that relates
coefficients of operators at higher twist. All final results are of course
entirely frame independent.

For spin-averaged DIS the hadronic tensor has the structure
\begin{align}
T_{\mu\nu} = & \left(-g_{\mu \nu} +\frac{q_\mu q_\nu}{q^2} \right) T_1(x,Q^2)
 + \left( p_\mu + \frac{q_\mu}{2x} \right) \left( p_\nu + \frac{q_\nu}{2x} \right) T_2(x,Q^2), 
\end{align}
where 
\begin{equation} \label{TThat}
T_{\mu\nu}(p,q) = \frac{1}{2} \sum_{\mathrm{spin}} \langle p|\hat{T}_{\mu\nu} (q)  |p\rangle, \qquad \hat{T}_{\mu\nu}  (q)  =i \int d^4 z e^{i q\cdot z} \mathrm{T} [J_\mu(z), J_\nu (0)] \, .
\end{equation}
The scalar structure functions $T_i$ can be projected out of $T_{\mu\nu}$ using
\begin{align} \label{T1T2}
  T_1(Q^2,x) &= -\frac12 \Big( g^{\mu\nu} -
   \frac{4x^2}{Q^2 + 4 m_p^2 x^2}\, p^\mu p^\nu\Big)
  T_{\mu\nu} \,,\nn\\
  T_2(Q^2,x) &= -\frac{2x^2}{Q^2+ 4 m_p^2 x^2} \Big( g^{\mu\nu} 
   - \frac{12x^2}{Q^2 + 4 m_p^2 x^2}\, p^\mu p^\nu\Big)
  T_{\mu\nu} \,.
\end{align}
The expansion of $T_1$ and $T_2$ has been carried out up to twist-4 with the
Wilson coefficients determined at tree level in
Refs.~\cite{Jaffe:1981td,Jaffe:1982pm,Ellis:1982cd}.  To simplify our
calculations we will make use of the fact that the projections in
Eq.~(\ref{T1T2}) commute with taking the proton matrix element, and hence can be
applied directly to $\hat T_{\mu\nu}$ to give $\hat T_1$ and $\hat T_2$, where
$\frac12 \sum_{\rm spin} \langle p | \hat T_i | p\rangle = T_i(Q^2,x)$. Thus we
consider the expansion of $\hat T_1$ and $\hat T_2$ in scalar chiral-even
operators, by writing
\begin{align} \label{Ti1}
  \hat T_i  &= \sum_j \, \int\! [d\omega_k]\: C_{j}^{[i]}(\omega_k)  O_j(\omega_k) \,.
\end{align}
Here $[d\omega_k] = d\omega_1\cdots d\omega_n$ is the integration measure over
the independent parton momenta $\omega_k$ carried by the Wilson coefficients
$C_j^{[i]}$ and the operators $O_j$. The superscript $[i]$ indicates that the
Wilson coefficients for the two tensor structures will in general differ. We
also consider a basis of RPI operators $\textbf{Q}_j$ by writing
\begin{align} \label{Ti2}
  \hat T_i &= \sum_j \, \int\! [d\hat\omega_k]\: \hat C_{j}^{[i]}(\hat\omega_k)
   \textbf{Q}_j(\hat \omega_k) \,.
\end{align}
Unlike the $O_j$'s the $\textbf{Q}_j$'s do not contain contributions of a definite order
in the power counting.  Using the RPI $\textbf{Q}_j$ operators we can test if there are
relations between the Wilson coefficients $C_j^{[i]}$ of the $O_j$'s. A
connection would mean, for example, that the one-loop coefficient for a twist-4
operator is determined by a coefficient at twist-2 at all orders in
$\alpha_s$.

We first write down a gauge invariant basis of chiral-even quark operators that
are homogeneous in the power counting. This can be done using the general basis
in Eq.~(\ref{GB}) with all directions $n_i =n$.  Furthermore, since the DIS
matrix element is forward, we have $\langle p | [\cP^\mu O] | p\rangle = 0$ for
any operator $O$. Thus we are free to integrate $\perp$-label momentum operators
by parts, and hence can ignore all terms with $\cP_\perp^\dagger$'s in
Eq.~(\ref{GB}).  (If we consider our analysis to be of the general scalar
operators with one collinear direction, then this is the only simplification that
we make which relies on the form of the final matrix element.) For simplicity we
also drop the square-brackets from inside $O^{(2f)}$ in Eq.~(\ref{GB}). A
minimal basis of chiral-even parity-even Dirac structures between the
$n$-collinear quark fields is easily constructed using the properties of the
SCET $\chi_n$ fields. We have i) just $\{\bnslash\}$ when there are no vector
indices on fields, ii) no elements at all when there is one vector index, and
iii) just $\{\bnslash g_\perp^{\mu\nu},\,
i\epsilon_{\perp}^{\mu\nu}\bnslash\gamma_5 \}$ or $\{\bnslash
g_\perp^{\mu\nu},\, \bnslash\gamma_\perp^\mu \gamma_\perp^\nu\}$ for two vector
indices on fields. Here ii) is the standard fact that the spin-averaged case
does not have twist-3 terms. (For polarized DIS it does not suffice to only
consider the scalar operators.) For the four-quark operators we can have
$\Gamma_1\otimes \Gamma_2 = \{ \bnslash\otimes\bnslash ,\, \bnslash\gamma_5
\otimes\bnslash\gamma_5\}$ and color structures $1\otimes 1$ or $T^A\otimes
T^A$. Thus the basis is
\begin{align} \label{DISGI}
 O_{1} &= \bar{\chi}_{n,\omega_1}  \frac{\bnslash}{2}\,
  \chi_{n,\omega_2} \, ,
& O_{2} &=  \bar{\chi}_{n,\omega_1}  \frac{\bnslash}{2}\, 
  \mathcal{P}_{\perp}^2 \,  \chi_{n,\omega_2} \, , 
  \\[3pt]
 O_{3a} &=  \bar{\chi}_{n,\omega_1}  \frac{\bnslash}{2}\, 
   ( g  \mathcal{B}_{n\perp}^\mu)_{\omega_3}  \mathcal{P}_\mu^{\perp }
      \chi_{n,\omega_2} \, ,\hspace{-2cm}
 & O_{3b} &=  \bar{\chi}_{n,\omega_1}  \frac{\bnslash}{2}\,  \mathcal{P}_\mu^{\perp }
   ( g  \mathcal{B}_{n\perp}^\mu)_{\omega_3}
      \chi_{n,\omega_2} \, ,
 \nonumber \\[3pt]
  O_{4a}  &= \bar{\chi}_{n,\omega_1} \frac{\bnslash}{2}\,
  ( g  \mathcal{B\!\!\!\slash}_{n\perp})_{\omega_3} \mathcal{P}\!\!\!\!\slash_\perp
   \chi_{n,\omega_2}  \, ,   
   & O_{4b}  &= \bar{\chi}_{n,\omega_1} \frac{\bnslash}{2}\,
  \mathcal{P}\!\!\!\!\slash_\perp ( g  \mathcal{B\!\!\!\slash}_{n\perp})_{\omega_3} 
   \chi_{n,\omega_2}  \, , 
  \hspace{-1.7cm} \nn\\[3pt]
  O_{5} &=  \bar{\chi}_{n,\omega_1}   \frac{\bnslash}{2}\,
   (g \mathcal{B\!\!\!\slash}_{n\perp})_{\omega_3}
   ( g  \mathcal{B\!\!\!\slash}_{n\perp})_{\omega_4}  \chi_{n,\omega_2}
   \, , 
& O_{6} &=  \bar{\chi}_{n,\omega_1}   \frac{\bnslash}{2}\, {\rm Tr} \big[
   (g \mathcal{B\!\!\!\slash}_{n\perp})_{\omega_3}
   ( g  \mathcal{B\!\!\!\slash}_{n\perp})_{\omega_4} \big] \chi_{n,\omega_2}
   \, ,  \nonumber \\[3pt]
O_{7} &=  \bar{\chi}_{n,\omega_1}  \frac{\bnslash}{2}\,    
   (g \mathcal{B}_{n\perp}^\mu)_{\omega_3}( g  \mathcal{B}^{\perp
     n}_\mu )_{\omega_4}  \chi_{n,\omega_2} \, ,
& O_{8} &= \bar{\chi}_{n,\omega_1}  \frac{\bnslash}{2}\,   {\rm Tr}\big[ 
   (g \mathcal{B}_{n\perp}^\mu)_{\omega_3}( g  \mathcal{B}^{\perp
     n}_\mu )_{\omega_4} \big]  \chi_{n,\omega_2}   \, ,  
 \nonumber \\[3pt]
 O_{9} &=\Big[\bar{\chi}_{n,\omega_1} \frac{\bar{n}\!\!\!\slash}{2} 
 \chi_{n,\omega_2}\Big]
  \Big[\bar{\chi}_{n,\omega_3} \frac{\bar{n}\!\!\!\slash}{2}  \chi_{n,\omega_4}\Big]  \,,
 & O_{10} &= \Big[\bar{\chi}_{n,\omega_1} \frac{\bar{n}\!\!\!\slash}{2}  \gamma_5  
 \chi_{n,\omega_2}\Big]
  \Big[ \bar{\chi}_{n,\omega_3}  \frac{\bar{n}\!\!\!\slash}{2} \gamma_5 
   \chi_{n,\omega_4} \Big] \, ,
 \nonumber \\[3pt]
 O_{11} &=   \Big[\bar{\chi}_{n,\omega_1}
  \frac{\bar{n}\!\!\!\slash}{2}  \gamma_5  T^A \chi_{n,\omega_2}\Big]
  \Big[ \bar{\chi}_{n,\omega_3}  \frac{\bar{n}\!\!\!\slash}{2} \gamma_5  
    T^A \chi_{n,\omega_4} \Big] ,
%
 & O_{12} &= \Big[\bar{\chi}_{n,\omega_1} \frac{\bar{n}\!\!\!\slash}{2}  T^A 
 \chi_{n,\omega_2}\Big]
  \Big[ \bar{\chi}_{n,\omega_3}  \frac{\bar{n}\!\!\!\slash}{2} T^A
   \chi_{n,\omega_4} \Big] \, . \nonumber
\end{align}
Recall that in an operator like $O_2$ the position space analog of
$\cP_\perp^\mu$ is to translate all gluon and quark fields in
$\chi_{n,\omega_2}$ in $x_\perp$, differentiate twice with respect to
$x_\perp^\mu$, and then set $x_\perp=0$. The basis shown in Eq.~(\ref{DISGI})
can be used to describe twist-4 effects in DIS at any order in $\alpha_s$.  Note
that we have already discussed and taken into account the quark and gluon
equations of motion in the general result in Eq.~(\ref{GB}) and hence already in
Eq.~(\ref{DISGI}). For $O_{5,7}$ there are two color structures associated with
the product of ${\cal B}_n$'s, but these are picked out by consider Wilson
coefficients $C_{5,7}$ that are odd or even in the exchange
$\omega_3\leftrightarrow \omega_4$. The forward proton matrix element of these
operators will be proportional to an overall $\delta$-function, which is
$\delta(\omega_1-\omega_2)$ for $O_{1,2}$, $\delta(\omega_1+\omega_3-\omega_2)$
for $O_{3a,3b,4a,4b}$, $\delta(\omega_1+\omega_3+\omega_4-\omega_2)$ for
$O_{5-8}$, and $\delta(\omega_1+\omega_3-\omega_2-\omega_4)$ for $O_{9-12}$.

Next we derive the analogous results for the RPI basis of chiral-even operators.
From Eq.~(\ref{TThat}) the hadronic tensor operator $\hat T_{\mu\nu}$ depends on
$q^\mu$ which we use as our reference vector. To construct this basis we cannot
use $n^\mu$ or $\bn^\mu$. Comparing Eqs.~(\ref{TThat}) and Eq.~(\ref{T1T2}) we
see that it suffices to construct a basis of scalar operators for the expansion
of $g^{\mu\nu} \hat T_{\mu\nu}$ and $p^\mu p^\nu \hat T_{\mu\nu}$.  The forward
proton matrix element of the expansion of these operators then yields an
expansion for the observables $T_1$ and $T_2$.  Thus, for the scalar basis we
allow any number of $q$'s to appear, but only zero or two $p$'s. This implies
that at twist-2 there is only one RPI bilinear quark operator
\begin{align} \label{disQ1}
\textbf{Q}_{1} &= \bar{\Psi}_{n,\hat{\omega}_1}  q\!\!\!\slash\, \Psi_{n,\hat{\omega}_2} \, .
\end{align} 
At twist-3 there are no scalar chiral-even RPI operators. The candidate
operators $\bar{\Psi}_{n} i\partial_n\!\!\! \!\!\!\slash\ \Psi_{n}$ and
$\bar{\Psi}_{n,,\hat\omega_1} q \!\!\!\slash(q \cdot i\partial_n)
\Psi_{n,,\hat\omega_2}$ are ruled out by the equations of motion in
Eqs.~(\ref{EOM1}) and (\ref{EOM1a}). Another possible operator is $\bar{\Psi}_{n} p
\!\!\!\slash(p \cdot \partial_n) \Psi_n$ but it starts at twist-6, since $(p
\cdot \partial_n) \sim O(\lambda^2)$, being suppressed either by an $n\mcdot p$
or $n\mcdot \partial_n$, and $p\!\!\!\slash$ adds another factor 2 to the power
counting when it is squeezed between the $n$-collinear fermion fields $\chi_n$.
All the operators with $\mathcal{G}_n^{\mu\nu}$, like for example
$\bar{\Psi}_{n}\gamma_\mu q_\nu \mathcal{G}^{\mu\nu}_{n}\Psi_{n}$, have
expansions whose lowest term is twist-4 because the Dirac structure of the
twist-3 component of this operator vanishes between the $n$-collinear fermion
fields, since $\bar\chi_n \nslash\chi_n =0$.  Thus the power suppressed terms
start at twist-4 in agreement with the homogeneous basis in Eq.~(\ref{DISGI}).
Writing out the RPI operators different from zero at twist-4 and not connected
by operator relations we have
\begin{align} \label{RPIDI}
%
 %
\textbf{Q}_{2}    &=  \bar{\Psi}_{n,\hat{\omega}_1}\gamma_\mu q_\nu
 \mathcal{G}^{\mu\nu}_{n,\hat{\omega}_3}\Psi_{n,\hat{\omega}_2} \, , 
%
& \textbf{Q}_{3}    &=  \bar{\Psi}_{n,\hat{\omega}_1} \qslash \gamma_\mu \gamma_\nu
 ig \mathcal{G}^{\mu\nu}_{n,\hat{\omega}_3}\Psi_{n,\hat{\omega}_2} \, ,  
\\[3pt]
%
 \textbf{Q}_{4}   &= -g^2 \bar{\Psi}_{n,\hat{\omega}_1} q\!\!\!\slash \,
\gamma_\mu q_\nu   \mathcal{G}^{\mu\nu}_{ n,\hat{\omega}_3} \gamma_\alpha q_\beta
\mathcal{G}^{\alpha\beta}_{n,\hat{\omega}_4} \Psi_{n,\hat{\omega}_2} \, ,
%
  & \textbf{Q}_{5}     &= -g^2 \bar{\Psi}_{n,\hat{\omega}_1} q\!\!\!\slash \,
\gamma_\mu q_\nu  \gamma_\alpha q_\beta {\rm Tr} [  \mathcal{G}^{\mu\nu}_{ n,\hat{\omega}_3}
\mathcal{G}^{\alpha\beta}_{n,\hat{\omega}_4}] \Psi_{n,\hat{\omega}_2}   \, ,
\nonumber \\[3pt]
%
   \textbf{Q}_{6}     &=  -g^2 \bar{\Psi}_{n,\hat{\omega}_1} q\!\!\!\slash \,
q_\mu q_\nu   (\mathcal{G}_{n,\hat{\omega}_3})^\mu_{\:\:\alpha}
\mathcal{G}^{\alpha\nu}_{n,\hat{\omega}_4} \Psi_{n,\hat{\omega}_2} \, ,
 & \textbf{Q}_{7}    &=  -g^2 \bar{\Psi}_{n,\hat{\omega}_1} q\!\!\!\slash \,
q_\mu q_\nu {\rm Tr} [  (\mathcal{G}_{n,\hat{\omega}_3})^\mu_{\:\:\alpha}
\mathcal{G}^{\alpha\nu}_{n,\hat{\omega}_4}] \Psi_{n,\hat{\omega}_2}  \, , 
\nonumber \\[3pt]
%
 \textbf{Q}_{8}    &=  \big[\bar{\Psi}_{n,\hat{\omega}_1} q \!\!\!\slash   \Psi_{n,\hat{\omega}_2}\big] 
  \big[\bar{\Psi}_{n,\hat{\omega}_3} q\!\!\!\slash  \Psi_{n,\hat{\omega}_4}\big]\, ,
%
  & \textbf{Q}_{9}    &=  \big[\bar{\Psi}_{n,\hat{\omega}_1} q \!\!\!\slash   \gamma_5 \Psi_{n,\hat{\omega}_2}\big] \!
 \big[\bar{\Psi}_{n,\hat{\omega}_3} q\!\!\!\slash  \gamma_5 \Psi_{n,\hat{\omega}_4}\big] \, , 
 \nonumber \\[3pt]
%
 \textbf{Q}_{10}    &=   \big[\bar{\Psi}_{n,\hat{\omega}_1} q \!\!\!\slash T^A \Psi_{n,\hat{\omega}_2}\big] 
 \big[\bar{\Psi}_{n,\hat{\omega}_3} q\!\!\!\slash T^A 
  \Psi_{n,\hat{\omega}_4}\big]    \,, 
 & \textbf{Q}_{11}    &=  
\big[\bar{\Psi}_{n,\hat{\omega}_1} q \!\!\!\slash  T^A \gamma_5 \Psi_{n,\hat{\omega}_2}\big] 
 \big[\bar{\Psi}_{n,\hat{\omega}_3} q\!\!\!\slash T^A \gamma_5
  \Psi_{n,\hat{\omega}_4}\big]    \,. \nn 
%
\end{align}
One can think of other possible operators at twist-4, but all of them are either
ruled out by the equations of motion and operator relations, or start at higher
twist.  For example, there are not operators with both $p^\mu$ and
$\mathcal{G}^{\mu\nu}_{\hat{\omega}_3}$, like $\bar{\Psi}_{n, \hat{\omega}_1 }
\gamma_\mu p_\nu \mathcal{G}^{\mu\nu}_{n, \hat{\omega}_3}\Psi_{n,
  \hat{\omega_2}} $, because they all start at higher twist. We have integrated
by parts making all derivatives act to the right, since here our interest is in
forward matrix elements, and we removed $i\partial_{\nu} {\cal G}_{n,\hat
  \omega}^{\mu\nu}$ with the gluon equation of motion in Eq.~(\ref{EOM2}). The
operator $\bar\Psi_{n,\hat\omega_1} \qslash (i \partial_n \mcdot\, i
\partial_n ) \Psi_{n,\hat{\omega}_2}=\bar\Psi_{n,\hat\omega_1} \qslash\, i
\slash\!\!\!\partial_n i\mbox{$\slash\!\!\!\partial_n$} \Psi_{n,\hat{\omega}_2}
$, and is removed by the quark equation of motion in Eq.~(\ref{EOM1}). For the
operators with two ${\cal G}'s$ only the structures in $\textbf{Q}_{4-7}$ have expansions
that start at twist-4. For example, $\mathcal{G}^{\mu\alpha}_{n, \hat{\omega}_3}
\mathcal{G}_{n, \hat{\omega}_4 \alpha\nu}$ at LO is proportional to $(
g\mathcal{B}_\perp^\alpha)_{\omega_3} (g\mathcal{B}^\perp_\alpha)_{\omega_4}
n^\mu n^\nu$ so closing the indexes with $\gamma^\mu$ or $\gamma^\nu$ generates
a $n\!\!\!\slash$ that next to $\chi_n$ gives zero.

It is less obvious that operators with one ${\cal G}_n$ and one $i\partial_n$
are redundant and can be eliminated from the RPI basis. Consider the operator
\begin{align} 
    \textbf{Q}_{*}   &= \bar{\Psi}_{n,\hat{\omega}_1} 
 q\!\!\!\slash \,q_\mu 
  ig \mathcal{G}^{\mu\nu}_{n,\hat{\omega}_3}  i\partial^n_{ \nu} 
\Psi_{n,\hat{\omega}_2}  \,  .
\end{align}
To remove it we use a manipulation discussed by Jaffe and Soldate in
Ref.~\cite{Jaffe:1982pm}. First we write
\begin{align} \label{T1DIS}
\mathcal{G}^{\mu\nu}_{n} i\partial^n_{
  \nu}=\mathcal{G}^{\mu\nu}_{n} i\hat {\cal D}^n_{ \nu}
-\mathcal{G}^{\mu\nu}_{n}[ 1/(q\cdot i\partial_n) q_\alpha ig
{\cal G}^{\alpha}_{n\nu}] \,,
\end{align}
and note that the term with two ${\cal G}_n$'s can be ignored since it is
already in our basis. Next using the definition (\ref{SF}) we can write
\begin{align} \label{CJ1}
 q_\mu
ig\mathcal{G}^{\mu\nu}_{n} i\hat {\cal D}^n_{ \nu}= q^\mu [i\hat
{\cal D}^n_{ \mu}, i\hat {\cal D}_n^{ \nu}]i\hat {\cal D}^n_{ \nu} = \frac12 \big\{ \!-\!
q^\mu[i\hat {\cal D}^n_{ \nu},[i\hat {\cal D}^n_{ \mu}, i\hat {\cal D}_n^{
  \nu}]]-(i\hat {\cal D}_n)^2 iq\mcdot\partial_n + iq\mcdot\partial_n (i\hat {\cal
  D}_n)^2\big\}
\,. 
\end{align}
The double commutator term is turned into a four-quark operator by the gluon
equations of motion in Eq.~(\ref{EOM2}). For the remaining terms we write
$(i\hat {\cal D}_n)^2= i\:\, \slash\!\!\!\! \hat {\cal D}_n\ \:\,
\slash\!\!\!\!\!\!i\hat {\cal D}_n + \frac{i}{2} \sigma_{\mu\nu} {\cal
  G}_{n}^{\mu\nu}\!$, where the $\sigma_{\mu\nu}$ term gives $\textbf{Q}_3$, and
terms involving $(\ \: \slash\!\!\!\!\!\! i\hat {\cal D}_n)^2$ are turned into
the operators $\textbf{Q}_2$, $\textbf{Q}_3$, $\textbf{Q}_4$, and $\textbf{Q}_6$
by the quark equation of motion in Eq.~(\ref{EOM1}).  (They are not simply set
to zero, since $[\ \: \slash\!\!\!\!\!\! i\hat {\cal D}_n]$ does not commute
with $\delta(\hat\omega -2q\cdot i\partial_n)$.)  Finally, we can also rule out
the only other non-trivial operator $\bar{\Psi}_{n, \hat{\omega_1}} 
q\!\!\!\slash i \partial_n\!\!\!\!\!\!\slash\ \gamma_\mu
q_\nu\mathcal{G}^{\mu\nu}_{n, \hat{\omega_3}} \Psi_{n, \hat{\omega_2}} $. Using
the gluon equation of motion we write
\begin{align} \label{Opsum}
 \bar{\Psi}_{n, \hat{\omega_1}} q\!\!\!\slash i
\partial\!\!\!\slash \gamma_\mu q_\nu\mathcal{G}^{\mu\nu}_{n,  \hat{\omega_3}}
\Psi_{n,  \hat{\omega_2}} 
 + \bar{\Psi}_{n,  \hat{\omega_1}}  q\!\!\!\slash
\gamma_\mu i \partial\!\!\!\slash \mathcal{G}^{\mu\nu}_{n,  \hat{\omega_3}}
q_\nu   \Psi_{n,  \hat{\omega_2}} 
= -2\textbf{Q}_* + \ldots\, ,
\end{align}
where the ellipsis denotes operators with two ${\cal G}_n$'s or four-quark fields
that are part of the basis.
The Bianchi identity in Eq.~(\ref{BI}) gives another relation for the two
operators on the LHS of Eq.~(\ref{Opsum}) and implies that they can be written in terms of
$\textbf{Q}_3$, $\textbf{Q}_{*}$, $\textbf{Q}_4$, and $\textbf{Q}_6$.  Thus both
the operators $ \bar{\Psi}_{n,\hat\omega_1} q\!\!\!\slash i
\partial_n\!\!\!\!\!\slash\ 
\gamma_\mu q_\nu \mathcal{G}^{\mu\nu}_{n,\hat\omega_3}\Psi_{n,\hat\omega_2} $
and $\bar{\Psi}_{n,\hat\omega_1} q\!\!\!\slash \gamma_\mu i
\partial_n\!\!\!\!\!\!\slash\ \, \mathcal{G}^{\mu\nu}_{n,\hat\omega_3} q_\nu
\Psi_{n,\hat\omega_2}$ are redundant. Finally we note that the order of the
${\cal G}_n$'s in an operator like $\textbf{Q}_6$ is not important, since we can
always symmetrize or antisymmetrize its Wilson coefficient in $\hat\omega_3$ and
$\hat\omega_4$. Note that when considering the transformation of the operators
under charge conjugation one must consider both the operator and its Wilson
coefficient. We discuss an example below in Eq.~(\ref{GW}).

The number of independent RPI operators in Eq.~(\ref{DISGI}) is smaller than in the
basis of homogeneous operators in Eq.~(\ref{RPIDI}), implying that there exist
further constraints on the Wilson coefficients of the homogeneous basis at
twist-4.  To find the constraints we must expand the operators in
Eq.~(\ref{RPIDI}) in terms of those in Eq.~(\ref{DISGI}). We start with
$\textbf{Q}_4$ through $\textbf{Q}_{11}$ which are in one-to-one correspondence
with operators in the homogeneous basis,
\begin{align} \label{disQ513}
 \textbf{Q}_{4} &= \frac{\omega_3\omega_4}{4(n\mcdot q)} \,O_{5} \,
, 
& \textbf{Q}_{5} &=   \frac{\omega_3\omega_4}{4(n \mcdot q)}\: O_6 \, , 
&\textbf{Q}_{6} &=  -\frac{\omega_3\omega_4}{4(n \mcdot q)}\: O_7 \, , \nonumber   \\[4pt]
 \textbf{Q}_{7} &=  \frac{-\omega_3\omega_4}{4(n \mcdot q)}\: O_8 \, ,
& \textbf{Q}_{8} &= \frac{1}{(n\mcdot q)^2}\,O_{9} \, , 
&\textbf{Q}_{9} &= \frac{1}{(n\mcdot q)^2} \,O_{10} \, ,  \nonumber  \\[4pt]
 \textbf{Q}_{10} &= \frac{1}{(n\mcdot q)^2} \,O_{12} \, , 
&\textbf{Q}_{11} &= \frac{1}{(n\mcdot q)^2} \,O_{11} \, .
\end{align}
Here the order of the $\hat\omega_i$ subscripts in operators on the left exactly
matches up with the $\omega_i$ subscripts on the right.  For the remaining
operators whose expansions start at twist-4 and for $Q_1$ that starts at twist-2,
we have
\begin{align} \label{disQ123}
\textbf{Q}_{2}(\hat\omega_1,\hat\omega_3,\hat\omega_2) &=
 \frac{1}{(n\mcdot q)^2}\,  \bigg[
  \frac{\omega_3}{2\omega_2} O_{4a}(\omega_1,\omega_3,\omega_2) 
 +  \frac{\omega_3}{2\omega_1} O_{4b}(\omega_1,\omega_3,\omega_2)\\
 &\qquad\qquad\quad
   - O_{3a}(\omega_1,\omega_3,\omega_2) + O_{3b}(\omega_1,\omega_3,\omega_2) \bigg]  + \ldots
 \, ,
 \nn \\[2pt]
\textbf{Q}_{3}(\hat\omega_1,\hat\omega_3,\hat\omega_2) &= \frac{2}{(n \mcdot q)^2}\,
 \bigg[ \frac{\omega_1}{\omega_2} O_{4a}(\omega_1,\omega_3,\omega_2) + O_{4b}(\omega_1,\omega_3,\omega_2) - 2 O_{3b}(\omega_1,\omega_3,\omega_2) \bigg]
 + \ldots \, , \nonumber\\[2pt]
 \textbf{Q}_{1}(\hat\omega_1,\hat\omega_2)
 &= 
 \frac{1}{n\mcdot q} O_1(\omega_1,\omega_2) + \frac{\bn\mcdot q}{(n\mcdot q)^2} \bigg[
  \Big\{ \frac{-1}{\omega_1 \omega_2}+\frac{d}{d\omega_1}\frac{1}{\omega_1} +
  \frac{d}{d\omega_2}\frac{1}{\omega_2} \Big\} O_2(\omega_1,\omega_2) 
 \nn \\[2pt]
 &+ \Big\{ \frac{2}{(\omega_a \!-\! \omega_2)^2} -
  \frac{d}{d\omega_2}\frac{2}{\omega_a \!-\! \omega_2} \Big\}   
 \Big\{ O_{3a}(\omega_1,\omega_a\!-\! \omega_2,\omega_a)
  - O_{3b}(\omega_1,\omega_a\!-\! \omega_2,\omega_a)
 \Big\} \nonumber \\[2pt]
  &+   \Big\{ \frac{-2}{(\omega_1 \!-\! \omega_a)^2}
   -\frac{d}{d\omega_1}\frac{2}{\omega_1 \!-\! \omega_a} \Big\}
  \Big\{ O_{3a}(\omega_a,\omega_1\!-\! \omega_a,\omega_2)
  - O_{3b}(\omega_a,\omega_1\!-\!  \omega_a,\omega_2)
   \Big\}
  \nonumber\\[2pt]
  &+  \Big\{ \frac{-1}{\omega_1 \omega_2} +
  \frac{d}{d\omega_1}\frac{1}{\omega_1} \Big\}
  O_{4a}(\omega_a,\omega_1\!-\!\omega_a,\omega_2)
  +  \frac{d}{d\omega_2}\frac{1}{\omega_a} \,
  O_{4a}(\omega_1,\omega_a\!-\!\omega_2,\omega_a)  
  \nonumber \\[2pt]
   &+ \Big\{ \frac{-1}{\omega_1 \omega_2}+\frac{d}{d\omega_2}\frac{1}{\omega_2}
   \Big\}  O_{4b}(\omega_1,\omega_a\!-\!\omega_2,\omega_a)   
   +\frac{d}{d\omega_1}\frac{1}{\omega_a} \,O_{4b}(\omega_a,\omega_1\!-\!\omega_a,\omega_2)   
   \bigg]   + \cdots \nonumber \,.
\end{align}
Here the ellipses indicate terms involving operators $O_{5-12}$ that have
already occurred in $\textbf{Q}_{4-11}$ and hence they are no longer important
for determining the linear independent combinations.  It is interesting to note
that expanding the operator $\textbf{Q}_*$ gives the same combination of $O_{3a}$ and
$O_{3b}$ that appears in $\textbf{Q}_2 - 4\omega_1/\omega_3 \textbf{Q}_3$, so
even if we had not eliminated $\textbf{Q}_*$ from the RPI basis, the implications for the
homogeneous basis would be the same.

The three RPI operators in Eq.~(\ref{disQ123}) have expansions in terms of six
homogeneous operators $O_1$, $O_2$, $O_{3a}$, $O_{3b}$, $O_{4a}$, and $O_{4b}$,
so there are three RPI relations.  The Wilson coefficients of these six
homogeneous operators are determined by three coefficients, $\hat C_{1,2,3}$ in
the RPI basis. It is convenient to trade $\hat C_{1,2,3}$ for the three
coefficients $C_1$, $C_{3a}$, and $C_{3b}$. The remaining coefficients $C_{2}$,
$C_{4a}$, and $C_{4b}$ are then determined by RPI. We find
\begin{align} \label{disRPIrelation}
C_2(\omega_1,\omega_2) &= -\frac{\bn\mcdot q}{n\mcdot q} 
  \Big\{ \frac{1}{\omega_1 \omega_2}+\frac{1}{\omega_1}\frac{d}{d\omega_1} +
  \frac{1}{\omega_2}\frac{d}{d\omega_2} \Big\} C_1(\omega_1,\omega_2) \, ,
  \nonumber \\[4pt]
C_{4a}(\omega_1,\omega_3,\omega_2) &=
  -\frac12 C_{3a}(\omega_1,\omega_3,\omega_2) 
  - \frac{\omega_1}{2\omega_2} C_{3b}(\omega_1,\omega_3,\omega_2) 
 + \frac{\bar{n} \mcdot q}{n \mcdot q\, \omega_2\omega_3}\,
C_1(\omega_1,\omega_2\minus\omega_3) \nn\\[2pt]
&\quad -\frac{\bar{n} \mcdot q\, (\omega_2\plus\omega_3)}{n \mcdot q\, \omega_3(\omega_2)^2}
\,C_1(\omega_1\plus\omega_3,\omega_2)
  \, , \nonumber \\[4pt]
C_{4b}(\omega_1,\omega_3,\omega_2) & = 
  - \frac{\omega_2}{2\omega_1}   C_{3a}(\omega_1,\omega_3,\omega_2) 
  -\frac12 C_{3b}(\omega_1,\omega_3,\omega_2) 
 + \frac{\bar{n} \mcdot q\,(\omega_1\minus\omega_3)}{n \mcdot q\, \omega_3(\omega_1)^2}
 \,  C_1(\omega_1,\omega_2\minus\omega_3) \nn\\[2pt]
&\quad -\frac{\bar{n} \mcdot q}{n \mcdot q\,\omega_1\omega_3}
 \,
C_1(\omega_1\plus\omega_3,\omega_2) \, .
\end{align}
We have cross-checked the relation for $C_2$ with a tree-level matching
computation. Note that $C_2(\omega_1,\omega_2)$ multiplies a matrix element that
gives $\delta(\omega_1-\omega_2)$, while $C_{4a,4b}(\omega_1,\omega_3,\omega_2)$
multiplies a $\delta(\omega_1+\omega_3-\omega_2)$, and that we have used these
$\delta$-functions at various intermediate steps. That is, the result in
Eq.~(\ref{disRPIrelation}) applies for a basis of operators, whose matrix
elements have vanishing total derivatives.

Our operator bases can be compared to the flavor singlet and parity even basis of
Jaffe and Soldate in Ref.~\cite{Jaffe:1982pm} which has one operator at twist-2,
and 12 operators at twist-4.\footnote{The notation in Eq.~(\ref{DISGI}) suggests
  that all quark bilinears are flavor singlet contractions if $\chi_n$ has
  multiple flavor components. To incorporate other possibilities for the flavor
  indices is straightforward~\cite{Jaffe:1982pm}.  We consider $\chi_n$ as a
  doublet of SU(2) flavor, or a triplet of SU(3) flavor, with elements
  $\chi_n^f$.  For photon currents one has a charge matrix in flavor space in
  each QCD current, which is $\hat Q=\text{diag}(2/3,-1/3,-1/3)$ for SU(3).
  Thus, at leading order in the electromagnetic interactions one must simply
  introduce a $\hat Q^2$ in all bilinear-quark operators, $O^1$ through $O^8$ in
  Eq.~(\ref{DISGI}).  When counting the four-quark operators $O^{9}$ to $O^{12}$
  induced by photons we double the number of operators because there are two
  possibilities, $\hat Q^2\otimes 1$ and $\hat Q\otimes \hat Q$.  In this
  notation the flavor singlet contraction for the four-quark operators is
  $1\otimes 1$.  For the RPI basis of operators the analysis of flavor
  structures is identical, and hence flavor does not modify the constraints in
  Eq.~(\ref{disRPIrelation}).}  There is a simple correspondence between the 11
operators in our RPI basis in Eq.~(\ref{disQ1},\ref{RPIDI}) and the QCD
operators in their basis. The correspondence is one-to-one for $\textbf{Q}_1$,
the four-quark operators $\textbf{Q}_{8-11}$, and the operators
$\textbf{Q}_{2,3}$ that have one ${\cal G}_n^{\mu\nu}$. For the operators with
two ${\cal G}_n^{\mu\nu}$'s we have four operators compared to their six, but
the difference is accounted for by the way in which the twist towers are
enumerated. We used continuous $\hat\omega_i$'s where even and odd symmetry
under the interchange $\hat\omega_3\leftrightarrow \hat\omega_4$ encodes two
possible color structures with $f^{ABC}$ and $d^{ABC}$, while
Ref.~\cite{Jaffe:1982pm} uses a discrete basis with integer powers of
$(i\bn\cdot D_n)$, where the choice of which operators to eliminate by
integration by parts implies that the two color structures yield different
operators. Our homogeneous basis has 14 operators up to twist-4, and most
closely corresponds to an enumeration of an operator basis in terms of the
so-called ``good'' quark and gluon fields. The good quark and gluon fields have
been discussed in Refs.~\cite{Kogut:1969xa,Jaffe:1996zw,Bashinsky:1998if}. In
this basis the power counting is manifest. From the three RPI relations in
Eq.~(\ref{disRPIrelation}) the number of independent short distance Wilson
coefficients is 11, and so encodes the same amount of information as the OPE
basis from Ref.~\cite{Jaffe:1982pm}.  Note that there is no room in the
traditional OPE in DIS for a correspondence with higher order operators with
ultrasoft fields. In our language, the validity of the OPE for DIS with generic
$x$ implies that ultrasoft degrees of freedom are not needed, and one can
consider that fluctuations from that region are reabsorbed into the collinear
fields.

When the basis of bilinear quark operators is considered in the forward proton
matrix element it can be reduced even further as discussed in detail in
Ref.~\cite{Ellis:1982cd}. In this process it is found that the matrix elements
of operators like $O_2$, $O_{4a}$, and $O_{4b}$ do not provide independent
information.  Hence at this level the RPI relations in
Eq.~(\ref{disRPIrelation}) do not appear to have practical implications.


\subsection{Deep Inelastic Scattering for Gluons at Twist-4} \label{sec:DISg}

Next let us consider the minimal basis for pure gluon DIS operators up to twist-4.
We proceed in a similar manner to our construction for quarks, first writing the
homogeneous basis and then the RPI basis to check if reparametrization
invariance provides constraints on the homogeneous operators.  The homogeneous
basis is 
\begin{align} \label{GI-gluon}
 O_{1} &=  {\rm Tr}\big[  ( g  \mathcal{B}_{n\perp \,})_{\omega_1} \mcdot
   ( g  \mathcal{B}_{n\perp})_{\omega_2}  \big]\, ,  \\[3pt]
       O_{2} &=  {\rm Tr}\big[  ( g  \mathcal{B}_{n\perp \,}^\mu)_{\omega_1}   \mathcal{P}^2_\perp 
   ( g  \mathcal{B}^{\perp}_{n \, \mu})_{\omega_2} \big] \, ,\nonumber \\[3pt]
    O_{3,4} &=  {\rm Tr}\big[  ( g  \mathcal{B}_{n\perp \,}^\mu)_{\omega_1}   ( g  \mathcal{B}_{n\perp \,}^\nu)_{\omega_2}  \mathcal{P}_\perp^\alpha
   ( g  \mathcal{B}_{n\perp}^\beta)_{\omega_3} \big] 
  \Gamma^{1,2}_{\mu \nu \alpha \beta} \, ,    \nonumber \\[3pt]
 O_{5,6} &=   {\rm Tr}\big[  ( g  \mathcal{B}_{n\perp \,}^\mu)_{\omega_1}  ( g  \mathcal{B}_{n\perp \,}^\nu)_{\omega_2}    ( g  \mathcal{B}_{n\perp \,}^\alpha)_{\omega_3} 
   ( g  \mathcal{B}_{n\perp}^\beta)_{\omega_4} \big] 
  \Gamma^{1,2}_{\mu \nu \alpha \beta}\ \, , \nonumber \\[3pt]
 O_{7,8} &=   {\rm Tr}\big[  ( g  \mathcal{B}_{n\perp \,}^\mu \!)_{\omega_1}
 \!( g  \mathcal{B}_{n\perp \,}^\nu \!)_{\omega_2} \big]
 {\rm Tr} \big[   ( g  \mathcal{B}_{n\perp \,}^\alpha \!)_{\omega_3} \!
   ( g  \mathcal{B}_{n\perp}^\beta )_{\omega_4} \big]
  \Gamma^{1,2}_{\mu \nu \alpha \beta}\, , \nonumber    \\[3pt]
  O_{9} &=    {\rm Tr}\big[   ( g  \mathcal{B}_{n\perp \,}^\mu)_{\omega_1}   \mathcal{P}^\perp_\mu  \mathcal{P}^\nu_\perp
   ( g  \mathcal{B}^{\perp}_{n \, \nu})_{\omega_2} \big] \, , \nonumber
\end{align} 
where $\Gamma^{1,2}_{\mu \nu \alpha \beta}=\{g_{\mu\nu} g_{\alpha \beta},g_{\mu
  \alpha} g_{\nu \beta} \}$ and the traces are over color. Recall that the
equations of motion (\ref{eomg}) were used to eliminate the operators $gn\cdot
\cB_n$ and $in\mcdot \partial_n (g \cB_{n\perp}^\mu)$. Again since the basis is
designed for taking forward matrix elements we are free to integrate by parts
and hence we do not consider $\cP_\perp^{\mu\dagger}$. There is a third tensor
structure, $\Gamma_{\mu\nu\alpha\beta}^3= g_{\mu\beta}g_{\alpha\nu}$, that can
also be considered for $O_{3-8}$, but which can always be eliminated.  For
$O_{3,4}$ this is done using integration by parts and the cyclic trace, giving
\begin{align}
  {\rm Tr}\big[  ( g  \mathcal{B}_{n\perp \,}^\mu)_{\omega_1}  
   ( g  \mathcal{B}_{n\perp \,}^\nu)_{\omega_2}  \mathcal{P}_\perp^\alpha
   ( g  \mathcal{B}_{n\perp}^\beta)_{\omega_3} \big] 
  \Gamma^{3}_{\mu \nu \alpha \beta}
 &= - O_4(\omega_2,\omega_3,\omega_1) - O_3(\omega_3,\omega_1,\omega_2) \,.
\end{align}
For $O_{5-8}$ the cyclic property of the trace suffices to eliminate
$\Gamma_{\mu\nu\alpha\beta}^3$ in an analogous manner.  The operator $ {\rm
  Tr}[( g \mathcal{B}_{n\perp \,}^\mu)_{\omega_1} \mathcal{P}_\perp^\alpha ( g
\mathcal{B}_{n\perp \,}^\nu)_{\omega_2} ( g \mathcal{B}_{n\perp
}^\beta)_{\omega_3} ]$ is also not needed in the basis because it can be put
into the form of the operators $O_{3}$ and $O_{4}$. This is done by acting with
the $\cP_\perp^\alpha$ on the two $\cB_\perp$'s to the right, using the cyclic property of
the trace, and again noting that $O_{3,4}$ encode all orderings for the $\omega_i$ subscripts.

For forward spin averaged matrix elements the RPI basis of gluon operators up to
twist-4 is
\begin{align} \label {RPIgluonDIS}
 \textbf{Q}_1 &= q_\nu  q_\beta {\rm Tr} \big[
 ig\mathcal{G}^{\mu\nu}_{n,\hat{\omega}_1}
 ig\mathcal{G}^{\alpha\beta}_{n,\hat{\omega}_2}
  \big] g_{\mu\alpha} \, , \\[4pt]
 \textbf{Q}_2 &=    {\rm Tr} \big[ 
   ig\mathcal{G}^{\mu\nu}_{n,\hat{\omega}_1}  
    ig\mathcal{G}^{\alpha\beta}_{n,\hat{\omega}_2}\big] 
  g_{\mu\alpha} g_{\nu \beta}
 \, , \nonumber \\[4pt]
   \textbf{Q}_3 &=  q_\nu  q_\sigma  {\rm Tr}
 \big[ ig \mathcal{G}^{\mu\nu}_{n,\hat{\omega}_1}
   ig \mathcal{G}^{\alpha\beta}_{n,\hat{\omega}_2}  
   ig \mathcal{G}^{\rho\sigma}_{n,\hat{\omega}_3} \big] g_{\mu\alpha} g_{\beta\rho}
 \, , \nonumber \\[4pt]
\textbf{Q}_4 &=   q_\nu q_\beta q_\sigma
  {\rm Tr} \big[ ig \mathcal{G}^{\mu\nu}_{n,\hat{\omega}_1} ig
  \mathcal{G}^{\alpha\beta}_{n,\hat{\omega}_2} i\partial_\mu ig
  \mathcal{G}^{\rho\sigma}_{n,\hat{\omega}_3} \big] g_{\alpha\rho} 
  \, , \nonumber \\[4pt]
\textbf{Q}_{5,6} &= q_\nu q_\beta q_\sigma q_\lambda {\rm Tr}
 \big[ ig \mathcal{G}^{\mu\nu}_{n,\hat{\omega}_1}
   ig  \mathcal{G}^{\alpha\beta}_{n,\hat{\omega}_2}  
   ig \mathcal{G}^{\rho\sigma}_{n,\hat{\omega}_3}  
   ig \mathcal{G}^{\tau\lambda}_{n,\hat{\omega}_4} \big]
  \Gamma^{1,2}_{\mu\alpha\rho\tau} 
  \, , \nonumber \\[4pt]
\textbf{Q}_{7,8} &= q_\nu q_\beta q_\sigma q_\lambda {\rm Tr} 
\big[ ig  \mathcal{G}^{\mu\nu}_{n,\hat{\omega}_1}
 ig \mathcal{G}^{\alpha\beta}_{n,\hat{\omega}_2} \big] {\rm Tr} \big[
 ig \mathcal{G}^{\rho\sigma}_{n,\hat{\omega}_3}
 ig \mathcal{G}^{\tau\lambda}_{n,\hat{\omega}_4}
  \big]\Gamma^{1,2}_{\mu\alpha\rho\tau} 
 \, . \nonumber 
\end{align}
Here we remove a possible operator $q_\nu q_\beta {\rm Tr} \big[
(ig\mathcal{G}_{n,\hat{\omega}_1}\!)^{\nu}_{\:\alpha} (i\partial)^2 ig
\mathcal{G}^{\alpha\beta}_{n,\hat{\omega}_2} \big]$ by writing $(i\partial)^2
\mathcal{G}^{\alpha\beta}_{n,\hat{\omega}_2}=(i\partial_\mu) i\partial^\mu
\mathcal{G}^{\alpha\beta}_{n,\hat{\omega}_2}$, and then using the Bianchi
identity in Eq.~(\ref{BI}) to rewrite this operator in terms of operator with
two ${\cal G}_{n}$'s, plus $(i\partial_\mu) i\partial^\alpha
\mathcal{G}^{\beta\mu}_{n,\hat{\omega}_2}$ and $(i\partial_\mu) i\partial^\beta
\mathcal{G}^{\mu\alpha}_{n,\hat{\omega}_2}$. The last two terms are removed by
the gluon equation of motion.  There is no
need to include the analog of $\textbf{Q}_4$ with the $i\partial_\mu$ acting on
$ig {\cal G}_{n,\hat\omega_2}^{\alpha\beta}$, because it is related to
$\textbf{Q}_4$ by integration by parts up to a term, $i\partial_\mu ig {\cal
  G}_{n,\hat\omega_1}^{\mu\nu}$ that reduces to other operators through the
gluon equation of motion. Again the cyclic nature of the trace allows one to
remove $\Gamma_{\mu\alpha\rho\tau}^3$ for $\textbf{Q}_{5-8}$. 

In order to consider the effect of charge conjugation on these basis one must
consider the transformation of
\begin{align} \label{GW} 
 \int [d \hat\omega_j] \hat {C}_i (\hat\omega_j)\, \textbf{Q}_i (\hat\omega_j) \,,
 \qquad \text{or} \qquad
 \int [d \omega_j] {C}_i (\omega_j)\, O_i (\omega_j) \,,
\end{align}
where $\hat {C}_i $ is the Wilson coefficient associated with $ \textbf{Q}_i$,
and $C_i$ the Wilson coefficient associated with $O_i$.  We can impose
constraints on $\hat {C}_i(\hat\omega_j)$ and $C_i(\omega_j)$ such that
(\ref{GW}) is $C$-invariant.  For example, note that under charge conjugation
$\textbf{Q}_3$ transforms into $- q_\nu q_\sigma {\rm Tr} \big[ ig
\mathcal{G}^{\mu\nu}_{n,\hat{\omega}_3} ig
\mathcal{G}^{\alpha\beta}_{n,\hat{\omega}_2} ig
\mathcal{G}^{\rho\sigma}_{n,\hat{\omega}_1} \big] g_{\mu\alpha} g_{\beta\rho}$,
so to make it $C$-invariant we impose that $\hat {C}_3(\hat\omega_1,
\hat\omega_2, \hat\omega_3)= - \hat {C}_3(\hat\omega_3, \hat\omega_2,
\hat\omega_1)$. Similar considerations apply to the homogeneous basis. For
example, the combinations
$O_3(\omega_1,\omega_2,\omega_3)-O_3(\omega_2,\omega_1,\omega_3)$ and
$O_4(\omega_1,\omega_2,\omega_3)+O_4(\omega_1,\omega_3,\omega_2)+O_3(\omega_3,\omega_2,\omega_1)$
are even under charge conjugation.

Next we must expand the RPI basis in Eq.~(\ref{RPIgluonDIS}) in terms of the
homogeneous basis in Eq.~(\ref{GI-gluon}) to find possible constraints.  We
first expand $\textbf{Q}_{5-8}$, they have only operators with four $g
\mathcal{B^\mu_\perp}$'s, that is $O_{5-8}$,
\begin{align} \label{expDISG1}
 \textbf{Q}_{5,6} &= \frac{\omega_1\omega_2\omega_3\omega_4}{16}\: O_{5,6}\, , 
& \textbf{Q}_{7,8} &= \frac{\omega_1\omega_2\omega_3\omega_4}{16}\: O_{7,8}\, .
\end{align}
Next we expand $\textbf{Q}_{3,4}$ to find
\begin{align} \label{expDISG2}
%
\textbf{Q}_3 & = \frac{\omega_1 \omega_3}{4 (n\cdot q)} \big[
 -O_4(\omega_1, \omega_2, \omega_3)
 -O_4(\omega_3, \omega_1, \omega_2) - O_3(\omega_2,
\omega_3, \omega_1) \big] + \cdots \,,
 \nonumber \\[4pt]
\textbf{Q}_4 &= \frac{\omega_1 \omega_2 \omega_3}{8} \Big[
O_4(\omega_1,\omega_2,\omega_3) - \frac{\omega_3}{\omega_1}
O_3(\omega_2,\omega_3,\omega_1) \Big] + \ldots \,,
\end{align}
where we integrate over the repeated $\omega_a$ variable.  The ellipses in
Eq.~(\ref{expDISG2}) indicate terms involving operators $O_{5-8}$ that have
already occurred in $\textbf{Q}_{5-8}$ and hence are no longer important for
determining the linear independent combinations. Eq.~(\ref{expDISG2}) implies
that $O_3$ and $O_4$ have Wilson coefficients that are independent of other
operators in the basis.
When we expand the remaining RPI operators
$\textbf{Q}_{1,2}$, we may also have terms with $O_{1,2,9}$ which have two
$g\mathcal{B^\mu_\perp}$'s. We find
\begin{align}
\textbf{Q}_{1}(\hat\omega_1,\hat\omega_2) &= 
  \frac{\omega_1\omega_2}{4} O^1(\omega_1,\omega_2) +   
  \frac{(\bar{n} \mcdot q)}{4 (n \mcdot q)} \Big(2 -\frac{d}{d\omega_1} \omega_1
  -\frac{d}{d\omega_2} \omega_2\Big) O^2(\omega_1,\omega_2) + \ldots
  \, ,\nn\\
 \textbf{Q}_2(\hat\omega_1,\hat\omega_2) &= \ldots \,,
\end{align}
where the ellipsis indicates terms involving operators $O_{3-8}$ that have
already occurred in $\textbf{Q}_{3-8}$.  The fact that $O^9$ does not occur in
the expansion of any of the RPI operators indicates that it is ruled out by RPI
(explaining why we listed it last in the basis).  Furthermore, the operators
$O^1$ and $O^2$ only enter in the combination obtained from expanding
$\textbf{Q}^1$, and so their Wilson coefficients are related by
\begin{align}
  C_2(\omega_1,\omega_2) 
 &= \frac{\bn\mcdot q}{  n\mcdot q} \Big(2 + \omega_1 \frac{d}{d\omega_1}
    + \omega_2 \frac{d}{d\omega_2} \Big)
   \frac{ C_1(\omega_1,\omega_2) }{\omega_1\omega_2} 
  \nn\\[4pt]
 &=  \frac{\bn\mcdot q}{\omega_1\omega_2\,  n\mcdot q}
     \Big( \omega_1 \frac{d}{d\omega_1}
    + \omega_2 \frac{d}{d\omega_2} \Big) C_1(\omega_1,\omega_2) \,.
\end{align}
For the gluon DIS operators the RPI relations are similar to that for the quark
basis, namely it is the collinear operators with $\cP_\perp$'s that are
constrained. This was also observed in Ref.~\cite{Arnesen:2005nk} for the
heavy-to-light currents at second order in the power counting. Overall there are
eight homogeneous operators for spin-averaged gluon DIS up to twist-4, and seven
independent Wilson coefficients.

An analysis of twist-4 gluon matrix elements was done in
Ref.~\cite{Bartels:1999km} using leading-order Feynman diagram, based on the
methods of Ref.~\cite{Ellis:1982cd}. To the best of our knowledge, the complete
linear independent bases of twist-4 pure glue operators given in
Eq.~(\ref{GI-gluon}) and (\ref{RPIgluonDIS}) have not been given earlier in the
literature.


\subsection{Two Jet production:  $n$-$n'$ operators } \label{sec:2jets}

An important application for operators with two-collinear directions, $n$-$n'$,
is the study of two jet phenomena and event shapes. The effective theory SCET
has been used to study jets at leading order in the power expansion and various
orders in the $\alpha_s$ expansion in Refs.~\cite{Bauer:2002ie, Bauer:2003di,
  Lee:2006nr, Lee:2006fn, Trott:2006bk, Fleming:2007xt, Schwartz:2007ib,
  Fleming:2007qr, Becher:2008cf,Jain:2008gb,Hoang:2008fs}.  Another interesting application
is to describing parton showers with SCET~\cite{Bauer:2006qp,Bauer:2006mk},
where both leading and subleading operators with two-collinear directions play
some role.  In this section we study the leading and first power suppressed
quark operators with two-collinear directions.  For two jet processes it is
convenient to use the center-of-momentum (CM) frame where the two jets are back
to back. In this frame we can take $n'\!=\! \bn$ so that $n'\mcdot n=2$. Our
main interest will be in the operators that do not vanish in this frame, however
part of our discussion touches on the additional operators that do.

To be concrete we consider operators that appear in two jet production from a
virtual photon of momentum $q^\mu$ in $e^+e^-\to J_nJ_{n'}$. In QCD the
fundamental hadronic operator is the current $J^\mu = \bar\psi \gamma^\mu \psi$,
which is conserved $\partial_\mu J^\mu=0$ or $q_\mu J^\mu=0$, is odd under
charge-conjugation, and transforms as a vector under parity and time-reversal.
To describe high-energy jet production this current is matched onto a series of
SCET currents $J_\ell^{(k)}(\omega_i)\sim \lambda^k$ with Wilson coefficients
$C_\ell(\omega_i)$,
\begin{align} \label{Jsum0}
  J^\mu = \sum_{n,n'} \sum_{k=0}^\infty \sum_\ell \int\! [ \prod_i d\omega_i]
   \: C_\ell(\omega_i)\:  \big[J^{(k)}_\ell(\omega_i) \big]^\mu_{\text{2-jet}} \,.
\end{align}
Here $k$ denotes the power in $\lambda$, the subscript $\ell$ denotes members of
the basis at a given order, and the $\omega_i$ are the set of gauge invariant
momentum fractions upon which the operator depends. We also sum over all
collinear directions $n$ and $n'$, and the appropriate ones for a given computation
are picked out by the jet-momenta in the states. Because of this sum we are free
to swap $n\leftrightarrow n'$ when considering symmetry implications. The
C, P, and T symmetry properties of $C_\ell(\omega_i) J_\ell^{(k)}(\omega_i)$ are the same as
$J^\mu$, and they also satisfy current conservation, $q_\mu
[J_\ell^{(k)}(\omega_i)]^\mu=0$.  Finally, since the matching takes place at a
hard scale where perturbation theory is valid, the SCET operators should have
the same $LL+RR$ chirality as $J^\mu$. 

We first construct a basis of SCET operators that is homogeneous in the power
counting and with even chirality.  For the construction of this basis it is
convenient to define
\begin{align}
g_T^{\mu\nu} &= g^{\mu\nu} - \frac{q^\mu q^\nu}{q^2} \,,
 \qquad\qquad\qquad\quad\!
   \gamma_T^\mu = \gamma^\mu - \frac{q^\mu \qslash}{q^2} \,,  
  \\[3pt]
  r_-^\mu &= \frac{n\mcdot q}{2} \, n^{\prime\mu} 
  -  \frac{n'\mcdot q}{2}\,n^\mu
  \, ,
 \qquad\qquad
  r_+^\mu = \frac{n\mcdot q}{2} \, n^{\prime\mu}
  + \frac{n'\mcdot q}{2}\, n^\mu  \, ,
   \nn
\end{align}
where $r_-^\mu$ is odd under $n\leftrightarrow n'$ and $r_+$ is even. We also
define $s_\pm^\mu$ as $r_\pm^\mu$ with $n\to\bn$ and $n'\to \bn'$. Four of these
objects are transverse to $q^\mu$, $q_\mu g_T^{\mu\nu}=0$, $q_\mu
\gamma_T^\mu=0$, and $q\cdot r_- =q\cdot s_- = 0$, which is helpful for
satisfying current conservation.  For constructing the homogeneous basis it
suffices to consider the vectors $\{r_-, q,s_-,s_+\}$ in place of
$\{n,\bar{n},n',\bn'\}$. When we specialize to the CM frame, $q_{n\perp}^\mu \!=
q_{n'\perp}^\mu\!=\!0$, $s_\pm^\mu = \mp r_\pm^\mu$, and the vector $r_+^\mu\!=\!q^\mu$,
and hence $r_+^\mu$ and $s_\pm^\mu$ do not need to be considered.

In a general frame the LO operator is $\bar{\chi}_{\bar{n},\omega_1} \Gamma^\mu
\bar{\chi}_{n,\omega_2}$ with $\Gamma^\mu= \{ \gamma_T^\mu, r_-^\mu\, \qslash,
r_-^\mu\, \slash\!\!\! r_-, g_T^{\mu\nu} r^+_\nu\, \qslash ,$ $r_-^\mu\,
\slash\!\!\! r_+, g_T^{\mu\nu} r^+_\nu\, \slash\!\!\! r_-, g_T^{\mu\nu}
r^+_\nu\, \slash\!\!\! r_+ \}$ plus terms where $r_+$ or $r_-$ are replaced by
$s_\pm$. No terms with $q^\mu$ are allowed by current conservation. Things
become much simpler if we focus on operators that are non-zero in the CM frame.
In the CM frame $\bar{\chi}_{n',\omega_1} \qslash \bar{\chi}_{n,\omega_2} =0$,
$\bar{\chi}_{n',\omega_1} \slash\!\!\! r_- \bar{\chi}_{n,\omega_2} =0$, and the
vectors $r_+$ and $s_\pm$ become redundant, so there is only one operator at lowest
order
\begin{align} \label{2jLO}
J_0^{(0)} =& 
 \bar{\chi}_{n',\omega_1} \gamma_T^\mu \chi_{n,\omega_2} \, .
\end{align}
Here $\omega_i = \{ \omega_1,\omega_2\}$ and for brevity we suppress the index
$\mu$ on the LHS. 

To construct a homogeneous basis at NLO we again consider only operators which
are non-vanishing in the CM frame.  In the CM frame we can take the total
transverse momentum of the jet equal to zero, so we have the relations $
\bar{\chi}_{n',\omega_1} \Gamma_{\mu} \cP_{ \perp}^{\,\,\mu} {\chi}_{n,\omega_2}
= \bar{\chi}_{n',\omega_1} \Gamma_{\mu} \cP_{ \perp}^{ \dagger \,\,\mu}
{\chi}_{n,\omega_2} =0$, with $\Gamma_{\mu}$ any gamma structure, and hence do
not need to consider operators with a single $\cP_\perp$. Again all operators
with a $\qslash$ or $ \slash\!\!\! r_-$ vanish, as do those with $q\cdot (g
\cB_{n\perp})$ and $r_-\cdot (g \cB_{n\perp})$, and the analogs with $n\to n'$.
Operators with three $\gamma$'s can all reduce to operators with a single
$\gamma$ plus terms that are zero in the CM frame. This implies that at NLO
there are only two operators
\begin{align}\label{2jNLO} 
J_1^{(1)} &=    r_{-}^\mu\, \bar{\chi}_{n',\omega_1}   \gamma_{\nu}
 (g \mathcal{B}_{n\perp}^{\nu})_{\omega_3} 
 {\chi}_{n,\omega_2} \,,\nn\\[4pt]
J_2^{(1)} &=  r_{-}^\mu\,  \bar{\chi}_{n',\omega_1}   \gamma_{\nu}
 (g \mathcal{B}_{n' \perp}^{\nu} )_{\omega_3}  
 {\chi}_{n,\omega_2} \,.
\end{align}
Linear combinations of these two SCET currents can both be made odd under charge
conjugation by imposing appropriate conditions on their coefficients under
$\omega_1\leftrightarrow -\omega_2$.

To see if there are constraints on the Wilson coefficients we write down a basis
of RPI operators up to NLO.  The objects $\gamma_T^\mu$ and $g_T^{\mu\nu}$ are
invariant under RPI and can be used for this construction, but the object
$r_\pm^\mu$ cannot. We find the basis
\begin{align} \label{TwoJetRPI}
\textbf{J}_0^{(0)} & =   \bar{\Psi}_{n',\hat{\omega}_1} 
  \gamma_T^\mu \Psi_{n,\hat{\omega}_2} 
  \,, \\[3pt]
\textbf{J}^{(1)}_{1}  &=  
   \bar{\Psi}_{n' ,\hat{\omega}_1} g^T_{\mu\alpha}\gamma_{\beta} 
 ig\mathcal{G}^{\alpha \beta}_{n,\hat{\omega}_3}   \Psi_{n,\hat{\omega}_2}
  \,,
&\textbf{J}^{(1)}_{2}  &=  
 \bar{\Psi}_{n' ,\hat{\omega}_1} g^T_{\mu\alpha}\gamma_{\beta}   ig\mathcal{G}^{\alpha \beta}_{n',\hat{\omega}_3}\Psi_{n,\hat{\omega}_2} \, , \nonumber
 \nn\\[3pt]
\textbf{J}^{(1)}_{3}  &=    g^T_{\mu\lambda} \bar{\Psi} _{n', \hat{\omega}_1} 
   \gamma_\alpha q_{\beta} \big[
 i \partial_n^{\lambda} ig \mathcal{G}_{n,\hat{\omega}_3}^{\alpha \beta} 
  \big] \Psi _{n, \hat{\omega}_2}
 \,,
& \textbf{J}^{(1)}_{4}  &=    g^T_{\mu\lambda} \bar{\Psi} _{n', \hat{\omega}_1} 
   \gamma_\alpha q_{\beta} \big[
  i \partial_{n'}^{\lambda} ig \mathcal{G}_{n',\hat{\omega}_3}^{\alpha \beta}\,
  \big] \Psi _{n, \hat{\omega}_2}
  \,,\nn\\[3pt]
\textbf{J}^{(1)}_{5}  &=    g^T_{\mu\lambda} \bar{\Psi} _{n', \hat{\omega}_1} 
   \gamma_\alpha q_{\beta} \,
 ig \mathcal{G}_{n,\hat{\omega}_3}^{\alpha \beta}  i \partial_n^{\lambda} 
   \Psi _{n, \hat{\omega}_2} 
 \,,
& \textbf{J}^{(1)}_{6}  &=    g^T_{\mu\lambda} \bar{\Psi} _{n', \hat{\omega}_1} 
 (-i\! \overleftarrow \partial\!{}_{n'}^{\lambda})\gamma_\alpha q_{\beta} \,
ig \mathcal{G}_{n',\hat{\omega}_3}^{\alpha \beta}\,   
     \Psi _{n, \hat{\omega}_2}
  \,.\nn
\end{align}
Here we do not write down RPI operators which vanish in the CM frame when
expanded, such as $\bar{\Psi}_{n'} g^T_{\mu\nu} i\partial_n^\nu \Psi_{n}$ or
operators with only the Dirac structure $\qslash$.  This set also includes three
$\gamma$ operators since in $\textbf{J}^{(1)}_1$ replacing
$\Gamma_{\mu\alpha\beta}= g_{\mu\alpha}^T \gamma_\beta$ by
$\Gamma_{\mu\alpha\beta}=\gamma_T^\mu \gamma_\alpha\gamma_\beta$ gives an
operator that vanishes in the CM frame, and any other order for the $\gamma$'s
is then redundant. The same is true for $\Gamma_{\mu\alpha\beta}=\gamma^T_\mu
\qslash \gamma_\alpha q_\beta - q^2 g^T_{\mu\alpha} \gamma_\beta$. Analogous
arguments rule out three $\gamma$ terms replacing the tensor in
$\textbf{J}_2^{(1)}$.
There are no others operators with $i\partial_{n}^\lambda$ or
$i\partial_{n'}^\lambda$ besides $\textbf{J}^{(1)}_{3-6}$ at LO. To see why,
notice that for the operators with $i\partial_{n}^\lambda$ and
$i\partial_{n'}^\lambda$, only the contraction with $g^T_{\mu\lambda}$ has the
potential to give a LO term. Momentum conservation requires $q^\lambda\!=
\!i\partial_{n}^\lambda+i\partial_{n'}^\lambda$ and because $q^\lambda
g^T_{\mu\lambda}=0$, we can exchange $i\partial_{n'}^\lambda$ and $
i\partial_{n}^\lambda$. The operators $\textbf{J}^{(1)}_{3-6}$ correspond to
keeping $i\partial_{n}^\lambda$ when we have a ${\cal
  G}_{n,\hat\omega}^{\alpha\beta}$, and $i\partial_{n'}^\lambda$ when we have a
${\cal G}_{n',\hat\omega}^{\alpha\beta}$.

The number of operators in Eq.~(\ref{TwoJetRPI}) is greater than that in the
homogeneous basis, and when expanded $\textbf{J}_{0,1,2}\to J_{0,1,2}$. Thus the
operators in Eqs.~(\ref{2jLO}) and (\ref{2jNLO}) are not connected by RPI. For
two jet production the constraints imposed by considering the CM frame are
strong enough that RPI provides no further information. (RPI could still
constrain the homogeneous basis of operators in a general frame, but does not
have practical implications for determining the basis of operators for an
analysis to be carried out with homogeneous operators in the CM frame.)

\subsection{Three Jet Production:  $n_1$-$n_2$-$n_3$ operators } \label{sec:3jets}

Here we analyze operators for three jet production. As in the two jet case, we
consider production of jets from $e^+ e^-$ scattering through a virtual photon.
To construct the minimal SCET basis needed for a matching we could proceed like
in the previous cases, by writing down both the most general homogeneous basis
and RPI basis consistent with the symmetry of the process, and expanding the RPI
basis to find possible connections. In the two jet processes the interesting
terms in the homogeneous basis are made of only two operators up to NLO.
However, for three jets the homogeneous basis has many operators at LO since we
have three distinct directions $n_1$, $n_2$ and $n_3$. With only two directions
we could greatly reduced the number of operators by focusing on the ones that
do not vanish in the center of momentum frame, meaning those that do not vanish
when $n_1\to n$, $n_2\to \bn$, and $q_\perp\to 0$. This choice rules out many
operators because of the relations $\bar{\chi}_{\bn} \Gamma_\perp n\!\!\!\slash
\chi_n =\bar{\chi}_{\bn} \Gamma_\perp \bar{n}\!\!\!\slash \chi_n =0$, where
$\Gamma_\perp$ is a Dirac structure without $\bnslash$ or $\nslash$ factors.
With three directions there is more freedom, for example the perpendicular
direction of $n_1$-$\bar{n}_1$ is not the same as the one of $n_2$-$\bar{n}_2$,
or of $n_3$-$\bar{n}_3$.  Now to construct the homogeneous basis, we can use
$n_1$, $\bar{n}_1$, $n_2$, $\bar{n}_2$, $n_3$, $\bar{n}_3$, $\gamma^\mu$, so in
the three jet case we have a bigger set of objects available.

On the other hand the RPI basis still has a reasonable number of objects, namely
$\Psi_{n_i}$, $\mathcal{G}_{n_i}^{\mu\nu}$, $\partial^\mu_{n_i}$, $\gamma^\mu$, and
$q^\mu$. The $\partial^\mu_{n_i}$ operators and $q^\mu$ are connected by
momentum conservation,
$i\partial^\mu_{n_1}+i\partial^\mu_{n_2}+i\partial^\mu_{n_3}=q^\mu$, so one of
them can be eliminated.  Hence we expect that reparametrization invariance will
give a large number of connections on the homogeneous basis, so many in fact
that it is not even convenient to write down the homogeneous basis. It is much
quicker to just write only the RPI basis and expand it to determine a basis of
allowed homogeneous operators.

The RPI basis for three jets at LO is made of two quarks fields and a gluon field
(we do not consider here the case with pure gluon jets). $n_1$ and $n_2$ will be
the directions of the quark and antiquark jets, and $n_3$ will be the direction
of the gluon jet. As for the two jet case, because of current conservation, the
only objects that can carry the vector index and are RPI invariant are
$g_T^{\mu\nu}$ and $\gamma_T^\mu$. The RPI basis is
\begin{align}   \label{3jGIB}
 \textbf{J}_1 & =   \bar{\Psi}_{n_1,\hat{\omega}_1} 
   \gamma_\nu\, q\!\!\!\slash\, \gamma_\mu^T q_\sigma 
   ig \mathcal{G}_{n_3,\hat{\omega}_3}^{\nu\sigma}
   \Psi_{n_2,\hat{\omega}_2} \, ,
& \textbf{J}_2 & =  \bar{\Psi}_{n_1,\hat{\omega}_1} 
  \gamma^T_\mu   q\!\!\!\slash \,\gamma_\nu q_{\sigma}
  ig \mathcal{G}_{n_3,\hat{\omega}_3}^{\nu\sigma}  \Psi_{n_2,\hat{\omega}_2} \,
  ,\nonumber \\[3pt]
    \textbf{J}_{3}& =     \bar{\Psi}_{n_1,\hat{\omega}_1} 
   \gamma_\sigma \gamma_\nu \gamma^T_\mu
     ig \mathcal{G}_{n_3,\hat{\omega}_3}^{\nu\sigma}
  \Psi_{n_2,\hat{\omega}_2}
  & \textbf{J}_4 & =  \bar{\Psi}_{n_1,\hat{\omega}_1} 
  \gamma^T_\mu \gamma_\nu \gamma_\sigma
  ig \mathcal{G}_{n_3,\hat{\omega}_3}^{\nu\sigma}  \Psi_{n_2,\hat{\omega}_2}  \,
  ,
   \nonumber \\[3pt]
 \textbf{J}_5 & = g^T_{\mu\alpha}  \bar{\Psi}_{n_1,\hat{\omega}_1} 
   \gamma_\nu q_\sigma \, i\! \overleftarrow{\partial}\!{}_{n_1}^{\alpha}
  ig \mathcal{G}_{n_3,\hat{\omega}_3}^{\nu\sigma}  \Psi_{n_2,\hat{\omega}_2}    \, ,
    & \textbf{J}_6 & =  g^T_{\mu\alpha}  \bar{\Psi}_{n_1,\hat{\omega}_1} 
    \gamma_\nu q_\sigma   
   ig \mathcal{G}_{n_3,\hat{\omega}_3}^{\nu\sigma}
  i\partial_{n_2}^{\alpha}  \Psi_{n_2,\hat{\omega}_2} \, , 
  \nonumber   \\[3pt]
    \textbf{J}_{7}& =  g^T_{\mu\alpha}   \bar{\Psi}_{n_1,\hat{\omega}_1} 
    q\!\!\!\slash \,  \gamma_\nu
  \gamma_\sigma  i\!\overleftarrow{\partial}\!{}_{n_1}^{\alpha}  
  ig \mathcal{G}_{n_3,\hat{\omega}_3}^{\nu\sigma}
  \Psi_{n_2,\hat{\omega}_2} \, , 
 &  \textbf{J}_8 & =   g^T_{\mu\alpha}    \bar{\Psi}_{n_1,\hat{\omega}_1} 
    q\!\!\!\slash  \, \gamma_\nu \gamma_\sigma
  ig \mathcal{G}_{n_3,\hat{\omega}_3}^{\nu\sigma} 
  i\partial_{n_2}^{\alpha} \Psi_{n_2,\hat{\omega}_2} \, , 
   \nonumber \\[3pt]
  \textbf{J}_9 & =  g^T_{\mu\nu}  \bar{\Psi}_{n_1,\hat{\omega}_1} 
  q\!\!\!\slash \,
  ig \mathcal{G}_{n_3,\hat{\omega}_3}^{\nu\sigma}  
  i\!\! {\overleftrightarrow{\partial}}_{\!\!12\sigma} \Psi_{n_2,\hat{\omega}_2} \, ,
 & \textbf{J}_{10} & =  \bar{\Psi}_{n_1,\hat{\omega}_1} 
  \gamma^T_\mu   q\!\!\!\slash\, \gamma_\nu  
  ig \mathcal{G}_{n_3,\hat{\omega}_3}^{\nu\sigma}  
  i\!\! {\overleftrightarrow{\partial}}_{\!\!12\sigma}\Psi_{n_2,\hat{\omega}_2} 
\, ,\nonumber \\[3pt]
\textbf{J}_{11} & =   \bar{\Psi}_{n_1,\hat{\omega}_1} 
  \gamma^T_\mu  q_\nu  ig \mathcal{G}_{n_3,\hat{\omega}_3}^{\nu\sigma} 
  i\!\! {\overleftrightarrow{\partial}}_{\!\!12\sigma} \Psi_{n_2,\hat{\omega}_2} 
 \, .
\end{align}
For the first four operators we chose the Dirac structures $\{\gamma_\nu\,
q\!\!\!\slash\, \gamma_\mu^T q_\sigma$, $\gamma^T_\mu q\!\!\!\slash \,\gamma_\nu
q_{\sigma}$, $\gamma_\sigma \gamma_\nu \gamma^T_\mu$, $\gamma^T_\mu \gamma_\nu
\gamma_\sigma \}$ in order to simplify the transformation of the basis under
charge conjugation. Since $\{ \gamma_\mu^T,\qslash\}=0$ the sum of the first two
structures gives $-g_{\mu\nu}^T\, \qslash q_\sigma$, and using the antisymmetry
of ${\cal G}_{n_3}^{\nu\sigma}$ the sum of the last two gives $4 g_{\mu\nu}^T
\gamma_\sigma$, so structures with a $g^T_{\mu\nu}$ are redundant. Other three
$\gamma$ operators are also redundant.  We have used the equations of motion and
Bianchi identity in Eqs.~(\ref{EOM1},\ref{BI}) to eliminate
$i\partial\!\!\!\slash_{n_i}$, and momentum conservation to eliminate
$i\partial^\mu_{n_3} = q^\mu-i\partial_{n_1}^\mu -i\partial_{n_2}^\mu$. For the
operators $\textbf{J}_{9-11}$ we have a derivative contracted with $ig {\cal
  G}_{n_3}^{\nu\sigma}$, and we can use the gluon equation of motion,
$i\partial_{n_3 \sigma}
\mathcal{G}_{n_3}^{\sigma\nu}=(q_\sigma-i\partial_{n_1\sigma}-i\partial_{n_2\sigma})
\mathcal{G}_{n_3}^{\sigma\nu} = \ldots$, where the ellipsis denotes higher twist
terms, to eliminate $(i\partial_{n_1}^\sigma +   i\partial_{n_2}^\sigma)$ and
leave only $i\!\!  \overleftrightarrow{\partial}_{\!\!12}^\sigma \equiv
(i\!\overleftarrow{\partial}_{n_1}^\sigma \!-\!  i\partial_{n_2}^\sigma)$.  Note
that we cannot use the trick used in DIS for $\textbf{Q}_{*}$, to eliminate
$\textbf{J}_{9-11}$, because here $i\partial_{n_{1,2}}^\alpha$ and
$\mathcal{G}^{\nu\sigma}_{n_3}$ have different collinear directions. Operators
with two or more derivatives are redundant for the construction of the LO basis
of RPI currents with one-vector index $\mu$, and hence do not need to be
considered.

We can match the three jet RPI basis of currents with the basis of homogeneous
SCET currents by writing
\begin{align} \label{Jsum3}
  \sum_{n_1,n_2,n_3} \sum_\ell \int\! [ \prod_i d\hat{\omega}_i]
   \: \hat{C}_\ell(\hat{\omega}_i)\:  [\textbf{J}_\ell(\hat{\omega}_i)]^\mu=
   \sum_{n_1,n_2,n_3}\sum_\ell \int\! [ \prod_i d\omega_i]
   \: C_\ell(\omega_i)\:  [\mathcal{J}_\ell(\omega_i)]^\mu_{\text{3-jet}} + \ldots\,.
\end{align}
On the RHS the integration variable was changed using $\hat\omega_i=n\mcdot q\,
\omega_i$ and any additional $n\mcdot q$ factors were absorbed into the Wilson
coefficients $C_\ell(\omega_i)$. We can determine the currents
$[\mathcal{J}_l(\omega_i)]_{\text{3-jet}}^\mu$ of the homogeneous basis, whose
form is as in Eq.~(\ref{GB}), by just expanding the currents (\ref{3jGIB}) using
Eqs.~(\ref{Psin}) and (\ref{Gsin}). This yields the homogeneous operator basis
\begin{align} \label{J3jethomo}
\mathcal{J}_1    & =
  \bar{\chi}_{n_1 ,\omega_1}  
    (g\mathcal{B}\!\!\!\slash^{\perp}_{n_3} )_{\omega_3}
   q\!\!\!\slash \, \gamma_T^\mu\,  \chi_{n_2,\omega_2} 
   \, ,  \\[4pt]
\mathcal{J}_2    & =
 \bar{\chi}_{n_1 ,\omega_1} \gamma_T^\mu \, q\!\!\!\slash
   (g\mathcal{B}\!\!\!\slash^\perp_{n_3} )_{\omega_3} \chi_{n_2,\omega_2} 
   \, ,  \nonumber \\[4pt]
\mathcal{J}_{3}   & =   \omega_3\,
\bar{\chi}_{n_1 ,\omega_1}   
   (g\mathcal{B}\!\!\!\slash^\perp_{n_3} )_{\omega_3} n\!\!\!\slash_3 \gamma_T^\mu  \,
   \chi_{n_2,\omega_2}
   \, , \nonumber \\[4pt]
\mathcal{J}_4   & = \omega_3\,
 \bar{\chi}_{n_1 ,\omega_1} \gamma_T^\mu  \, n\!\!\!\slash_3 
   (g\mathcal{B}\!\!\!\slash^\perp_{n_3} )_{\omega_3} \chi_{n_2,\omega_2} 
   \, ,  \nonumber \\[4pt]
\mathcal{J}_5   & = \omega_1\,
 \bar{\chi}_{n_1 ,\omega_1}   n_{1T}^\mu 
   (g\mathcal{B}\!\!\!\slash^\perp_{n_3} )_{\omega_3}  
   \chi_{n_2,\omega_2}
  \, ,  \nonumber \\[4pt]
\mathcal{J}_6   & =\omega_2\,
 \bar{\chi}_{n_1 ,\omega_1}  n_{2T}^\mu  
   (g\mathcal{B}\!\!\!\slash^\perp_{n_3} )_{\omega_3} 
   \chi_{n_2,\omega_2} 
   \, ,  \nonumber \\[4pt]
\mathcal{J}_{7}    & = \omega_1\,  
 \bar{\chi}_{n_1 ,\omega_1}  n_{1T}^\mu \, q\!\!\!\slash \,  n\!\!\!\slash_3 
  (g\mathcal{B}\!\!\!\slash^\perp_{n_3} )_{\omega_3}
  \chi_{n_2,\omega_2} \, ,
 \nonumber \\[4pt]
\mathcal{J}_8 & = \omega_2\,
 \bar{\chi}_{n_1 ,\omega_1}  n_{2T}^\mu \, q\!\!\!\slash \, n\!\!\!\slash_3 
  (g\mathcal{B}\!\!\!\slash^\perp_{n_3} )_{\omega_3} 
  \chi_{n_2,\omega_2} \, ,  
  \nonumber \\[4pt]
\mathcal{J}_9  & = \omega_3\,
 \bar{\chi}_{n_1 ,\omega_1} \, q\!\!\!\slash  
   \Big[ n_3^{\nu} (g\mathcal{B}^{\mu}_{n_3\perp } )_{\omega_3} 
   - n_{3T}^\mu (g  \mathcal{B}_{n_3\perp}^\nu )_{\omega_3} \Big]
  ( n_{2\nu} \omega_2 \!+\! n_{1\nu} \omega_1) 
   \chi_{n_2,\omega_2}
   \, ,  \nonumber \\[4pt]
\mathcal{J}_{10} &= \omega_3\,
 \bar{\chi}_{n_1 ,\omega_1}   \gamma_T^\mu  \, q\!\!\!\slash  
  \Big[ n_3^\nu (g\mathcal{B}\!\!\!\slash^\perp_{n_3} )_{\omega_3} 
   - n\!\!\!\slash_3  (g  \mathcal{B}^\nu_{n_3\perp} )_{\omega_3}
   \Big]  ( n_{2\nu} \omega_2 \!+\! n_{1\nu} \omega_1) 
  \chi_{n_2,\omega_2} \, ,  
  \nonumber \\[4pt]
\mathcal{J}_{11}   & = 
\bar{\chi}_{n_1 ,\omega_1} \gamma_T^\mu 
   (g   \mathcal{B}^\nu_{n_3\perp} )_{\omega_3}
    ( n_{2\nu} \omega_2 \!+\! n_{1\nu} \omega_1) \chi_{n_2,\omega_2}  
   \, ,  \nn
\end{align}
where $n_{1T}^\mu=n_1^\mu-q^\mu( n_1 \mcdot q)/q^2$, $n_{2T}^\mu=n_2^\mu-q^\mu(
n_2 \mcdot q)/q^2$ and $q\!\!\!\slash = (\bar{n}_3 \cdot q)
n\!\!\!\slash^\mu_3/2+(n_3 \cdot q) \bar{n}\!\!\!\slash^\mu_3/2$. To simplify
the results we did not bother to write out the terms with $q_\mu (g \cB_{
  n_3\perp}^\mu)$ in Eq.~(\ref{J3jethomo}), which are terms that vanish in a
frame where $q_{\perp n_3}=0$.  In some cases we have absorbed RPI factors in
the Wilson coefficients $C_\ell(\omega_i)$ when carrying out the expansion.

The tree level matching from QCD to SCET for three jets comes from matching two
Feynman diagrams in QCD onto the operator basis in Eq.~(\ref{J3jethomo}), and is
done at the hard scale $\mu=Q$. This gives
\begin{align} \label{Cimatch3jet}
 C_1 &= C_6 = \frac{-2}{n_1\mcdot n_3\, \omega_1\omega_3}\,,
 & C_2 & = -C_5 =\frac{-2}{n_2\mcdot n_3\, \omega_2\omega_3}\,,
 & C_{3,4} &= C_{7-11} =0\, .
\end{align}
The results for these Wilson coefficients are invariant under type-III RPI as
expected.

The above   matching  computation can  be   compared with   the  tree level SCET
computations for parton showers  in Ref.~\cite{Bauer:2006mk}, where three  final
state jets are considered. To compare the calculations we take the two stages of
matching of Ref.~\cite{Bauer:2006mk} both at  $\mu=Q$, and we split the
operators in Eqs.~(27,28)  of  Ref.~\cite{Bauer:2006mk} into  two parts,  ${\cal
  O}_3={\cal O}_{3a}+ {\cal O}_{3b}$  and ${\cal O}_3^{(2)} ={\cal O}_{3a}^{(2)}
+ {\cal   O}_{3b}^{(2)}$.  The matching computation  of Ref.~\cite{Bauer:2006mk}
used a frame $q_{n_2\perp}=0$ for ${\cal O}_{3a}$  and ${\cal O}_{3a}^{(2)}$ and
a frame $q_{n_1\perp}=0$  for  ${\cal O}_{3b}$  and  ${\cal O}_{3b}^{(2)}$. With
these frame choices, we confirm that $C_1 {\cal J}_1 + C_6 {\cal J}_6 = O_{3a} +
O_{3a}^{(2)}$ and  $C_2 {\cal J}_2  + C_5 {\cal  J}_5  = O_{3b} + O_{3b}^{(2)}$,
providing a cross-check on the results in Eq.~(\ref{Cimatch3jet}).

\subsection{Two Jets from Gluon Fusion: $g g \rightarrow q \bar{q}$ operators}
\label{sec:gfusion}

Next we consider the example of the production of two quark jets from gluon
fusion, which is relevant for the LHC. In this application we will see that RPI
substantially constrains the number and structure of operators. This basis of
operators have not yet been constructed. The factorization theorem for $pp\to
2\,{\rm jets}$ has been discussed in Ref.~\cite{Kidonakis:1998bk}, and were also
considered recently in Ref.~\cite{Bauer:2008jx} using SCET. SCET has also been
used to resum electroweak Sudakov logarithms by solving RGE equations for four
quark collinear operators in Refs.~\cite{Chiu:2007dg,Chiu:2007yn,Chiu:2008vv},
and to consider Higgs production from $pp$ collisions~\cite{Ahrens:2008qu}.

We consider the incoming gluons to be collinear in different directions, which is
appropriate for the high energy collision of energetic protons at the LHC, and we
assume that the final state jets have a large perpendicular momentum relative to
the beam axis. Hence the final jets are described by two additional collinear
directions, making four in total. Unlike our previous examples, here there is
not an external $q^\mu$ vector, the hard interaction takes place entirely
between strongly interacting particles. Hence this is an example of the case ii)
discussed above Eq.~(\ref{deltaw}).

Similarly to the three jets case, it is convenient to directly write the RPI
basis without first writing the homogeneous basis, because the presence of four
collinear directions imply that there are a large number of homogeneous
operators, many of which are restricted by RPI. Due to the absence of an
external hard vector $q^\mu$ in this process, in the definition of the currents
we make use the RPI delta function factors of Eq.~(\ref{DEL}),
$\hat\Delta_{km}$. The general formula for matching the RPI operators onto
homogeneous operators is
\begin{align} \label{Jsum2}
  i\!\!\!\! \sum_{n_1,n_2,n_3, n_4}\!\! \sum_\ell \int\! [\prod_{i,j} d\hat{\omega}_{ij}] 
   \: \hat{C}_\ell(\hat{\omega}_i)\:  [\prod_{km} \hat\Delta_{km}]
   \textbf{Q}_\ell
  & = i\!\!\!\! \sum_{n_1,n_2,n_3, n_4} \!\! \sum_\ell \int\! [\prod_i d\omega_i]
   \: C_\ell(\omega_i)\:  [O_\ell(\omega_i)]_{g g \rightarrow q \bar{q}} \nn\\
  & +
   \ldots \,,
\end{align}
where we use the same manipulations needed to get Eq.~(\ref{AAA}).  Note that
here we have divided the RPI operators into the $\delta$-functions in $\hat
\Delta_{km}$ which depend on $\hat\omega_{km}$, and the remainder of the
operator $\textbf{Q}_\ell$ that does not.  The starting point for building a
basis for $\textbf{Q}_\ell$ is the object $\bar{\Psi}_{n_1}
\mathcal{G}_{n_3}^{\mu\nu} \mathcal{G}_{n_4}^{\alpha\beta} \Psi_{n_2}$.  We
assume a $LL+RR$ chirality for the quarks which is suitable when strong
interactions produce massless quarks, and hence include either $\gamma^\lambda$
or $\gamma^\lambda \gamma^\sigma\gamma^\tau$. Since the overall operator is a
scalar, all the vector indices on the field strengths and on the Dirac structure
must be contracted with $g_{\mu\nu}$'s or $i\partial_{n_i}^\mu$'s. We can use
the equations of motion and Bianchi identity in Eqs.~(\ref{EOM1},\ref{EOM2},\ref{BI}) to
eliminate terms with $i\partial\!\!\!\slash_{n_i}$ in any operator, and terms with
$\partial_{n_3 \mu} \mathcal{G}_{n_3}^{\mu\nu} $ or $\partial_{n_4 \mu}
\mathcal{G}_{n_4}^{\mu\nu} $. In addition, momentum conservation implies
$i\partial_{n_1}^\mu + i\partial_{n_2}^\mu + i\partial_{n_3}^\mu +
i\partial_{n_4}^\mu =0$, and we will use this to eliminate all operators with an
$i\partial_{n_1}$.  This leaves twenty operators for the RPI basis
\begin{align} \label{ggqq}
   \textbf{Q}_1 & =    \bar{\Psi}_{n_1} 
  \gamma_\beta g_{\nu \alpha}
  \, ig\mathcal{G}_{n_3}^{\mu\nu}  i\partial_{n_4\mu}
  ig\mathcal{G}_{n_4}^{\alpha\beta} 
   \Psi_{n_2} \, ,
%
&  \textbf{Q}_2 & =    \bar{\Psi}_{n_1} 
  \gamma_\mu g_{\nu\alpha}   i \partial_{n_3\beta} 
  \,ig\mathcal{G}_{n_3}^{\mu\nu}   
  ig\mathcal{G}_{n_4}^{\alpha\beta} 
 \Psi_{n_2} \, , 
 \nn \\[4pt]
  \textbf{Q}_3 & =    \bar{\Psi}_{n_1} 
  \gamma_\mu g_{\nu\alpha} 
  \,ig\mathcal{G}_{n_3}^{\mu\nu}   
  ig\mathcal{G}_{n_4}^{\alpha\beta} 
  i \partial_{n_2\beta}  \Psi_{n_2} \, , 
%
&   \textbf{Q}_4 & =    \bar{\Psi}_{n_1} 
  \gamma_\beta g_{\nu \alpha}
  \, ig\mathcal{G}_{n_3}^{\mu\nu}
  ig\mathcal{G}_{n_4}^{\alpha\beta} 
  i\partial_{n_2\mu}  \Psi_{n_2} \, ,
  \nn\\[4pt]
   \textbf{Q}_5 & =  \bar{\Psi}_{n_1} 
 \gamma_\nu \gamma_\alpha \gamma_\beta
  \, ig\mathcal{G}_{n_3}^{\mu\nu}     i\partial_{n_4\mu}
  ig\mathcal{G}_{n_4}^{\alpha\beta}
   \Psi_{n_2} \, , 
%
& \textbf{Q}_6 & =   \bar{\Psi}_{n_1} 
 \gamma_\mu \gamma_\nu \gamma_\alpha    i\partial_{n_3\beta}
 \,ig \mathcal{G}_{n_3}^{\mu\nu}
 ig \mathcal{G}_{n_4}^{\alpha\beta}  
\Psi_{n_2} \, , \nn\\[4pt]
 \textbf{Q}_7 & =   \bar{\Psi}_{n_1} 
 \gamma_\mu \gamma_\nu \gamma_\alpha
 \,ig \mathcal{G}_{n_3}^{\mu\nu}
 ig \mathcal{G}_{n_4}^{\alpha\beta}  
   i\partial_{n_2\beta} \Psi_{n_2} \, ,
%
&   \textbf{Q}_8 & =  \bar{\Psi}_{n_1} 
 \gamma_\nu \gamma_\alpha \gamma_\beta
  \, ig\mathcal{G}_{n_3}^{\mu\nu}   
  ig\mathcal{G}_{n_4}^{\alpha\beta}
   i\partial_{n_2\mu}  \Psi_{n_2} \,,
 \nn\\[4pt]
  \textbf{Q}_{9} & =  \bar{\Psi}_{n_1} 
  \gamma_\alpha 
  \, ig\mathcal{G}_{n_3}^{\mu\nu}   
  i\partial_{n_4\mu} ig\mathcal{G}_{n_4}^{\alpha\beta}
  i\partial_{n_2\nu} i\partial_{n_2\beta}    \Psi_{n_2}   
\,,
  \nn\\[4pt]
  \textbf{Q}_{10} & =  \bar{\Psi}_{n_1} 
  \gamma_\mu i\partial_{n_3\alpha}
  \, ig\mathcal{G}_{n_3}^{\mu\nu}   
  ig\mathcal{G}_{n_4}^{\alpha\beta}
  i\partial_{n_2\nu} i\partial_{n_2\beta}   \Psi_{n_2} \, .
\end{align}
The other ten operators $\textbf{Q}_{11-20}$ have the same structure as
Eq.~(\ref{ggqq}) but with a trace over color for the gluon operators, for
example $ \textbf{Q}_{11} = \bar{\Psi}_{n_1} \gamma_\beta g_{\nu \alpha} {\rm
  Tr}[ \, ig\mathcal{G}_{n_3}^{\mu\nu} i\partial_{n_4\mu}
ig\mathcal{G}_{n_4}^{\alpha\beta} ] \Psi_{n_2} $.  Note that $\textbf{Q}_{1-10}$
have ${\cal G}_{n_3}$ to the left of ${\cal G}_{n_4}$, so one might think that
there are ten more operators with the ${\cal G}$'s in the other order.  However,
in Eq.~(\ref{Jsum2}) we sum over $n_{3,4}$ and integrate over $d\hat\omega_3
d\hat\omega_4$, and hence include operators obtained from the interchange
$n_3\leftrightarrow n_4$, $\hat\omega_3\leftrightarrow \hat{\omega}_4$. Recall
that the directions $n_i$ are only determined by the matrix elements. So if we
consider a matrix element with gluons in the $n$ and $n'$ direction then there
is a contribution from $n_3\!=\!n$, $n_4\!=\!n'$, and from $n_3\!=\!n'$,
$n_4\!=\!n$. Other possible operators might be $ \bar{\Psi}_{n_1} \gamma_\alpha
\,i\partial_{n_3\beta} ig\mathcal{G}_{n_3}^{\mu\nu} i\partial_{n_4\mu}
ig\mathcal{G}_{n_4}^{\alpha\beta} i\partial_{n_2\nu} \Psi_{n_2} $,
$\bar{\Psi}_{n_1} \gamma_\mu i\partial_{n_3\alpha} \,
ig\mathcal{G}_{n_3}^{\mu\nu} i\partial_{n_4\nu}
ig\mathcal{G}_{n_4}^{\alpha\beta} i\partial_{n_2\beta} \Psi_{n_2}$ and similarly
with the trace. We can use the Bianchi identity~(\ref{BI}) to rule them out. For
example, in the first operator we have implicitly already used the Bianchi
identity for the $i\partial_{n_4\mu} ig\mathcal{G}_{n_4}^{\alpha\beta}$ term because
we did not write operators with $i\partial\!\!\!\slash_{n_4}
ig\mathcal{G}_{n_4}^{\alpha\beta}$. But we can apply the Bianchi identity  to
$i\partial_{n_3\beta} ig\mathcal{G}_{n_3}^{\mu\nu}$, that is not connected with
$\gamma$'s. In this way we can write this operator in terms of $ \textbf{Q}_{1}
$, $ \textbf{Q}_{4} $ and operators with three gluon fields. Note that we do not
need to consider operators with $i\partial_{n} \cdot i\partial_{n'}$ since all these
contracted derivatives are contained in the $\hat\Delta_{km}$'s.

A natural frame for analyzing $gg\to q\bar q$ is the CM frame with the choices
$\bn_1\!=\!n_2$, $\bn_2\!=\!n_1$, $\bn_3\!=\!n_4$, $\bn_4\!=\!n_3$. We expand
the currents (\ref{ggqq}) with an eye towards using them in this frame. Actually,
only the condition $\bn_3\!=\!n_4$, $\bn_4\!=\!n_3$ is necessary to find the
following operators
\begin{align} \label{Jhomoggqq}
   O_1 & = \omega_4 \,
  \bar{\chi}_{n_1 ,\omega_1}     n\!\!\!\slash_4\,    (g  \mathcal{B}^\mu_{n_3\perp}
     )_{\omega_3}  
    (g  \mathcal{B}^\perp_{ n_4 \mu} )_{\omega_4} \chi_{n_2,\omega_2} 
   \, , \\[4pt]
   O_2 & = \omega_3 \,
   \bar{\chi}_{n_1 ,\omega_1}  n\!\!\!\slash_3\,  (g  \mathcal{B}^\mu_{n_3\perp}
     )_{\omega_3} 
     (g  \mathcal{B}^\perp_{ n_4 \mu} )_{\omega_4} \chi_{n_2,\omega_2} 
   \, , \nonumber \\[4pt]
   O_3 & =  \omega_2\,
    \bar{\chi}_{n_1 ,\omega_1}  (g\mathcal{B}\!\!\!\slash_{n_3\perp})_{\omega_3}  
   (g  n_2 \mcdot  \mathcal{B}^\perp_{n_4} )_{\omega_4}  \chi_{n_2,\omega_2}  
    \, , \nonumber \\[4pt]
  O_4 & =   \omega_2\, 
   \bar{\chi}_{n_1 ,\omega_1}    (g  n_2 \mcdot  \mathcal{B}^\perp_{n_3} )_{\omega_3}
    (g\mathcal{B}\!\!\!\slash_{n_4\perp} )_{\omega_4}  \chi_{n_2,\omega_2}    
   \, , \nonumber \\[4pt]
     O_5 & = \omega_4\,
    \bar{\chi}_{n_1 ,\omega_1} n\!\!\!\slash_4 (g\mathcal{B}\!\!\!\slash_{n_3\perp} )_{\omega_3} 
     (g\mathcal{B}\!\!\!\slash_{n_4\perp} )_{\omega_4}  \chi_{n_2,\omega_2} 
   \, ,\nonumber \\[4pt]
    O_6 & = \omega_3\, 
   \bar{\chi}_{n_1 ,\omega_1}  n\!\!\!\slash_3 (g\mathcal{B}\!\!\!\slash_{n_3\perp} )_{\omega_3}  
   (g\mathcal{B}\!\!\!\slash_{n_4\perp} )_{\omega_4}  \chi_{n_2,\omega_2}  
   \, , \nonumber \\[4pt]
    O_7 & = \omega_2\omega_3\omega_4\,
   \bar{\chi}_{n_1 ,\omega_1}    n\!\!\!\slash_3\,    n\!\!\!\slash_4\,
    (g\mathcal{B}\!\!\!\slash_{n_3\perp} )_{\omega_3}  
    (g  n_2 \mcdot \mathcal{B}^\perp_{ n_4})_{\omega_4}   \chi_{n_2,\omega_2}
    \, , \nonumber \\[4pt]
  O_8 & =\omega_2\omega_3\omega_4\,
   \bar{\chi}_{n_1 ,\omega_1}  n\!\!\!\slash_3\,   n\!\!\!\slash_4 
    (g  n_2 \mcdot \mathcal{B}^\perp_{ n_3} )_{\omega_3}
     (g\mathcal{B}\!\!\!\slash_{n_4\perp} )_{\omega_4} \chi_{n_2,\omega_2}
   \, ,  \nonumber \\[4pt]
     O_9 & = (\omega_2)^2\omega_4\,
     \bar{\chi}_{n_1 ,\omega_1}  n\!\!\!\slash_4\,  (g  n_2 \mcdot \mathcal{B}^\perp_{ n_3} )_{\omega_3} 
     (g  n_2 \mcdot \mathcal{B}^\perp_{ n_4} )_{\omega_4}
        \chi_{n_2,\omega_2}
\, , \nonumber \\[4pt]
    O_{10} & = (\omega_2)^2\omega_3\,
      \bar{\chi}_{n_1 ,\omega_1}
    n\!\!\!\slash_3\,  
  (g  n_2 \mcdot \mathcal{B}^\perp_{ n_3} )_{\omega_3} 
  ( g  n_2 \mcdot \mathcal{B}^\perp_{ n_4} )_{\omega_4} 
   \chi_{n_2,\omega_2}
  \, . \nonumber
\end{align}
$ O_{11-20}$ have the same structure of (\ref{Jhomoggqq}) but with a trace over
color of the two gluon operators.  $ O_i$ is given by the expansion of $
\textbf{Q}_i$ for $i\!=\!1,2,5,6$, by the expansion of a suitable linear
combination of $ \textbf{Q}_i$ and $ \textbf{Q}_{i-1}$ for $i\!=\!3,6$, and of $
\textbf{Q}_i$ and $ \textbf{Q}_{i-3}$ for $i\!=\!4,8$. $ O_{9/10}$ are given by
the expansion of a suitable linear combination of $ \textbf{Q}_{9/10}$, $
\textbf{Q}_{1/2}$ and $ \textbf{Q}_{4/3}$. In some cases we have absorbed
reparametrization invariant prefactors that appear in the expansion into the
Wilson coefficients $C_l(\omega_i)$.  By using momentum conservation it is
possible to reduce these ten operators to just four independent operators at
leading order in SCET.\footnote{We thank W.~Waalewijn for his explicit
  derivation of this point.}

It is straightforward to carry out the matching from QCD onto the SCET operators
in Eq.~(\ref{Jhomoggqq}). At tree level there are three Feynman diagrams. The
amplitude squared is also known analytically at one-loop~\cite{Ellis:1985er},
and a full matching computation at this order involves regulating infrared
singularities in the same way for the loops in QCD and SCET before subtracting.
The only point to be careful about is the sum over the $n_i$'s in
Eq.~(\ref{Jsum2}), since definite values for these $n_i$'s should be determined
by the states. For example, if we consider the tree level $gg\to q\bar q$ matrix
element of $O_1$ with perpendicular polarization for the gluons then
\begin{align} \label{Meltggqq}
 &  \Big\langle q_{n_1'}(p_1') \bar q_{n_2'}(p_2) \Big| \sum_{n_i} \int
  [d\omega_j] i C_1(\omega_1,\omega_3,\omega_4,\omega_2) O_1(\omega_j) 
  \Big| g^\perp_{n_3'}(p_3') g^\perp_{n_4'}(p_4') \Big\rangle \nn\\
 &= ig^2  C_1(\omega_1',\omega_3',\omega_4',\omega_2') \omega_4'\, \big[\epsilon_{ n_3'\perp
  }^{\mu A} \epsilon^B_{n_4'\perp\mu}\big] \big[\bar u_{n_1'} n'\!\!\!\!\slash_4 T^A T^B
 u_{n_2'}\big] \nn\\
 & \quad +ig^2  C_1(\omega_1',\omega_4',\omega_3',\omega_2') \omega_3'\, \big[\epsilon_{n_3'\perp
   }^{\mu A} \epsilon^B_{n_4'\perp\mu}\big] \big[ \bar u_{n_1'} n'\!\!\!\!\slash_3 T^B T^A
 u_{n_2'} \big]
  \,.
\end{align}
The two terms come from the cases $n_{3,4}=n_{3,4}'$ and $n_{3,4}=n_{4,3}'$
respectively. Therefore to determine the $C_\ell$'s it suffices to compute terms
contributing to the color structure $T^A T^B$ in QCD, which at tree level gives
\begin{align}
 C_1&= \frac{-1}{(n_3\mcdot n_4) \omega_3\omega_4} \, ,
&C_2&= \frac{1}{(n_3\mcdot n_4) \omega_3\omega_4}  \nonumber \, ,
& C_3&= \frac{2 }{(n_2 \mcdot n_4)\, \omega_2\omega_4} \, , 
%
& C_5&= \frac{1}{(n_2 \mcdot n_4)\, \omega_2\omega_4}  
 \,  ,\nonumber \\[4pt]
C_{4} &=C_{6-20} =0 \, .
\end{align}
Note that the results for the $C_\ell$'s are invariant under type-III RPI
transformations as expected, and that in the frame used for our computation
$n_3\cdot n_4=2$. We have confirmed that a consistent result is obtained by
considering the $T^BT^A$ terms. Eq.~(\ref{Meltggqq}) expresses the interesting
fact that with distinct collinear directions for all final state particles, only
the color ordered QCD amplitudes are needed for the matching which determines
the SCET Wilson coefficients.

\section{Conclusion} \label{sect:conclusion}

In SCET the momenta of collinear particles are decomposed with light-like
vectors $n^\mu$ and $\bar{n}^\mu$, where $\vec n$ is close to the direction of
motion. The vectors $n^\mu$ and $\bn^\mu$ are required to define collinear
operators that have a definite order in the power counting. However, there is a
freedom in defining $n$ and $\bn$, which leads to reparametrization constraints.
The decomposition of operators in the theory must satisfy these constraints in
order to be consistent.  This reparametrization invariance gives nontrivial
relations among the Wilson coefficients of collinear operators occurring at
different orders in the power counting, and for situations with multiple
collinear directions gives constraints on the form of operators making up a
complete basis.

In this paper we have constructed objects that are invariant under both
reparametrization transformations and collinear gauge transformations, a
superfields $\Psi_{n_i}$ for fermions and a superfield
$\mathcal{G}^{\mu\nu}_{n_i}$ for gluons. Here the subscript $n_i$ denotes an
equivalence class of light-like vectors under RPI.  The superfields are
invariant under collinear gauge transformations through a reparametrization
invariant Wilson line $\mathcal{W}_{n_i}$ that is the generalization of the
usual $W_{n_i}$. We constructed RPI operators out of these superfields by
introducing reparametrization invariant $\delta$-functions. The
$\delta$-functions act on the RPI operators to pick out large momenta, and are
convoluted with hard Wilson coefficients that must be computed by matching
computations.  The power of the RPI operators is that they encode information
about the minimal basis of Wilson coefficients. However, they do not have a
definite power counting order. By expanding them in $\lambda$ one obtains a
minimal basis of operators with a good power counting, where all constraints on
the Wilson coefficients are made explicit. The final basis of operators with a
good power counting involves a two-component field $\chi_{n_i}$ for quarks, a
field $\cB_{n_i\perp}^\mu$ for the two physical gluon polarizations,
derivatives $\cP_{n_i\perp}^\mu$, and delta functions
$\delta(\omega-\bnP_{n_i})$ that pick out the large momenta of these collinear
fields. That is it. Other field components such as $n_i\mcdot \cB_{n_i}$, and
other derivatives such as $in_i\cdot \partial_{n_i}$, are eliminated from the
purely collinear operator basis using the equations of motion.

This procedure was applied to several processes. We studied spin-averaged DIS
for quarks at twist-4, as a means of testing our setup in a framework where the
power suppressed basis of operators is well understood. We then constructed a
minimum basis of pure glue operators for DIS at twist-4.  These applications
involve a single collinear direction. For processes with multiple collinear
directions we considered operator bases for jet production.  Useful constraints
from RPI were not found for the first power suppressed operators in $e^+e^- \to
2\,{\rm jets}$. On the other hand, already at leading order in the power
counting, RPI provided important constraints on the complete basis of operators
for $e^+e^- \to 3\,{\rm jets}$ with three distinct collinear directions. RPI was
also very useful in constructing a complete basis of operators for gluon fusion
producing two quark initiated jets, where there are four collinear directions.
In this case the process of interest is $pp\to 2\,{\rm jets}$, which will be
studied at the LHC. We expect the complete bases of operators constructed here
will be a useful ingredient in the study of factorization theorems for this
process. The steps we used to construct complete basis will also be useful when
considering factorization for processes with more jets in the final state.  In
general we found that RPI becomes more powerful for processes involving more
jets, essentially because the number of vectors $n_i$ and $\bn_i$ proliferates
faster than the number of objects that must be considered to build the RPI
basis.

An interesting observation discussed in section~\ref{sec:gfusion} is that when
matching from QCD onto SCET operators describing multiple collinear directions
$n_i$, the Wilson coefficient is determined by the color ordered QCD amplitude.
Since results for multi-leg QCD amplitudes are often expressed in a color
ordered form, this should simplify the matching of QCD amplitudes onto SCET.

This work was supported in part by the Director, Office of Science, Office of
Nuclear Physics of the U.S.\ Department of Energy under the Contract
DE-FG02-94ER40818. IS was also supported in part by the DOE OJI program and by
the Sloan Foundation.


\appendix

\section{Invariance to the choice of hard-vector $q^\mu$}
\label{app_qqp}

From the construction in section~\ref{sect:RPIGI}, a natural question arises
about the special role of $q^\mu$ in Eq.~(\ref{DEL}). What happens if there is
more than one possible choice for $q^\mu$ in a given process? Say we have a
$q^\mu$ and a $q^{\prime \mu}$ with Wilson coefficients that can depend on
$q^2$, $q^{\prime 2}$, and $q\cdot q'$, where $q_\perp\sim q'_\perp\sim
\lambda$.  It turns out that in this case any linear combination of $q^\mu$ and
$q^{\prime\mu}$ in Eq.~(\ref{DEL}) is equally good, and is equivalent to any
other choice. Hence, one choice suffices. To prove this we consider the
expansion of the reparametrization invariant variable
\begin{align}
  \xi \equiv \frac{2q\mcdot q'}{q^2} 
    \pm \sqrt{ \Big(\frac{2q\mcdot q'}{q^2}\Big)^2 - \frac{4q^{\prime 2}}{q^2} }
    \ = \frac{n\cdot q'}{n\cdot q} + {\cal O}(q_\perp^2)
   \,,
\end{align}
where we take the plus sign if the expansion is done with $n\cdot q'/n\cdot q >
\bn\cdot q'/\bn\cdot q$ and the minus sign otherwise.  Now use this variable to
define
\begin{align}
 q' \cdot i\partial -\xi \, (q\cdot i\partial) \equiv \hat Q^{[2]}_{\rm INV} \,,
\end{align}
where the operator $\hat Q^{[2]}_{\rm INV}$ is RPI and its expansion starts at
order $\lambda^2$. Thus
\begin{align} \label{qqswap}
  &\int d\omega\:  C(\omega)\: \delta(\omega - q'\cdot i\partial) 
   = \int d\omega\: C(\omega)\: 
   \delta\big(\omega -\xi\: q\cdot i\partial - \hat Q^{[2]}_{\rm INV} \big) 
  \nn\\[5pt]
  &\quad = \int d\omega' \: C(\xi \omega')\: \delta(\omega'-q\cdot i\partial 
    - \hat Q^{[2]}_{\rm INV}/\xi \big) 
    \nn\\
  &\quad = \int d\omega' \:\Big[ \tilde C(\omega')\: \delta(\omega' - q\cdot
  i\partial) + \tilde B(\omega') \: \hat Q_{\rm INV}^{[2]} \:
  \delta^\prime (\omega'-q\cdot i\partial) +\ldots \Big]
  \,,
\end{align}
where in the second line we changed the dummy variable to $\omega'=\xi\omega$,
In the last line both terms are RPI, and the ellipsis denotes higher order terms
which are also RPI order by order in $\lambda$. Eq.~(\ref{qqswap}) demonstrates
that we can swap the parameter $q'\to q$ in the $\delta$-function, since the
change is compensated by a change of notation in the leading order Wilson
coefficient $C\to \tilde C$. Given that we imagine starting with a complete
basis of RPI operators built with $\delta(\omega-q'\mcdot i\partial)$ or with
$\delta(\omega'-q\mcdot i\partial)$, the higher terms in the series in
Eq.~(\ref{qqswap}), like $\tilde B$, also simply change a Wilson coefficient in
our basis.  Thus, the choice of $q$ or $q'$ in the $\delta$-function just
corresponds to a different choice of the basis for the invariant operators, and
one choice suffices.

\bibliographystyle{iain}
\bibliography{masterbib}

\end{document}